\documentclass[english,aps,prl,reprint]{revtex4-2}
\usepackage[T1]{fontenc}
\usepackage[latin9]{inputenc}
\usepackage{mathrsfs}
\usepackage{bm}
\usepackage{amsmath}
\usepackage{amssymb}
\usepackage{graphicx}

\makeatletter
\usepackage[all]{xy}
\usepackage{variables,new_variables,mathtools}
\makeatother

\usepackage{babel}
\begin{document}
\title{Foundations of the Variational Discrete Action Theory}
\author{Zhengqian Cheng and Chris A. Marianetti}
\affiliation{Department of Applied Physics and Applied Mathematics, Columbia University,
New York, NY 10027}
\date{\today}
\begin{abstract}
Variational wave functions and Green's functions are two important
paradigms for solving quantum Hamiltonians, each having their own
advantages. Here we detail the Variational Discrete Action Theory
(VDAT), which exploits the advantages of both paradigms in order to
approximately solve the ground state of quantum Hamiltonians. VDAT
consists of two central components: the sequential product density
matrix (SPD) ansatz and a \textit{discrete action} associated with
the SPD. The SPD is a variational ansatz inspired by the Trotter decomposition
and characterized by an integer $\discn$, recovering many well known
variational wave functions, in addition to the exact solution for
$\discn=\infty$. The discrete action describes all dynamical information
of an effective \textit{integer time} evolution with respect to the
SPD. We generalize the path integral to our integer time formalism,
which converts a dynamic correlation function in integer time to a
static correlation function in a compound space. We also generalize
the usual many-body Green's function formalism to integer time, which
results in analogous but distinct mathematical structures, yielding
integer time versions of the generating functional, Dyson equation,
and Bethe-Salpeter equation. We prove that the SPD can be exactly
evaluated in the multi-band Anderson impurity model (AIM) by summing
a finite number of diagrams. For the multi-band Hubbard model, we
prove that the self-consistent canonical discrete action approximation
(SCDA), which is the integer time analogue of the dynamical mean-field
theory, exactly evaluates the SPD for d=$\infty$. VDAT within the
SCDA provides an efficient yet reliable method for capturing the local
physics of quantum lattice models, which will have broad applications
for strongly correlated electron materials. More generally, VDAT should
find applications in various many-body problems in physics.
\end{abstract}
\maketitle

\section{Introduction}

The quantum many-body problem represents a forefront in most areas
of physics, and determining the ground state of the Hamiltonian is
a primary objective. Variational wave functions are a key paradigm
for solving the ground state of a Hamiltonian. Simple variational
approaches such as the Hartree-Fock approximation provide a baseline
for understanding many-body systems at a modest computational cost.
There are many more sophisticated variational wave functions, such
as the Jastrow wave function\cite{Jastrow19551479,Capello2005026406}
or unitary coupled cluster\cite{Bartlett1989133,Kutzelnigg1991349,Taube20063393},
but most approaches do not have a natural mechanism for trading off
between accuracy and computational cost, which will be a key idea
addressed in this paper. 

Another viewpoint for addressing the many-body problem is to start
from the formally exact density matrix and perform the Trotter-Suzuki
decomposition\cite{Trotter1959545,Suzuki1976183,Suzuki1993432}, yielding
the Euclidean path integral, which may then be approximately evaluated
using quantum Monte-Carlo or using a diagrammatic approach. A prominent
example of the former is auxiliary field quantum Monte-Carlo (AFQMC)\cite{Hubbard195977,Blankenbecler19812278},
which requires a relatively fine discretization of imaginary time
in order to achieve converged ground state properties. If one is only
seeking ground state properties, no dynamical information needs to
be extracted from the Green's function, which motivates the possibility
of obtaining highly precise ground state properties from an extremely
coarse discretization of imaginary time. This suggestion can be realized
by creating a variational density matrix ansatz based upon the Trotter-Suzuki
decomposition, and in this paper we propose the sequential product
density matrix (SPD). Given that the SPD is inspired by the Trotter-Suzuki
decomposition, it is naturally characterized by an integer $\discn$,
which controls the trade-off between accuracy and computational complexity. 

The SPD is an extremely generic ansatz which can recover a large number
of well known variational wave functions. An example at $\discn=1$
is the Hartree-Fock approximation, while examples at $\discn=2$ include
the Gutzwiller wave function\cite{Gutzwiller1963159,Gutzwiller1964923,Gutzwiller19651726}
and the Jastrow wave function\cite{Jastrow19551479,Capello2005026406}
(see Subsection \ref{subsec:Categorizing-Existing-WF-w-SPD} for a
complete list). Therefore, we already begin with decades of intuition
for the efficacy of this ansatz at small $\discn$, and one can easily
envision the potency of larger $\discn$. Many formal constructions
are useful for organizing ideas, but the question is whether or not
they can be evaluated in practice. A key development in this paper
is demonstrating that our proposed integer time Green's function and
its corresponding discrete action theory provide a rich formalism
for systematically evaluating observables under the SPD ansatz. Most
practically, this yields new theories that have not yet been discovered
and can be implemented in practice. More generally, this discrete
action theory provides a new way to think about variational wave functions
in the context of Green's function. The discrete action theory naturally
gives rise to its own version of the path integral, generating functional,
Dyson equation, and Bethe-Salpeter equation. Therefore, many of the
key ideas from Green's functions may be generalized to the discrete
action theory. The discrete action theory can then be used along with
the variational principle to give rise to the variational discrete
action theory (VDAT). Furthermore, the discrete action theory provides
many different avenues for precisely evaluating the SPD at a given
set of variational parameters. 

A decisive milestone of this paper is proving that a certain type
of SPD can be exactly evaluated in infinite dimensions, implying that
we can achieve a strict upper bound on the ground state energy in
this case. This result extends the well known fact that the Gutzwiller
wave function is exactly evaluated by the Gutzwiller approximation
for the Hubbard model in $d=\infty$\cite{Metzner1987121,Metzner19887382,Metzner1989324,Bunemann19977343},
proving that the Gutwziller-Baeriswyl\cite{Otsuka19921645} and Baeriswyl-Gutzwiller\cite{Dzierzawa19951993}
wave functions can be exactly evaluated in $d=\infty$, in addition
to an infinite number of more precise generalizations. As a result,
VDAT is a potent theory for efficiently and precisely evaluating the
Hubbard model in $d=\infty$. Indeed, we have demonstrated that VDAT
achieves highly precise results in the $d=\infty$ Hubbard model for
$\discn=3$\cite{Cheng2020short}. The VDAT method can immediately
be understood as a practically important tool given our knowledge
of the dynamical mean-field theory (DMFT)\cite{Georges199613,Kotliar200453,Vollhardt20121},
which allows for the numerically exact solution of the Hubbard model
in infinite dimensions. DMFT is the de facto standard for capturing
local physics in models of strongly correlated electrons, and plays
a key role in describing realistic strongly correlated materials in
the context of DFT+DMFT\cite{Kotliar2006865}. Given that VDAT precisely
captures the physics of infinite dimensions at a tiny fraction of
the cost of DMFT, VDAT might be transformational as an efficient replacement
for DMFT in the context of ground state properties. A DFT+VDAT theory
might finally yield a first-principles approach of strongly correlated
electron materials with a computational overhead not far beyond DFT
itself, yet contain all of the physics of DFT+DMFT, and more. 

In this study, we restrict ourselves to static observables at zero
temperature. Given that VDAT is a variational theory with an explicit
density matrix ansatz, one can also naturally study static observables
at finite temperatures, though doing so requires the direct evaluation
of the entropy of the SPD. Evaluating the entropy is highly nontrivial,
even in the relatively simple case of $\discn=2$ or the Gutzwiller
wave function\cite{Wang2010125105,Sandri2013205108,Lanata2015081108}.
Therefore, extending the VDAT formalism to finite temperatures will
be pursued in future work. Another important direction for VDAT would
be to study excited states, which are not naturally captured in a
variational theory. However, one can straightforwardly apply the Landau-Gutzwiller
quasiparticle approach\cite{Bunemann2003075103}, which would clearly
result in a Fermi liquid picture which can go beyond Gutzwiller. More
generally, there is a possibility that the incoherent part of the
spectrum may be recovered with further extensions given that VDAT
with $\discn>2$ precisely captures the insulating phase in the single-band
Hubbard model\cite{Cheng2020short}. 

The structure of this manuscript is as follows. In Section \ref{subsec:Variational-Theory-and-SPD},
we begin by motivating and introducing a generic SPD, and introduce
three important classes of SPD which are useful for prominent models
of interacting electrons. Additionally, we apply the SPD to the Hubbard
plaquette, where it can be exactly evaluated, to illustrate the convergence
of the SPD with respect to $\discn$. In Section \ref{sec:Integer-Time-Green's},
we introduce the notion of the integer correlation function, and demonstrate
how it can be evaluated using the integer time Wick's theorem. A pedagogical
example is given for the Anderson impurity model containing a single
bath site. In Section \ref{sec:The-Discrete-Action-Theory}, we introduce
the notion of a discrete action, and generalize the standard tools
of many-body physics to the integer time case, including the integer
time path integral, the discrete generating function, the discrete
Dyson equation, and the discrete Bethe-Salpeter equation. In Section
\ref{sec:Categorizing-Discrete-Action}, we introduce the canonical
discrete action, and use it to evaluate the SPD that is associated
with the Anderson impurity model. In Section \ref{sec:Self-consistent-Canonical-Discre},
we introduce the self-consistent canonical discrete action approximation
(SCDA), and we prove that it exactly evaluates the SPD-d in infinite
dimensions. Furthermore, in the case of $\discn=2$ we pedagogically
illustrate how the SCDA recovers the Gutzwiller approximation, and
we derive basis-independent, rotationally invariant Gutzwiller equations
for the multi-band Hubbard model. In Section \ref{sec:Parameterizing-the-SPD},
we discuss the general workflow of performing a VDAT calculation.
Finally, we provide an appendix which illustrates the Lie group properties
of the non-interacting density matrix (see Subsection \ref{subsec:Appendix-Lie-group}),
in addition to a second appendix which proves the integer time Wick's
theorem (see Subsection \ref{appendix:wicks_theorem}). Additionally,
we provide supplementary information which illustrates the evaluation
of the CDA for the case of a single orbital and $\discn=3$\cite{supplementary}.
It should be noted that there is short companion manuscript for this
paper, which highlights the basic aspects of the VDAT while presenting
key results on the Anderson impurity model and the $d=\infty$ Hubbard
model\cite{Cheng2020short}.

\section{Sequential Product Density Matrix (SPD)\label{subsec:Variational-Theory-and-SPD}}

\subsection{Motivating the SPD}

Here we motivate the notion of the sequential product density matrix
(SPD), which is the variational ansatz for the VDAT. First, let us
begin by recalling the variational principle at zero temperature.
Given some Hamiltonian $\hat{H}$, the ground state energy is obtained
by the constrained search over the density matrix ansatz as

\begin{equation}
\mathcal{E}=\min_{\hat{\rho}}\left\{ \langle\hat{H}\rangle_{\hat{\rho}}|\hat{\rho}\in\mathcal{C}\right\} ,
\end{equation}
where $\mathcal{C}$ denotes the space of all density matrices described
by the ansatz, and we used the notation $\langle\hat{O}\rangle_{\hat{\rho}}=\textrm{Tr}(\hat{\rho}\hat{O})/\textrm{Tr}(\hat{\rho})$
for the measurement of some operator $\hat{O}$ under a density matrix
$\hat{\rho}$.

We now consider a special case of the SPD ansatz which is dictated
by the form of the Hamiltonian, and we refer to this ansatz as the
Trotter SPD. The essence of the wave function version of the Trotter
SPD was anticipated several decades ago\cite{Dzierzawa19951993}.
To motivate the Trotter SPD, consider the Trotter-Suzuki Decomposition\cite{Trotter1959545,Suzuki1976183,Suzuki1993432}
of a finite temperature density matrix for a system with $L$ spin
orbitals 
\begin{align}
\exp(-\beta\hat{H}) & \approx\prod_{\tau=1}^{\discn}\exp(-\frac{\beta}{\discn}\hat{H}_{0})\exp(-\frac{\beta}{\discn}\hat{V}),
\end{align}
where $\hat{H}=\hat{H}_{0}+\hat{V}$ is the Hamiltonian, $\hat{H}_{0}$
is the non-interacting part, and $\hat{V}$ is interacting part. The
Trotter SPD ansatz can be obtained by replacing $-\beta/\discn$ with
variational parameters $\gamma_{\tau},g_{\tau}$ with $\tau=1\dots\discn$
as 
\begin{align}
 & \spd=\exp(\gamma_{1}\hat{H}_{0})\exp(g_{1}\hat{V})\dots\exp(\gamma_{\discn}\hat{H}_{0})\exp(g_{\discn}\hat{V}).
\end{align}
We see that the Trotter SPD is composed of pairs of non-interacting
and interacting projectors, which are sequentially multiplied together.
Considering the case of $\discn=2$
\begin{align}
 & \spd=\exp(\gamma_{1}\hat{H}_{0})\exp(g_{1}\hat{V})\exp(\gamma_{2}\hat{H}_{0})\exp(g_{2}\hat{V}),
\end{align}
where we can restrict to a purely Hermitian form (see Section \ref{subsec:Defining-the-SPD}
for a detailed discussion) as
\begin{align}
 & \spd=\exp(g_{1}\hat{V})\exp(\gamma_{1}\hat{H}_{0})\exp(g_{1}^{*}\hat{V}).
\end{align}
In the limit of large $\discn$ and $\beta$, the variational theorem
can select $\gamma_{\tau}=g_{\tau}=-\beta/\discn$ and recover the
exact density matrix. For a given finite $\discn$, the variational
principle will gaurantee that the Trotter SPD will generate superior
ground state results to any approach based on the standard Trotter-Suzuki
decomposition, such as auxiliary field quantum Monte-Carlo (AFQMC)\cite{Hubbard195977,Blankenbecler19812278,Hirsch19834059,Hirsch19862521}.

\begin{figure}
\includegraphics[width=1\columnwidth]{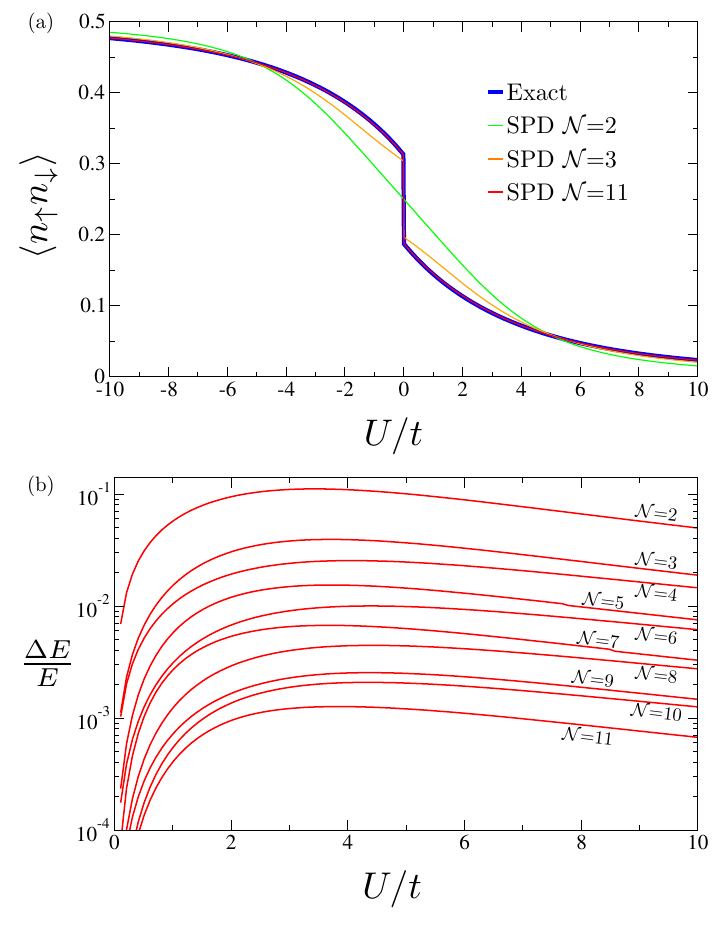}

\caption{\label{fig:plaq_docc_vs_u-1} Results for the Hubbard Plaquette, comparing
the Trotter SPD ansatz to the exact solution. (Panel a) Double occupancy
vs. $U/t$ for $\discn=$ 2, 3, and 11. (Panel b) Energy error vs.
$U/t$ for $\discn=$ 2-11.}
\end{figure}

To understand the convergence of the Trotter SPD with $\discn$, it
is useful to solve the Hubbard plaquette, which is the one dimensional
Hubbard model with four sites and translational symmetry, at half-filling
and zero temperature; where we can directly evaluate the exact solution.
We restrict our attention to the case of real variational parameters
(see Subsection \ref{subsec:Classification-of-SPD} for a general
discussion). The Hubbard model is given by $\hat{H}_{0}=t\sum_{\langle ij\rangle\sigma}\hat{a}_{i\sigma}^{\dagger}\hat{a}_{j\sigma}$
and $\hat{V}=U\sum_{i}\hat{n}_{i\uparrow}\hat{n}_{i\downarrow}$,
where $\langle ij\rangle$ denotes nearest neighbor sites. We now
compare the double occupancy, $\langle\hat{n}_{i\uparrow}\hat{n}_{i\downarrow}\rangle=\partial\langle\hat{H}\rangle/\partial U$,
of the exact solution and Trotter SPD ansatz at a given $\discn$
(see Figure \ref{fig:plaq_docc_vs_u-1}, panel $a$). For $\discn=2$,
which rigorously recovers the Gutzwiller wave function\cite{Gutzwiller1963159,Gutzwiller1964923,Gutzwiller19651726},
there is relatively large disagreement with the exact solution, completely
missing the discontinuity at $U/t=0$. Moving to $\discn=3$, we recover
the discontinuity, and move closer to the exact solution. For $\discn=11$,
there is almost no discernible difference with the exact solution,
and we see that the error monotonically decreases with increasing
$\discn$. Similarly, the error for the total energy monotonically
decreases with increasing $\discn$, as it must (see Figure \ref{fig:plaq_docc_vs_u-1},
panel $b$). This simple example illustrates the efficacy of the Trotter
SPD ansatz\cite{baeriswylisdif}. However, it should be emphasized
that one can only directly evaluate an SPD in the full Fock space
for a sufficiently small system. For macroscopic systems in the thermodynamic
limit, we will need a more advanced approach (see Sections \ref{sec:Integer-Time-Green's}
and \ref{sec:The-Discrete-Action-Theory}).

\subsection{Defining the SPD\label{subsec:Defining-the-SPD}}

The Trotter SPD defined in the preceding subsection is dictated based
on the form of the Hamiltonian being studied, whereas a general SPD
will not have such limitations. In general, one can enlarge both the
non-interacting and interacting projectors to include operators which
do not appear in the Hamiltonian itself, in addition to changing the
relative weights of existing operators. For the non-interacting projector,
we replace $\exp(\gamma_{\tau}\hat{H}_{0})$ from the Trotter SPD
with $\exp\left(\sppm_{\tau}\cdot\spom\right)$, where 
\begin{align}
\sppm_{\tau}\cdot\spom & \equiv\sum_{i=1}^{L}\sum_{j=1}^{L}[\sppm_{\tau}]_{ij}[\spom]_{ij}, & [\spom]_{ij}=\hat{a}_{i}^{\dagger}\hat{a}_{j},\label{eq:spom}
\end{align}
where $[\sppm_{\tau}]_{ij}$ are the variational parameters and $L$
is the number of spin orbitals. It should be noted that most general
non-interacting projector would include the terms $\hat{a}_{i}^{\dagger}\hat{a}_{j}^{\dagger}$
and $\hat{a}_{i}\hat{a}_{j}$, but we presently omit them for brevity.
For the interacting projector, we replace $\exp(g_{1}\hat{V})$ with
$\hat{P}_{\tau}$, which is a general Bosonic interacting projector
which will have various constraints (see Subsection \ref{subsec:Classification-of-SPD}).
Mathematically, the general SPD is then defined as
\begin{align}
 & \spd=\exp\left(\sppm_{1}\cdot\spom\right)\hat{P}_{1}\dots\exp\left(\sppm_{\discn}\cdot\spom\right)\hat{P}_{\discn}=\hat{\mathcal{P}}_{1}\dots\hat{\mathcal{P}}_{\discn},\\
 & \hat{\mathcal{P}}_{\tau}=\exp\left(\sppm_{\tau}\cdot\spom\right)\hat{P}_{\tau},\\
 & \hat{P}_{\tau}=\exp(\hat{\mathcal{V}}_{\tau})=\sum_{\Gamma\Gamma'}P_{\tau,\Gamma\Gamma'}\hat{X}_{\Gamma\Gamma'},\label{eq:interacting_projector_w_nu}
\end{align}
where $\Gamma,\Gamma'$ label the basis of the Fock space, $\hat{X}_{\Gamma\Gamma'}$
is a Hubbard operator, and $P_{\tau,\Gamma\Gamma'}$ are the variational
parameters. The SPD will always be constrained to be Hermitian and
semi-definite, and this can be achieved in two distinct ways, which
we denote as Gutzwiller-type (G-type) or Baeriswyl-type (B-type).
For $\discn=1,2,3$, the G-type and B-type SPD are 
\begin{align}
 & \spd_{G}^{\left(1\right)}= & \exp( & \sppm_{1}\cdot\spom),\label{eq:SPD_HF}\\
 & \spd_{G}^{\left(2\right)}= & \hat{P}_{1}\exp( & \sppm_{2}\cdot\spom)\hat{P}_{1}^{\dagger},\\
 & \spd_{G}^{\left(3\right)}= & \exp\left(\sppm_{1}\cdot\spom\right)\hat{P}_{1}\exp( & \sppm_{2}\cdot\spom)\hat{P}_{1}^{\dagger}\exp(\sppm_{1}^{\dagger}\cdot\spom),\\
 & \spd_{B}^{\left(1\right)}= &  & \hat{P}_{1},\\
 & \spd_{B}^{\left(2\right)}= & \exp\left(\sppm_{1}\cdot\spom\right) & \hat{P}_{1}\exp(\sppm_{1}^{\dagger}\cdot\spom),\\
 & \spd_{B}^{\left(3\right)}= & \hat{P}_{1}\exp\left(\sppm_{2}\cdot\spom\right) & \hat{P}_{2}\exp(\sppm_{2}^{\dagger}\cdot\spom)\hat{P}_{1}^{\dagger}.
\end{align}
For $\discn=1$ and G-type, $\bm{\gamma}_{1}$ is a Hermitian matrix.
For $\discn=2,3$ and G-type, $\bm{\gamma}_{2}$ is a Hermitian matrix,
while $\bm{\gamma}_{1}$ and $\hat{P}_{1}$ are unrestricted. For
$\discn=1,2$ and B-type, $\hat{P_{1}}$ is a Hermitian and semi-definite
operator, while $\bm{\gamma}_{1}$ is unrestricted. For $\discn=3$
and B-type, $\hat{P_{2}}$ is a Hermitian and semi-definite operator,
while $\bm{\gamma}_{2}$ and $\hat{P_{1}}$ are unrestricted. It should
be noted that the $\discn+1$ G-type can always recover the $\discn$
B-type, and vice versa. Throughout the manuscript, we will use the
G-type unless otherwise specified. 

It should be noted that the G-type and B-type ansatz are related by
an abstract ``dual'' transformation, whereby one ansatz can be obtained
from the other by interchanging the interacting projector with the
non-interacting projector. This same notion of a dual transformation
has been previously introduced in the context of the Gutzwiller, Baeriswyl,
Gutzwiller-Baeriswyl, and Baeriswyl-Gutzwiller wave functions\cite{Dzierzawa19951993},
in addition to the $\mathcal{K}$ and $\mathcal{X}$ formulations
of the off-shell effective energy theory\cite{Cheng2020081105}. 

\subsection{Classification of the SPD\label{subsec:Classification-of-SPD}}

\begin{figure}
\includegraphics[width=1\columnwidth]{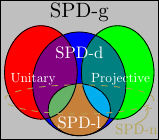}

\caption{\label{fig:schem_spd_intp} A schematic for the classification of
SPD's. The SPD-l has an interacting projector that is restricted to
a local subspace. The SPD-d has an interacting projector which is
composed of disjoint projectors. The SPD-g corresponds to a completely
general interacting projector. The SPD-n is an SPD in which $\hat{\mathcal{V}}_{\tau}$
is restricted to n-particle operators. A unitary (projective) SPD
is one in which the unrestricted projectors are unitary (Hermitian). }
\end{figure}

The generically defined Hermitian and semi-definite SPD encompasses
a broad variety of possibilities, and it is useful to consider various
categorization schemes. The first categorization we consider is partitioning
into projective, unitary, or general SPD. A projective SPD is the
subset where the unrestricted projectors are Hermitian, while a unitary
SPD is the subset where the unrestricted projectors are unitary (see
Figure \ref{fig:schem_spd_intp}). Most variational wave functions
in the early literature belong the projective subset of SPD's, while
examples of approaches which fall into the category of unitary SPD's
can be found in the context of quantum computing (see Subsection \ref{subsec:Categorizing-Existing-WF-w-SPD}
for a detailed discussion). In this manuscript, we largely focus on
projective SPD's, though we do explore simple cases of unitary SPD's
as well (see Subsection \ref{subsec:ex-AIM-nle2} for examples).

The second major categorization scheme is based on the restrictions
of the interacting projectors (see Figure \ref{fig:schem_spd_intp}).
The choice of interacting projector has a trade-off between computational
complexity and rate of convergence with respect to $\discn$. For
example, if the interacting projector is completely unrestricted,
one already obtains the exact solution for the B-type at $\discn=1$,
but this simply amounts to directly diagonalizing the target Hamiltonian
in the Fock space. While the particular Hamiltonian under consideration
will ultimately guide the choice of interacting projectors for the
SPD, there are certain interacting projectors which would be natural
choices for wide classes of Hamiltonians. If one is considering a
Hamiltonian where the interactions are local to some subspace, such
as the Anderson impurity model (AIM), a natural choice for the interacting
projector is 
\begin{align}
\hat{P}_{\tau} & =\sum_{\Gamma\Gamma'\in\mathcal{C}}P_{\tau,\Gamma\Gamma'}\hat{X}_{\Gamma\Gamma'},\label{eq:localP-1}
\end{align}
where $\Gamma,\Gamma'$ label the basis of the local subspace $\mathcal{C}$,
$\hat{X}_{\Gamma\Gamma'}$ is a Hubbard operator, and $P_{\tau,\Gamma\Gamma'}$
are the variational parameters. We refer to this particular category
of SPD as a local SPD (SPD-l). 

Another common scenario for models of interacting electrons is where
the interaction is local, but not restricted to a single subspace;
prominent examples include the Hubbard model and the periodic Anderson
impurity model. In such cases, it is natural to study an SPD where
the interacting projectors are composed of disjoint interacting projectors,
and we refer to such SPD as disjoint SPD (SPD-d). The SPD-d with $N$
disjoint regions has an interacting projector of 

\begin{align}
\hat{P}_{\tau}=\prod_{i=1}^{N}\hat{P}_{\tau,i}=\prod_{i=1}^{N}\left(\sum_{\Gamma\Gamma'}P_{\tau,i,\Gamma\Gamma'}\hat{X}_{i,\Gamma\Gamma'}\right).\label{eq:SPD-1-3}
\end{align}

Another possible way to categorize the interacting projector is to
restrict to n-particle interaction in $\hat{\mathcal{V}}_{\tau}$,
but apply no other limitations; and we refer to this as SPD-n, where
$n\ge2$ is the number of excitations. Existing examples of the SPD-2
are the Jastrow wave function\cite{Jastrow19551479,fazekas19881021,Yokoyama19903669,Capello2005026406},
unitary coupled cluster\cite{Bartlett1989133,Kutzelnigg1991349,Taube20063393},
and the adaptive variational algorithm of Grimsley \textit{et al}\cite{Grimsley20193007}. 

In presenting the most general form of the SPD, we introduce the possibility
of having an infinite number of variational parameters, which is excessive
in practice. While we will demonstrate that it is beneficial to have
variational parameters which deviate from the form of the Hamiltonian
itself, typically only a few degrees of freedom beyond the parameters
present in the Hamiltonian are needed. The beauty of a variational
theory is that no matter how the form of the projector is restricted,
one always obtains an upper bound on the energy, so long as the SPD
is evaluated exactly. 

In summary, the SPD presents a systematic variational ansatz for studying
quantum Hamiltonians. While the underlying idea behind the SPD was
appreciated decades ago in the context of the generalizations of the
Gutzwiller wave function\cite{Dzierzawa19951993}, the general idea
has not been fully exploited. A key advancement achieved by our discrete
action theory is that the SPD can be formally understood in terms
of an integer time Green's function, which has a perfect parallel
to the standard many-body Green's function formalism, allowing one
to generalize many of the existing ideas from many-body physics to
the discrete action theory. It should be emphasized that the discrete
action theory has very practical implications, such as allowing for
the exact evaluation of SPD-d in infinite dimensions (see Subsection
\ref{subsec:Proof-that-LSA} for the proof).

\subsection{Categorizing existing wave functions in terms of the SPD\label{subsec:Categorizing-Existing-WF-w-SPD}}

\begin{figure}
\includegraphics[width=1\columnwidth]{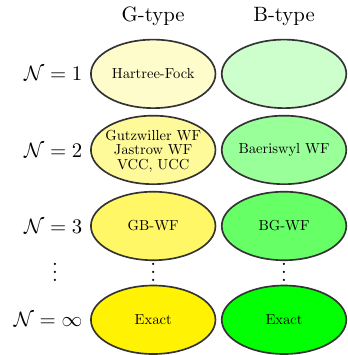}

\caption{\label{fig:schem_spd} Existing variational wave functions (WF) classified
in terms of the SPD, characterized by the number of integer time steps
$\discn$ and the type of SPD. The following acronyms are used: variation
coupled cluster (VCC), unitary coupled cluster (UCC), Gutzwiller-Baeriswyl
(GB), and Baeriswyl-Gutzwiller (BG).}
\end{figure}

In this subsection, we use the SPD to categorize existing variational
wave function approaches within the literature (see Figure \ref{fig:schem_spd}).
We begin with the case of projective SPD's, and first enumerate all
G-type SPD's. For $\discn=1$, we have the well known Hartree-Fock
approximation, given that Eq. \ref{eq:SPD_HF} will result in the
lowest energy single Slater determinant. For $\discn=2$, the SPD-d
recovers the Gutzwiller wave function\cite{Gutzwiller1963159,Gutzwiller1964923,Gutzwiller19651726};
the SPD-g recovers the variational coupled cluster (VCC) ansatz\cite{Bartlett198829,Kutzelnigg1991349};
the SPD-n recovers the Jastrow wave function\cite{Jastrow19551479,fazekas19881021,Yokoyama19903669,Capello2005026406}.
For $\discn=3$, the SPD-d recovers the Gutzwiller-Baeriswyl wave
function\cite{Otsuka19921645}. For the preceding two cases, variational
quantum Monte-Carlo is typically used to evaluate the ansatz\cite{Ceperley19773081,Otsuka19921645,Sorella2001024512,Sorella2005241103,Capello2005026406}.
For $\discn>3$, we are not aware of existing ansatz in the literature. 

We now consider projective SPD's of B-type. For $\discn=1$, we are
not aware of existing ansatz, which seems reasonable given that this
would often amount to a crude approximation (see Subsection \ref{subsec:ex-AIM-nle2}
for an illustration). For $\discn=2$, the SPD-d recovers a generalized
version of the Baeriswyl wave function\cite{Baeriswyl19870}. The
SPD-d is more general given that the interacting projector is fully
variational, whereas Baeriswyl made certain restrictions. For $\discn=3$,
the SPD-d recovers the Baeriswyl-Gutzwiller wave function\cite{Dzierzawa19951993}.
For $\discn>3$, we are not aware of existing ansatz in the literature.

We now consider the unitary SPD, and the most well known example is
SPD-g for $\discn=2$ in the case of the unitary coupled cluster (UCC)
approach\cite{Bartlett1989133,Kutzelnigg1991349,Taube20063393}. Due
to the complexity of the Hamiltonians which are being studied with
UCC, one cannot exactly evaluate the ansatz, resulting in applications
to very small systems or uncontrolled approximations. For $\discn>2$,
there has recently been interest in the context of quantum computing.
Farhi \textit{et al.} proposed a Trotter-like ansatz composed of multiple
unitary operations with the intent of evaluating it within a quantum
computer\cite{Farhi1411.4028}. These ideas were then extended and
examined in the context of small Hubbard models\cite{Wecker2015042303}.
A further generalization was made by Grimsley \textit{et al.}, where
they considered pairs of non-interacting and two-particle interacting
projectors\cite{Grimsley20193007}. All of these ansatz are pursued
under the assumption that a quantum computer can be used to evaluate
them. 

\subsection{Minimization of the total energy under the SPD\label{subsec:Minimization-of-the-SPD}}

A key task in using any variational ansatz is to minimize the total
energy with respect to the variational parameters. Given that there
will typically be numerous variational parameters, it is critical
to be able to compute the gradient of the energy with respect to the
variational parameters. In this subsection, we demonstrate how to
compute the gradient, which will showcase the emergence of integer
time correlation functions.

To begin, we parameterize the non-interacting and interacting projectors
as 
\begin{align}
\exp\left(\boldsymbol{\gamma}_{\tau}\left(\left\{ c_{\tau i}\right\} \right)\cdot\spom\right), &  & \hat{P}_{\tau}\left(\left\{ g_{\tau i}\right\} \right),
\end{align}
where we have chosen a parameterization such that the derivatives
have the following form
\begin{align}
 & \frac{\partial\exp\left(\boldsymbol{\gamma}_{\tau}\left(\left\{ c_{\tau i}\right\} \right)\cdot\spom\right)}{\partial c_{\tau i}}=\hat{K}_{\tau i}\exp\left(\boldsymbol{\gamma}_{\tau}\left(\left\{ c_{\tau i}\right\} \right)\cdot\spom\right),\\
 & \frac{\partial\hat{P}_{\tau}\left(\left\{ g_{\tau i}\right\} \right)}{\partial g_{\tau i}}=\hat{P}_{\tau}\left(\left\{ g_{\tau i}\right\} \right)\hat{W}_{\tau i},
\end{align}
where $\hat{K}_{\tau i}$ and $\hat{W}_{\tau i}$ are operators which
characterize the derivative. We note that this choice of parameterization
is analogous to that of Sorella\cite{Sorella2005241103}. We can now
compute the derivative of the energy as

\begin{align}
 & \frac{\partial}{\partial c_{\tau i}}\langle\hat{H}\rangle_{\spd}=\frac{\partial}{\partial c_{\tau i}}\frac{\text{\ensuremath{\text{Tr}(\spd\hat{H}})}}{\text{Tr}\left(\spd\right)}=\nonumber \\
 & \langle\hat{K}_{\tau i}\left(\tau-1\right)\hat{H}\left(\discn\right)\rangle_{\spd}-\langle\hat{K}_{\tau i}\left(\tau-1\right)\rangle_{\spd}\langle\hat{H}\left(\discn\right)\rangle_{\spd},\\
 & \frac{\partial}{\partial g_{\tau i}}\langle\hat{H}\rangle_{\spd}=\frac{\partial}{\partial g_{\tau i}}\frac{\text{\ensuremath{\text{Tr}(\spd\hat{H})}}}{\text{Tr}(\spd)}=\nonumber \\
 & \langle\hat{W}_{\tau i}\left(\tau\right)\hat{H}\left(\discn\right)\rangle_{\spd}-\langle\hat{W}_{\tau i}\left(\tau\right)\rangle_{\spd}\langle\hat{H}\left(\discn\right)\rangle_{\spd},
\end{align}
where $\hat{O}(\tau)$ is an operator in the \textit{integer time}
Heisenberg representation defined as 
\begin{align}
\hat{O}(\tau) & =\hat{U}_{\tau}\hat{O}\hat{U}_{\tau}^{-1}, & \hat{U}_{\tau}=\hat{\mathcal{P}}_{1}\dots\hat{\mathcal{P}}_{\tau}.\label{eq:heisenberg_rep_def-1}
\end{align}
This notion of integer time will be carefully introduced and explored
in Sections \ref{sec:Integer-Time-Green's} and \ref{sec:The-Discrete-Action-Theory}.
It should be noted that the second derivative can also be expressed
in terms of integer time correlation functions, resulting in a new
approximate saddle point for a given set of variational parameters.
The fact that these derivatives can be evaluated in terms of computable
correlation functions is critical to minimizing the total energy in
practice. 

\section{Integer Time Green's function Formalism\label{sec:Integer-Time-Green's}}

\subsection{Integer time correlation functions \label{subsec:Integer-Time-Correlation}}

While the SPD ansatz provides an intelligent route to approaching
the exact solution, this form is only useful if it can be efficiently
evaluated. Therefore, it will be essential to have a mathematical
formalism which is conducive for developing robust approximations.
We introduce the notation $\langle\hat{O}\rangle_{\hat{\rho}}=\textrm{Tr}(\hat{\rho}\hat{O})/\textrm{Tr}(\hat{\rho})$
for the measurement of some operator $\hat{O}$ under a density matrix
$\hat{\rho}$. We begin by considering the expectation value of $\hat{O}$
under the SPD

\begin{align}
\langle\hat{O}\rangle_{\spd} & =\frac{\textrm{Tr}(\exp\left(\sppm_{1}\cdot\spom\right)\hat{P}_{1}\dots\exp\left(\sppm_{\discn}\cdot\spom\right)\hat{P}_{\discn}\hat{O})}{\textrm{Tr}(\exp\left(\sppm_{1}\cdot\spom\right)\hat{P}_{1}\dots\exp\left(\sppm_{\discn}\cdot\spom\right)\hat{P}_{\discn})}.
\end{align}
This expectation value can be re-expressed in the \textit{integer
time} interaction representation as
\begin{align}
\langle\hat{O}\rangle_{\spd} & =\frac{\langle\hat{P}_{1,I}(1)\hat{P}_{2,I}(2)\dots\hat{P}_{\discn,I}(\discn)\hat{O}_{I}(\discn)\rangle_{\spd_{0}}}{\langle\hat{P}_{1,I}(1)\hat{P}_{2,I}(2)\dots\hat{P}_{\discn,I}(\discn)\rangle_{\spd_{0}}},\label{eq:int_time_correlation_function}
\end{align}
where
\begin{align}
 & \spd_{0}=\exp\left(\sppm_{1}\cdot\spom\right)\dots\exp\left(\sppm_{\discn}\cdot\spom\right),\\
 & \hat{O}_{I}\left(\tau\right)=\hat{U}_{\tau;I}\hat{O}\hat{U}_{\tau;I}{}^{-1},\\
 & \hat{U}_{\tau;I}=\exp\left(\sppm_{1}\cdot\spom\right)\dots\exp\left(\sppm_{\tau}\cdot\spom\right),
\end{align}
where $\spd_{0}$ is the non-interacting SPD, the subscript $I$ denotes
the integer time interaction representation, and $\tau=1,\dots,\discn$.
This interpretation of integer time evolution can be viewed as arising
from a discrete action (see Subsection \ref{subsec:Introducing-the-Discrete-Action}).

While this notion of integer time may appear artificial, it allows
for a systematic evaluation using the integer time generalization
of Wick's theorem (see Appendix \ref{appendix:wicks_theorem} for
a derivation). In the common scenario where the interacting projectors
$\hat{P}_{i}$ will be the exponential of some interacting operator
(see Eq. \ref{eq:interacting_projector_w_nu}), it is natural to Taylor
series expand such operators, yielding
\begin{align}
\langle\hat{O}\rangle_{\spd} & =\frac{\langle\textrm{T}\hat{P}_{1,I}(1)\hat{P}_{2,I}(2)\dots\hat{P}_{\discn,I}(\discn)\hat{O}_{I}(\discn)\rangle_{\spd_{0}}}{\langle\textrm{T}\hat{P}_{1,I}(1)\hat{P}_{2,I}(2)\dots\hat{P}_{\discn,I}(\discn)\rangle_{\spd_{0}}}\\
 & =\frac{\langle\textrm{T}\exp(\sum_{\tau=1}^{\discn}\hat{\mathcal{V}}_{\tau,I}(\tau))\hat{O}_{I}(\discn)\rangle_{\spd_{0}}}{\langle\textrm{T}\exp(\sum_{\tau=1}^{\discn}\hat{\mathcal{V}}_{\tau,I}(\tau))\rangle_{\spd_{0}}}\\
 & =\frac{\sum_{n=0}^{\infty}\frac{1}{n!}\langle\textrm{T}(\sum_{\tau=1}^{\discn}\hat{\mathcal{V}}_{\tau,I}(\tau))^{n}\hat{O}_{I}(\discn)\rangle_{\spd_{0}}}{\sum_{n=0}^{\infty}\frac{1}{n!}\langle\textrm{T}(\sum_{\tau=1}^{\discn}\hat{\mathcal{V}}_{\tau,I}(\tau))^{n}\rangle_{\spd_{0}}},\label{eq:integer_time_correlation_infinite}
\end{align}
where the integer time ordering operator $\textrm{T}$ first sorts
the operators according to ascending integer time, and then according
to the position in the original ordering of operators, and finally
the result is presented from left to right; additionally, the resulting
sign must be tracked when permuting Fermionic operators. It should
be noted that our time convention is opposite to the usual definition\cite{Mahan20000306463385}.
It is useful to illustrate the integer time ordering operator with
the example 
\begin{align}
 & \textrm{T}\hat{a}_{k,I}^{\dagger}(2)\hat{a}_{k',I}(1)\hat{a}_{k,I}(2)=-\hat{a}_{k',I}(1)\hat{a}_{k,I}^{\dagger}(2)\hat{a}_{k,I}(2).
\end{align}
Inspecting Eq. \ref{eq:integer_time_correlation_infinite}, it is
clear that there are an infinite number of terms to be evaluated;
and each term can be evaluated using the integer time Wick's theorem
(see Appendix \ref{subsec:Evaluating-a-Local-SPD-via-Wick}) in terms
of the non-interacting integer time Green's function 
\begin{align}
 & [\boldsymbol{g}_{0}]_{k\tau,k'\tau'}=\langle\text{T}\hat{a}_{k,I}^{\dagger}\left(\tau\right)a_{k',I}\left(\tau'\right)\rangle_{\spd_{0}},\label{eq:g0interactionrep}
\end{align}
where $k=1,\dots,L$ labels the spin-orbital index, and $\boldsymbol{g}_{0}$
is a matrix of dimension $L\discn\times L\discn$. 

Having reformulated the expectation value of some operator in terms
of integer time correlation functions, it becomes clear how to straightforwardly
apply the integer time version of Wick's theorem. This advancement
will already allow us to exactly evaluate the SPD-l in terms of a
finite number of diagrams (see Subsection \ref{subsec:Evaluating-a-Local-SPD-via-Wick}),
allowing for the efficient and robust solution of the Anderson impurity
model within VDAT\cite{Cheng2020short}. Recently, Baeriswyl employed
a perturbative approach to approximately evaluate a variant of the
projective G-type SPD-d at $\discn=3$ (as characterized from our
general conventions) for the two dimensional Hubbard model\cite{Baeriswyl2019235152}.
Baeriswyl's perturbative approach is recovered by our integer time
formulation in the special case of $\discn=3$ where we restrict Eq.
\ref{eq:integer_time_correlation_infinite} to second order, but our
approach can naturally be applied at arbitrary $\discn$ (it should
be noted that $\tau$ is a variational parameter in Baeriswyl's approach
while $\tau$ is an integer time in ours). 

\subsection{Evaluating the SPD-l via Wick's theorem\label{subsec:Evaluating-a-Local-SPD-via-Wick}}

Having developed the integer time formalism, we already have the tools
necessary to evaluate the SPD-l (see Eq. \ref{subsec:Variational-Theory-and-SPD}),
given that the interacting projector is confined to a subspace. Therefore,
only a finite number of terms are required to evaluate an expectation
value in this scenario, given as 
\begin{align}
 & \langle\hat{O}\rangle_{\spd}=\nonumber \\
 & \frac{\sum_{\left\{ \Gamma_{\tau}\Gamma'_{\tau}\right\} }(\prod_{\tau}P_{\tau,\Gamma_{\tau}\Gamma'_{\tau}})\langle\textrm{T}\prod_{\tau}\hat{X}_{\Gamma_{\tau}\Gamma'_{\tau},I}(\tau)\hat{O}_{I}(\discn)\rangle_{\spd_{0}}}{\sum_{\left\{ \Gamma_{\tau}\Gamma'_{\tau}\right\} }(\prod_{\tau}P_{\tau,\Gamma_{\tau}\Gamma'_{\tau}})\langle\textrm{T}\prod_{\tau}\hat{X}_{\Gamma_{\tau}\Gamma'_{\tau},I}(\tau)\rangle_{\spd_{0}}},
\end{align}
where each term can be evaluated using the integer time Wick's theorem
(see Appendix \ref{appendix:wicks_theorem}). 

Given this capability of evaluating the SPD-l, we can already address
many important Hamiltonians. For example, we can approximately solve
the Anderson impurity model (AIM)\cite{Cheng2020short}. For $\discn=2$,
the integer time formalism exactly evaluates the Gutzwiller wave function,
which has long been available in the literature, but our $\discn=3$
result had never been realized, and provides an accuracy comparable
to the numerically exact density matrix renormalization group results\cite{Barcza2019165130}
with relatively negligible computation cost. More generally, the above
evaluation of the SPD-l allows us to evaluate the multi-orbital AIM,
which we will address in future work.

We now consider pedagogical examples of the single orbital AIM with
one bath site for $\discn\le2$, evaluating unitary and projective
SPD's with G-type or B-type. Given that the exact solution can easily
be evaluated by diagonalizing the Hamiltonian in the Fock space, this
example provides a nice illustration of the evaluation of the SPD-l
using the integer time Wick's theorem.

\subsubsection{Illustrative examples for $\discn\le2$: Anderson impurity model
with one bath\label{subsec:ex-AIM-nle2}}

Here we study the AIM with only one bath orbital and particle-hole
symmetry. The Hamiltonian is given as
\begin{align}
\hat{H} & =t\hat{K}+U\Delta\hat{d},
\end{align}
where
\begin{align}
\hat{K}=\sum_{\sigma}(\hat{f}_{\sigma}^{\dagger}\hat{c}_{\sigma}+h.c.), &  & \Delta\hat{d} & =\prod_{\sigma}(\hat{f}_{\sigma}^{\dagger}\hat{f}_{\sigma}-\frac{1}{2}).\label{eq:delta_d}
\end{align}
We first consider the case of $\discn=1$ with a projective G-type
SPD-l, given as 
\begin{align}
\spd=\exp(\gamma\hat{K}), &  & \hat{P}_{1}=\hat{1},
\end{align}
where $\gamma$ is the non-interacting variational parameter. It will
be inconvenient to use $\gamma$ directly given that it is not bounded,
and therefore we can effectively reparameterize it with $\nu=\frac{1}{2}\tanh\left(\frac{\gamma}{2}\right)\in\left[-1/2,1/2\right]$
or $\gamma=2\text{\ensuremath{\tanh^{-1}\left(2\nu\right)}}$. As
we have spin symmetry, we only need to compute the non-interacting
integer time Green's function for a given spin
\begin{align}
\boldsymbol{g}_{\sigma;0}=\left(\begin{array}{cc}
\frac{1}{2} & \nu\\
\nu & \frac{1}{2}
\end{array}\right).
\end{align}
We can then use the integer time Wick's theorem to compute the necessary
integer time correlation functions required to evaluate the total
energy
\begin{align}
 & \langle\text{T}\hat{P}_{1,I}\left(1\right)\rangle_{\spd_{0}}=1,\\
 & \langle\text{T}\hat{P}_{1,I}\left(1\right)\Delta\hat{d}_{I}\left(1\right)\rangle_{\spd_{0}}=0,\\
 & \langle\text{T}\hat{P}_{1,I}\left(1\right)\hat{f}_{\sigma,I}^{\dagger}\left(1\right)\hat{c}_{\sigma,I}\left(1\right)\rangle_{\spd_{0}}=\nu.
\end{align}
The relevant expectation values can be obtained from the last time
step, which is the only time step for $\discn=1,$ as 
\begin{align}
 & \langle\hat{f}^{\dagger}\hat{c}\rangle_{\spd}=\nu, & \langle\Delta\hat{d}\rangle_{\spd}=0.
\end{align}
Finally, the total energy can be written as
\begin{align}
\mathcal{E} & =\min_{\nu\in[-\frac{1}{2},\frac{1}{2}]}\left(4t\nu\right)=-2t.
\end{align}
We observe that $\discn=1$ with a G-type SPD recovers the well known
Hartree-Fock approximation, where the energy is independent of the
Hubbard $U$ (given the chosen form of the interacting Hamiltonian). 

We now move on to $\discn=1$ with a B-type projective SPD-l, given
as 
\begin{align}
\spd=\exp(g\Delta\hat{d}), &  & \hat{P}_{1}=\exp(g\Delta\hat{d})=(\hat{1}+u\Delta\hat{d}).
\end{align}
Here we reparameterize the interacting variational parameter $g$
in terms of $u\in[-4,4]$. We proceed by constructing the non-interacting
integer time Green's function 

\begin{equation}
\boldsymbol{g}_{0;\sigma}=\left(\begin{array}{cc}
\frac{1}{2} & 0\\
0 & \frac{1}{2}
\end{array}\right),
\end{equation}
and compute all the necessary integer time matrix elements 

\begin{align}
 & \langle\textrm{T}\hat{P}_{1,I}(1)\rangle_{\spd_{0}}=1,\\
 & \langle\textrm{T}\hat{P}_{1,I}(1)\Delta\hat{d}_{I}(1)\rangle_{\spd_{0}}=\frac{u}{16},\\
 & \langle\textrm{T}\hat{P}_{1,I}(1)\hat{f}_{\sigma,I}^{\dagger}(1)\hat{c}_{\sigma,I}(1)\rangle_{\spd_{0}}=0.
\end{align}
The relevant expectation values can be obtained as 
\begin{align}
 & \langle\hat{f}^{\dagger}\hat{c}\rangle_{\spd}=0, & \langle\Delta\hat{d}\rangle_{\spd}=\frac{u}{16}.
\end{align}
The resulting total energy is then 
\begin{align}
\mathcal{E} & =\min_{u\in\left[-4,4\right]}\left(U\frac{u}{16}\right)=-\frac{U}{4}.
\end{align}
Here we see that the energy is independent of the hopping parameter,
given that this ansatz amounts to a collection of two decoupled atoms. 

We now move on to $\discn=2$ for the G-type projective SPD-l, given
as
\begin{align}
\spd=\exp(g\Delta\hat{d})\exp(\gamma\hat{K})\exp(g\Delta\hat{d}),
\end{align}
where the interacting projector is 
\begin{align}
\hat{P}_{1}=\hat{P}_{2}=\exp(g\Delta\hat{d})=(\hat{1}+u\Delta\hat{d}),
\end{align}
and $\gamma$ is reparameterized with $\nu$ as before, and there
is no restriction on $u$ given that it occupies the outermost position
in the SPD. The noninteracting integer time Green's function is then

\begin{align}
\boldsymbol{g}_{\sigma;0}=\left(\begin{array}{cccc}
\frac{1}{2} & \nu & \frac{1}{2} & -\nu\\
\nu & \frac{1}{2} & -\nu & \frac{1}{2}\\
-\frac{1}{2} & -\nu & \frac{1}{2} & \nu\\
-\nu & -\frac{1}{2} & \nu & \frac{1}{2}
\end{array}\right),\label{eq:exAIMg0n2}
\end{align}
where we have used integer time major ordering of the basis, where
integer time is the slow index and goes in ascending order and the
orbital is the fast index and goes from $f$ to $c$, resulting in
the four sequential indices $(f,1),(c,1),(f,2),(c,2)$. For example,
we have $[\boldsymbol{g}_{\sigma;0}]_{13}=\langle\textrm{T}\hat{f}_{I}^{\dagger}(1)\hat{f}_{I}(2)\rangle_{\spd_{0}}$,
etc. The necessary integer time correlation functions needed to compute
the total energy are

\begin{align}
 & \langle\text{T}\hat{P}_{1,I}\left(1\right)\hat{P}_{2,I}\left(2\right)\rangle_{\spd_{0}}=\frac{u^{2}}{16}+1,\\
 & \langle\text{T}\hat{P}_{1,I}\left(1\right)\hat{P}_{2,I}\left(2\right)\Delta\hat{d}_{I}\left(2\right)\rangle_{\spd_{0}}=\frac{u}{8},\\
 & \langle\text{T}\hat{P}_{1,I}\left(1\right)\hat{P}_{2,I}\left(2\right)\hat{f}_{\sigma,I}^{\dagger}\left(2\right)\hat{c}_{\sigma,I}\left(2\right)\rangle_{\spd_{0}}=\frac{-\nu}{16}(u^{2}-16).
\end{align}
The relevant expectation values can be obtained as 
\begin{align}
 & \langle\hat{f}^{\dagger}\hat{c}\rangle_{\spd}=\frac{16-u^{2}}{16+u^{2}}\nu, & \langle\Delta\hat{d}\rangle_{\spd}=\frac{2u}{u^{2}+16}.
\end{align}
The ground state energy can then be obtained as
\begin{align*}
\mathcal{E} & =\min_{\nu\in[-\frac{1}{2},\frac{1}{2}],u}\left(4t\nu\frac{16-u^{2}}{16+u^{2}}+U\frac{2u}{u^{2}+16}\right)\\
 & =\min_{u}\left(-2t\frac{16-u^{2}}{16+u^{2}}+U\frac{2u}{u^{2}+16}\right)\\
 & =-\frac{1}{4}\sqrt{64t^{2}+U^{2}}.
\end{align*}
Notice that in this simple single bath case, $\discn=2$ with the
G-type projective SPD-l provides the exact ground state energy. 

We now consider the B-type projective SPD-l for $\discn=2$, given
as

\begin{align}
\spd=\exp(\gamma\hat{K})\exp(g\Delta\hat{d})\exp(\gamma\hat{K}),
\end{align}
where the interacting projectors are 
\begin{align}
\hat{P}_{1}=\exp(g\Delta\hat{d})=(\hat{1}+u\Delta\hat{d}), &  & \hat{P}_{2}=\hat{1},
\end{align}
and we reparameterize $\gamma$ as in the previous case, though $\nu\in\left[-\infty,\infty\right]$
given that the non-interacting projector is in the outer position
of the SPD; in addition to reparameterizing $g$ in terms of $u$,
though here the interacting projector is in the center position of
the SPD and therefore $u\in[-4,4]$. We can proceed by constructing
the non-interacting integer time Green's function 

\begin{equation}
\boldsymbol{g}_{0;\sigma}=\left(\begin{array}{cccc}
\frac{1}{2} & \frac{2\nu}{4\nu^{2}+1} & \frac{1-4\nu^{2}}{8\nu^{2}+2} & 0\\
\frac{2\nu}{4\nu^{2}+1} & \frac{1}{2} & 0 & \frac{1-4\nu^{2}}{8\nu^{2}+2}\\
\frac{4\nu^{2}-1}{8\nu^{2}+2} & 0 & \frac{1}{2} & \frac{2\nu}{4\nu^{2}+1}\\
0 & \frac{4\nu^{2}-1}{8\nu^{2}+2} & \frac{2\nu}{4\nu^{2}+1} & \frac{1}{2}
\end{array}\right).
\end{equation}
We then compute all the necessary matrix elements 

\begin{align}
 & \langle\textrm{T}\hat{P}_{1,I}(1)\rangle_{\spd_{0}}=1,\\
 & \langle\textrm{T}\hat{P}_{1,I}(1)\Delta\hat{d}_{I}(2)\rangle_{\spd_{0}}=\frac{\left(1-4\nu^{2}\right)^{4}u}{16\left(4\nu^{2}+1\right)^{4}},\\
 & \langle\textrm{T}\hat{P}_{1,I}(1)\hat{f}_{\sigma,I}^{\dagger}(2)\hat{c}_{\sigma,I}(2)\rangle_{\spd_{0}}=\frac{2\nu}{4\nu^{2}+1}.
\end{align}
The relevant expectation values can then be obtained as 
\begin{align}
 & \langle\hat{f}^{\dagger}\hat{c}\rangle_{\spd}=\frac{2\nu}{4\nu^{2}+1}, & \langle\Delta\hat{d}\rangle_{\spd}=\frac{\left(1-4\nu^{2}\right)^{4}u}{16\left(4\nu^{2}+1\right)^{4}}.
\end{align}
The resulting total energy is then 
\begin{align}
\mathcal{E} & =\min_{u\in\left[-4,4\right],\nu}\left(t\frac{8\nu}{4\nu^{2}+1}+U\frac{\left(1-4\nu^{2}\right)^{4}u}{16\left(4\nu^{2}+1\right)^{4}}\right),\\
 & =\min_{\nu}\left(\frac{8\nu t}{4\nu^{2}+1}-\frac{\left(1-4\nu^{2}\right)^{4}U}{4\left(4\nu^{2}+1\right)^{4}}\right).
\end{align}
The saddle points are then found by individually solving the following
two equations

\begin{align}
\left(4\nu^{2}+1\right)^{3}t+2\nu\left(1-4\nu^{2}\right)^{2}U & =0, & \left(4\nu^{2}-1\right)=0.
\end{align}
For the former, we have

\begin{align}
U & =-\frac{\left(4\nu^{\star2}+1\right)^{3}t}{2\nu^{\star}\left(1-4\nu^{\star2}\right)^{2}}, & \mathcal{E}=\frac{\left(16\nu^{\star4}+56\nu^{\star2}+1\right)t}{8\left(4\nu^{\star3}+\nu^{\star}\right)}.
\end{align}
The ground state energy can then be written as

\begin{equation}
\mathcal{E}=\bigg\{\begin{array}{cc}
\frac{\left(16\nu^{\star4}+56\nu^{\star2}+1\right)t}{8\left(4\nu^{\star3}+\nu^{\star}\right)}, & U\ge6.75t\\
-2t, & U\le6.75t
\end{array}
\end{equation}
where for $U\le6.75t$, the B-type $\discn=2$ projective SPD gives
the same energy as the G-type $\discn=1$ projective SPD. 

\begin{figure}
\includegraphics[width=1\columnwidth]{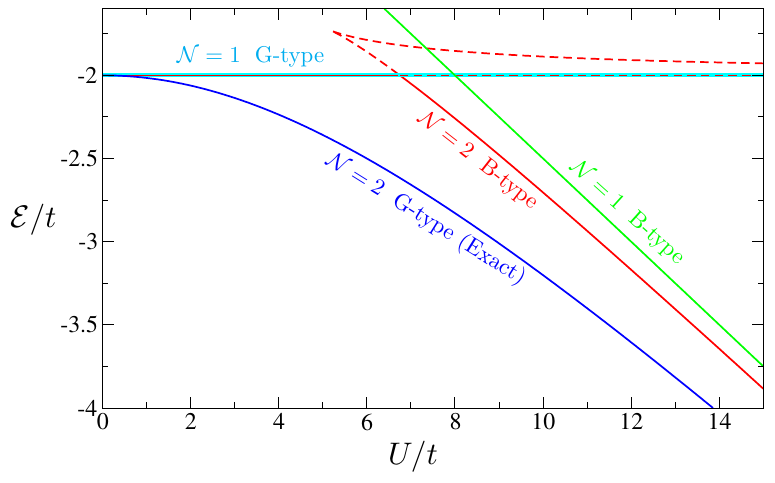}

\caption{\label{fig:aim1bath} Ground state energy as a function of $U/t$
for the single orbital AIM with one bath site using VDAT for $\discn=1,2$,
with both G-type and B-type projective SPD-l. Dotted lines denote
higher energy saddle points, while solid lines denote the ground state
energy for a given $\discn$.}
\end{figure}

We summarize the results for the ground state energy as a function
of $U/t$ in these four cases in Figure \ref{fig:aim1bath}. As noted
above, the $\discn=2$ G-type projective SPD (i.e. the Gutzwiller
wave function) gives the exact solution in this case. The $\discn=2$
B-type projective SPD is interesting given that is has multiple saddle
points, foreshadowing the possibility of becoming stuck in a false
minimum when minimizing the energy. Another point illustrated by this
plot is that larger $\discn$ always has a lower ground state energy,
as it must. 

All of the above cases have been for projective SPD-l, and now we
consider unitary SPD-l. For $\discn=1$, there is no difference between
the projective and unitary case, so we begin with $\discn=2$ G-type
with an SPD given as 
\begin{equation}
\spd=\exp(ig\Delta\hat{d})\exp(\gamma\hat{K})\exp(-ig\Delta\hat{d}),
\end{equation}
where $g$ is a real number and the unconstrained projector is unitary.
We reparameterize the interacting projector as

\begin{align}
\hat{P}_{1} & =1+iu\Delta\hat{d}=\hat{P}_{2}^{\dagger}.
\end{align}
The non-interacting integer time Green's function is the same as in
the projective case, given in Eq. \ref{eq:exAIMg0n2}. The integer
time correlation functions needed to compute the total energy are

\begin{align}
 & \langle\text{T}\hat{P}_{1,I}\left(1\right)\hat{P}_{2,I}\left(2\right)\rangle_{\spd_{0}}=\frac{u^{2}}{16}+1,\\
 & \langle\text{T}\hat{P}_{1,I}\left(1\right)\hat{P}_{2,I}\left(2\right)\Delta\hat{d}_{I}\left(2\right)\rangle_{\spd_{0}}=0,\\
 & \langle\text{T}\hat{P}_{1,I}\left(1\right)\hat{P}_{2,I}\left(2\right)\hat{f}_{\sigma,I}^{\dagger}\left(2\right)\hat{c}_{\sigma,I}\left(2\right)\rangle_{\spd_{0}}=\frac{-\nu}{16}(u^{2}-16).
\end{align}
The relevant expectation values can be obtained as 
\begin{align}
 & \langle\hat{f}^{\dagger}\hat{c}\rangle_{\spd}=\frac{16-u^{2}}{16+u^{2}}\nu, & \langle\Delta\hat{d}\rangle_{\spd}=0.
\end{align}
The ground state energy can then be obtained as
\begin{align}
\mathcal{E} & =\min_{\nu\in[-\frac{1}{2},\frac{1}{2}],u}\left(4t\nu\frac{16-u^{2}}{16+u^{2}}\right)=-2t,
\end{align}
which is identical to $\discn=1$ for the projective G-type case.
Similarly, for the unitary case of the B-type at $\discn=2$, the
result is identical to the projective B-type at $\discn=1$. 

The unitary case for the G-type at $\discn=3$ does indeed recover
the exact solution (details not shown), similar to the projective
case for the G-type at $\discn=2$. Therefore, we see that the projective
SPD is clearly superior to the unitary SPD in this Hamiltonian. This
same trend was found when comparing the variational coupled cluster
approach, which is a projective SPD, and the unitary coupled cluster
approach, which is a unitary SPD, in the context of the Lipkin Hamiltonian\cite{Harsha2018044107}.

\section{The Discrete Action Theory \label{sec:The-Discrete-Action-Theory}}

\subsection{Introducing and categorizing the discrete action\label{subsec:Introducing-the-Discrete-Action}}

We have illustrated that the SPD can be evaluated using only the non-interacting
integer time Green's function and the integer time Wick's theorem,
allowing for the efficient evaluation of the SPD-l (see Section \ref{sec:Integer-Time-Green's}).
However, this path forward will not be able to extend to more complex
scenarios such as SPD-d, where the interacting projector is not strictly
local, which will require the summation of an infinite number of diagrams.
A more sophisticated approach is needed, which motivates the introduction
of a discrete action. 

We begin by introducing both the integer time Heisenberg and Schrodinger
representations. As discussed in Subsection \ref{subsec:Integer-Time-Correlation},
the integer time evolution in the integer time interaction representation
is defined as
\begin{align}
 & \hat{O}_{I}\left(\tau\right)=\hat{U}_{\tau;I}\hat{O}\hat{U}_{\tau;I}{}^{-1},\\
 & \hat{U}_{\tau;I}=\exp\left(\sppm_{1}\cdot\spom\right)\dots\exp\left(\sppm_{\tau}\cdot\spom\right),
\end{align}
where $\tau=1,\dots,\discn$. The integer time Green's function under
an SPD in the integer time interaction representation is then
\begin{align}
[\boldsymbol{g}]_{k\tau,k'\tau'} & =\frac{\langle\text{T}\big(\prod_{\tau=1}^{\discn}\hat{P}_{\tau,I}(\tau)\big)\hat{a}_{k,I}^{\dagger}\left(\tau\right)\hat{a}_{k',I}\left(\tau'\right)\rangle_{\spd_{0}}}{\langle\text{T}\big(\prod_{\tau=1}^{\discn}\hat{P}_{\tau,I}(\tau)\big)\rangle_{\spd_{0}}},
\end{align}
where $\tau=1,\dots,\discn$ and $k=1,\dots,L$. Therefore, $\boldsymbol{g}$
is a matrix of dimension $L\discn\times L\discn$, and plays a similar
role to the usual many-particle Green's function. 

In the integer time Heisenberg representation, integer time evolution
of operators is given as 
\begin{align}
\hat{O}(\tau) & =\hat{U}_{\tau}\hat{O}\hat{U}_{\tau}^{-1}, & \hat{U}_{\tau}=\hat{\mathcal{P}}_{1}\dots\hat{\mathcal{P}}_{\tau}.\label{eq:heisenberg_rep_def}
\end{align}
The integer time Green's function under an SPD in the integer time
Heisenberg representation is then
\begin{align}
[\boldsymbol{g}]_{k\tau,k'\tau'} & =\langle\text{T}\hat{a}_{k}^{\dagger}\left(\tau\right)\hat{a}_{k'}\left(\tau'\right)\rangle_{\spd}.
\end{align}
In the integer time Schrodinger representation, integer time evolution
is defined as

\begin{equation}
\hat{O}_{S}(\tau)=\hat{O},
\end{equation}
where the time index now only serves the purpose of tracking which
integer time an operator is associated with, such that time ordering
can be performed. The integer time Green's function under an SPD in
the integer time Schrodinger representation is then
\begin{align}
[\boldsymbol{g}]_{k\tau,k'\tau'} & =\frac{\langle\text{T}\big(\prod_{\tau=1}^{\discn}\hat{\mathcal{P}}_{\tau,S}(\tau)\big)\hat{a}_{k,S}^{\dagger}\left(\tau\right)\hat{a}_{k',S}\left(\tau'\right)\rangle_{\hat{1}}}{\langle\text{T}\big(\prod_{\tau=1}^{\discn}\hat{\mathcal{P}}_{\tau,S}(\tau)\big)\rangle_{\hat{1}}},
\end{align}

A more general integer time correlation function under the SPD can
be represented in the Heisenberg, interaction, and Schrodinger representation,
respectively, as 

\begin{alignat}{1}
 & \langle\textrm{T}\hat{O}_{1}(\tau_{1})\dots\hat{O}_{M}(\tau_{M})\rangle_{\spd}\nonumber \\
 & =\frac{\langle\textrm{T}\big(\prod_{\tau=1}^{\discn}\hat{P}_{\tau,I}(\tau)\big)\hat{O}_{1,I}(\tau_{1})\dots\hat{O}_{M,I}(\tau_{M})\rangle_{\spd_{0}}}{\langle\textrm{T}\big(\prod_{\tau=1}^{\discn}\hat{P}_{\tau,I}(\tau)\big)\rangle_{\spd_{0}}}\\
 & =\frac{\langle\textrm{T}\big(\prod_{\tau=1}^{\discn}\hat{\mathcal{P}}_{\tau,S}(\tau)\big)\hat{O}_{1,S}(\tau_{1})\dots\hat{O}_{M,S}(\tau_{M})\rangle_{\hat{1}}}{\langle\textrm{T}\big(\prod_{\tau=1}^{\discn}\hat{\mathcal{P}}_{\tau,S}(\tau)\big)\rangle_{\hat{1}}}.
\end{alignat}
We therefore have the three standard pictures for describing integer
time correlations. 

We now introduce the most general integer time correlation function,
which is not necessarily associated with an SPD, and this is most
naturally expressed in the Schrodinger represenation as 
\begin{alignat}{1}
 & \frac{\langle\textrm{T}\hat{\mathcal{A}}\hat{O}_{1,S}(\tau_{1})\dots\hat{O}_{M,S}(\tau_{M})\rangle_{\hat{1}}}{\langle\textrm{T}\hat{\mathcal{A}}\rangle_{\hat{1}}},
\end{alignat}
where $\mathbb{\hat{\mathcal{A}}}$ is a \textit{discrete action}
(DA) which characterizes all possible integer time correlations for
a given $\discn$ and is defined as 
\begin{equation}
\hat{\mathcal{A}}=\sum_{\substack{\eta_{1}..\eta_{\discn}\\
\eta'_{1}...\eta'_{\discn}
}
}\mathcal{A}_{\eta_{1}..\eta_{\discn},\eta'_{1}..\eta'_{\discn}}\hat{X}_{\eta_{1}\eta'_{1},S}\left(1\right)...\hat{X}_{\eta_{\discn}\eta'_{\discn},S}\left(\discn\right),\label{eq:general_discrete_action_physical}
\end{equation}
where $\hat{X}_{\Gamma\Gamma'}=|\Gamma\rangle\langle\Gamma'|$ is
a Hubbard operator, $\{|\eta_{\tau}\rangle\}$ forms an orthonormal
basis for the Fock space, and $\mathcal{A}_{\eta_{1}..\eta_{\discn},\eta'_{1}..\eta'_{\discn}}$
is the discrete action represented in the given basis. In the case
of the SPD, the discrete action is a product of $\discn$ distinct
operators 
\begin{equation}
\hat{\mathcal{A}}=\hat{\mathcal{P}}_{1,S}(1)\dots\hat{\mathcal{P}}_{\discn,S}(\discn).\label{eq:simple_discrete_action_physical}
\end{equation}
Given a system with $L$ spin orbitals, a general discrete action
for $\discn$ integer time steps will contain at most $4^{L\discn}$
nonzero entries, which is exponentially larger than the discrete action
of an SPD where there are at most $4^{L}\discn$ nonzero entries.
The more general discrete action will prove useful in practical applications.

\begin{figure}
\includegraphics[width=1\columnwidth]{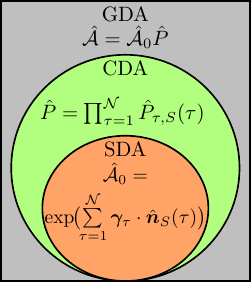}

\caption{\label{fig:schem_da} A schematic for the classification of discrete
actions. The SDA is the discrete action corresponding to an SPD. The
CDA is the discrete action where the corresponding non-interacting
discrete action is integer time mixed. The GDA is a completely general
discrete action. The SDA is a subset of the CDA, which is a subset
of the GDA.}
\end{figure}

We now categorize the common types of discrete actions, which can
naturally be broken down into three varieties (see Figure \ref{fig:schem_da}
for a schematic): sequential discrete actions (SDA), canonical discrete
actions (CDA), and general discrete actions (GDA). We start with GDA,
which is defined as
\begin{align}
 & \hat{\mathcal{A}}=\hat{\mathcal{A}}_{0}\hat{P},\hspace{1em}\hat{\mathcal{A}}_{0}=\hat{\mathcal{A}}_{G}(\boldsymbol{g}_{0}),\label{eq:gda}\\
 & \hat{\mathcal{A}}_{G}(\boldsymbol{g}_{0})=\exp([\boldsymbol{v}_{Q}(\boldsymbol{g}_{0})]_{k\tau,k'\tau'}\hat{a}_{k,S}^{\dagger}(\tau)\hat{a}_{k',S}(\tau')),\\
 & \hat{P}=P_{\eta_{1}..\eta_{\discn},\eta'_{1}..\eta'_{\discn}}\hat{X}_{\eta_{1}\eta'_{1},S}\left(1\right)...\hat{X}_{\eta_{\discn}\eta'_{\discn},S}\left(\discn\right),\label{eq:proj_full}\\
 & \boldsymbol{v}_{Q}(\boldsymbol{g}_{0})=\ln((\boldsymbol{g}_{0}^{-1}-1)^{-1}(\boldsymbol{g}_{Q}^{-1}-1))^{T},\label{eq:vq}\\
 & [\boldsymbol{g}_{Q}]_{k\tau,k'\tau'}=\langle\text{T}\hat{a}_{k,S}^{\dagger}\left(\tau\right)\hat{a}_{k',S}\left(\tau'\right)\rangle_{\hat{1}},\label{eq:gqsrep}
\end{align}
where the Einstein summation convention has been used, and the derivation
of Eq. \ref{eq:vq} is given in Subsection \ref{subsec:Integer-time-Path-integral}.
Therefore, we see that the GDA is composed of a non-interacting discrete
action $\hat{\mathcal{A}}_{0}$ and an interacting projector $\hat{P}$,
where both have no constraint with respect to the integer time structure.

The CDA is a GDA by restricting the interacting projector into a time
blocked form as
\begin{align}
\hat{P} & =\prod_{\tau=1}^{\discn}\hat{P}_{\tau,S}(\tau),
\end{align}
which will be very useful in evaluating the SPD-d (see Section \ref{sec:Self-consistent-Canonical-Discre}).
Finally, the SDA is the discrete action corresponding to an SPD, which
can be viewed as a CDA by restricting the non-interacting discrete
action into the following form
\begin{equation}
\hat{\mathcal{A}}_{0}=\exp(\sum_{\tau=1}^{\discn}\boldsymbol{\gamma}_{\tau}\cdot\spom_{S}(\tau)),
\end{equation}
where $\spom$ was defined in Eq. \ref{eq:spom}.

The form of the CDA and GDA requires the evaluation of an integer
time ordered expression with nontrivial integer time structure, which
is inconvenient to manipulate. Therefore, it is useful to develop
a more adept mathematical framework which is conducive to handling
such scenarios. This motivates the introduction of the compound space,
which is the topic of the next subsection. 

\subsection{Integer time path integral and the compound space\label{subsec:Integer-time-Path-integral}}

Here we prescribe a mathematical formalism to recast the integer time
ordered correlations of a general discrete action as a static measurement
under an effective density matrix in a compound quantum system. This
can be viewed as a reformulation of the path integral. In the usual
path integral formalism, one can interpret the action as the effective
energy of a classical system (though for Fermions one needs Grassmann
numbers, which have no classical counterpart) with the same spatial
structure and one more dimension for the time correlation\cite{Negele19980738200522}.
Thus, we have the well-known fact that $d$ dimensional quantum fields
correspond to a ($d$+1) dimensional classical system. In the following,
we reformulate this mapping, resulting in two key differences. First,
as the number of time steps is finite, the number of points within
the extra dimension is also finite. Second, for the evolution of each
time step, we need to retain an exact operator form given that the
$\mathcal{\hat{P}}_{i}$ can not be treated as an infinitesimally
small expansion from the identity matrix. As a result, we obtain a
compound quantum system with a Fock space of $\mathbb{H}_{c}=\otimes_{\tau=1}^{\discn}\mathbb{H}$,
where $\mathbb{H}$ is the Fock space of the original system. 

To begin, we must define how we represent operators from the original
system in the compound system. Each creation and destruction operator
will be attached to a given integer time index when promoted to the
compound space. For example, for a system with $L$ spin orbitals,
any operator can be built algebraically from the $2L$ operators $\hat{a}_{1}^{\dagger},\dots,\hat{a}_{L}^{\dagger}$
and $\hat{a}_{1},\dots,\hat{a}_{L}$. Given $\discn$ time steps,
any operator for the compound system can be built algebraically from
$2L\discn$ operators $\barhat[a]_{1}^{\dagger(1)},\dots,\barhat[a]_{L}^{\dagger(1)},\dots,\barhat[a]_{1}^{\dagger(\tau)},\dots,\barhat[a]_{L}^{\dagger(\tau)},\dots,\barhat[a]_{1}^{\dagger(\discn)},\dots,\barhat[a]_{L}^{\dagger(\discn)}$
and $\barhat[a]_{1}^{(1)},\dots,\barhat[a]_{L}^{(1)},\dots,\barhat[a]_{1}^{(\tau)},\dots,\barhat[a]_{L}^{(\tau)},\dots,\barhat[a]_{1}^{(\discn)},\dots,\barhat[a]_{L}^{(\discn)}$
, where the underscore denotes an operator in the compound space,
and the superscript $\tau$ associates the operator with integer time
$\tau$. These operators in the compound space must obey the Fermionic
anti-commutation relations, yielding 
\begin{align}
 & \left\{ \barhat[a]_{k}^{\dagger(\tau)},\barhat[a]_{k'}^{(\tau')}\right\} =\delta_{kk'}\delta_{\tau\tau'},\\
 & \left\{ \barhat[a]_{k}^{(\tau)},\barhat[a]_{k'}^{(\tau')}\right\} =\left\{ \barhat[a]_{k}^{\dagger(\tau)},\barhat[a]_{k'}^{\dagger(\tau')}\right\} =0.
\end{align}
The commutation relations indicate that the time index has the same
behavior as the orbital index in the compound space. In order to promote
some generic operator $\hat{O}=f(\hat{a}_{1}^{\dagger},\dots,\hat{a}_{L}^{\dagger},\hat{a}_{1},\dots,\hat{a}_{L})$
associated with a given $\tau$, we obtain the corresponding operator
in the compound space as $\barhat[O]^{(\tau)}=f(\barhat[a]_{1}^{\dagger(\tau)},\dots,\barhat[a]_{L}^{\dagger(\tau)},\barhat[a]_{1}^{(\tau)},\dots,\barhat[a]_{L}^{(\tau)})$.

\begin{figure}
\includegraphics[width=1\columnwidth]{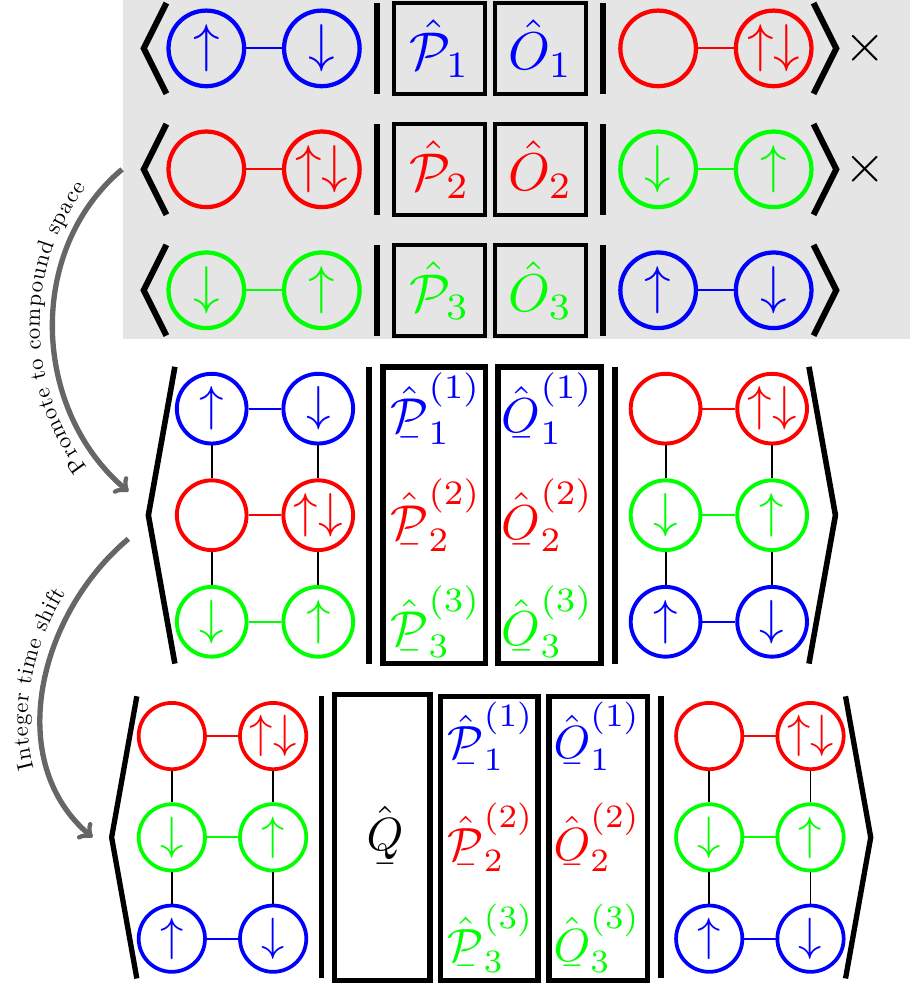}

\caption{\label{fig:schem_cs}A schematic of the integer time path integral
which derives a map to the compound space, using the example of the
Hubbard dimer for $\discn=3$. The first step denotes a particular
snapshot within Eq. \ref{eq:pathintegral1}. The second step merges
the state vectors from different time steps into the compound space,
while the operators are promoted, illustrating Eq. \ref{eq:pathintegral5}.
The final step illustrates the integer time shift operator $\barhat[Q]$,
which results in a diagonal matrix element, corresponding to Eq. \ref{eq:pathintegral7}.}
\end{figure}

Now we proceed to derive the mapping to the compound space in the
special case of the SPD (see Figure \ref{fig:schem_cs} for a schematic
of the following derivation). Considering a general integer time correlation
function under the SPD and performing $\discn$ identity insertions

\begin{widetext}
\begin{align}
 & \langle\textrm{T}\hat{O}_{1}(1)\dots\hat{O}_{\discn}(\discn)\rangle_{\spd}=\textrm{Tr}\ensuremath{(\hat{\mathcal{P}}_{1}\hat{O}_{1}\dots\hat{\mathcal{P}}_{\discn}\hat{O}_{\discn})}/\textrm{Tr}\ensuremath{(\hat{\mathcal{P}}_{1}\dots\hat{\mathcal{P}}_{\discn})}\\
 & =\frac{1}{C}\sum_{\Gamma_{1}\dots\Gamma_{\discn}}\langle\Gamma_{1}|\exp(\sppm_{1}\cdot\spom)\hat{P}_{1}\hat{O}_{1}|\Gamma_{2}\rangle\langle\Gamma_{2}|\exp(\sppm_{2}\cdot\spom)\hat{P}_{2}\hat{O}_{2}|\Gamma_{3}\rangle\dots\langle\Gamma_{\discn}|\exp(\sppm_{\discn}\cdot\spom)\hat{P}_{\discn}\hat{O}_{\discn}|\Gamma_{1}\rangle\label{eq:pathintegral1}\\
 & =\frac{1}{C}\sum_{\Gamma_{1}\dots\Gamma_{\discn}}\langle\Gamma_{1}|\left(\sum_{\Gamma\Gamma'}c_{1;\Gamma\Gamma'}\hat{X}_{\Gamma\Gamma'}\right)|\Gamma_{2}\rangle\langle\Gamma_{2}|\left(\sum_{\Gamma\Gamma'}c_{2;\Gamma\Gamma'}\hat{X}_{\Gamma\Gamma'}\right)|\Gamma_{3}\rangle\dots\langle\Gamma_{\discn}|\left(\sum_{\Gamma\Gamma'}c_{\discn;\Gamma\Gamma'}\hat{X}_{\Gamma\Gamma'}\right)|\Gamma_{1}\rangle\label{eq:pathintegral2}\\
 & =\frac{1}{C}\sum_{\Gamma_{1}\dots\Gamma_{\discn}}c_{1;\Gamma_{1}\Gamma_{2}}c_{2;\Gamma_{2}\Gamma_{3}}\dots c_{\discn;\Gamma_{\discn}\Gamma_{1}}\frac{\langle\Gamma_{1}\otimes\Gamma_{2}\dots\otimes\Gamma_{\discn}|\barhat[X]_{\Gamma_{1}\Gamma_{2}}^{(1)}\barhat[X]_{\Gamma_{2}\Gamma_{3}}^{(2)}\dots\barhat[X]_{\Gamma_{\discn}\Gamma_{1}}^{(\discn)}|\Gamma_{2}\otimes\Gamma_{3}\dots\otimes\Gamma_{1}\rangle}{\langle\Gamma_{1}\otimes\Gamma_{2}\dots\otimes\Gamma_{\discn}|\barhat[X]_{\Gamma_{1}\Gamma_{2}}^{(1)}\barhat[X]_{\Gamma_{2}\Gamma_{3}}^{(2)}\dots\barhat[X]_{\Gamma_{\discn}\Gamma_{1}}^{(\discn)}|\Gamma_{2}\otimes\Gamma_{3}\dots\otimes\Gamma_{1}\rangle}\label{eq:pathintegral3}\\
 & =\frac{1}{C}\sum_{\Gamma_{1}...\Gamma_{\discn}}\theta_{\Gamma_{2}\Gamma_{3}...\Gamma_{1}}\langle\Gamma_{1}\otimes\Gamma_{2}...\otimes\Gamma_{\discn}|\left(\sum_{\Gamma\Gamma'}c_{1;\Gamma\Gamma'}\barhat[X]_{\Gamma\Gamma'}^{(1)}\right)\left(\sum_{\Gamma\Gamma'}c_{2;\Gamma\Gamma'}\barhat[X]_{\Gamma\Gamma'}^{(2)}\right)...\left(\sum_{\Gamma\Gamma'}c_{\discn;\Gamma\Gamma'}\barhat[X]_{\Gamma\Gamma'}^{(\discn)}\right)|\Gamma_{2}\otimes\Gamma_{3}...\otimes\Gamma_{1}\rangle\label{eq:pathintegral4}\\
 & =\frac{1}{C}\sum_{\Gamma_{1}\dots\Gamma_{\discn}}\theta_{\Gamma_{2}\Gamma_{3}\dots\Gamma_{1}}\langle\Gamma_{1}\otimes\Gamma_{2}\dots\otimes\Gamma_{\discn}|\exp(\sppm_{1}\cdot\spomcs^{(1)})\barhat[P]_{1}^{(1)}\barhat[O]_{1}^{(1)}\dots\exp(\sppm_{\discn}\cdot\spomcs^{(\discn)})\barhat[P]_{\discn}^{(\discn)}\barhat[O]_{\discn}^{(\discn)}|\Gamma_{2}\otimes\Gamma_{3}\dots\otimes\Gamma_{1}\rangle\label{eq:pathintegral5}\\
 & =\frac{1}{C}\text{Tr}\left(\barhat[Q]\exp(\sppm_{1}\cdot\spomcs^{(1)})\barhat[P]_{1}^{(1)}\barhat[O]_{1}^{(1)}\dots\exp(\sppm_{\discn}\cdot\spomcs^{(\discn)})\barhat[P]_{\discn}^{(\discn)}\barhat[O]_{\discn}^{(\discn)}\right)\label{eq:pathintegral6}\\
 & =\frac{1}{C}\text{Tr}\left(\barhat[Q]\exp(\sppm_{1}\cdot\spomcs^{(1)})\dots\exp(\sppm_{\discn}\cdot\spomcs^{(\discn)})\barhat[P]_{1}^{(1)}\dots\barhat[P]_{\discn}^{(\discn)}\barhat[O]_{1}^{(1)}\dots\barhat[O]_{\discn}^{(\discn)}\right),\label{eq:pathintegral7}
\end{align}
where
\begin{align}
 & \theta_{\Gamma_{2}\Gamma_{3}..\Gamma_{1}}=\theta_{\Gamma_{2}\Gamma_{3}..\Gamma_{1}}^{-1}=\langle\Gamma_{1}\otimes\Gamma_{2}..\otimes\Gamma_{\discn}|\barhat[X]_{\Gamma_{1}\Gamma_{2}}^{(1)}\dots\barhat[X]_{\Gamma_{\discn-1}\Gamma_{\discn}}^{(\discn-1)}\barhat[X]_{\Gamma_{\discn}\Gamma_{1}}^{(\discn)}|\Gamma_{2}\otimes\Gamma_{3}..\otimes\Gamma_{1}\rangle,\\
 & \barhat[Q]=\sum_{\Gamma_{1}\dots\Gamma_{\discn}}\theta_{\Gamma_{2}\Gamma_{3}\dots\Gamma_{1}}|\Gamma_{2}\otimes\Gamma_{3}\dots\otimes\Gamma_{1}\rangle\langle\Gamma_{1}\otimes\Gamma_{2}\dots\otimes\Gamma_{\discn}|=\sum_{\Gamma_{1}\dots\Gamma_{\discn}}\barhat[X]_{\Gamma_{1}\Gamma_{\discn}}^{(\discn)}\barhat[X]_{\Gamma_{\discn}\Gamma_{\discn-1}}^{(\discn-1)}\dots\barhat[X]_{\Gamma_{2}\Gamma_{1}}^{(1)},\\
 & C=\text{Tr}(\spd)=\text{Tr}\bigg(\barhat[Q]\exp(\sum_{\tau=1}^{\discn}\sppm_{\tau}\cdot\spomcs^{(\tau)})\barhat[P]_{1}^{(1)}\dots\barhat[P]_{\discn}^{(\discn)}\bigg).
\end{align}
\end{widetext}In Eq. \ref{eq:pathintegral1}, the identity insertions
consist of states $|\Gamma_{\tau}\rangle$ built from applying the
creation operators in the chosen basis set to the empty state. In
Eq. \ref{eq:pathintegral2}, we express the operators in terms of
Hubbard operators $\hat{X}_{\Gamma\Gamma'}=|\Gamma\rangle\langle\Gamma'|$
in the same basis set. In Eq. \ref{eq:pathintegral3}, we insert unity
in order to connect the matrix elements with their corresponding operators
in the compound space, where each Hubbard operator is formed via the
promotion rules described above. In Eq. \ref{eq:pathintegral4}, the
quantity $\theta_{\Gamma_{2}\Gamma_{3}..\Gamma_{1}}=\pm1$, and we
have introduced a summation over all Hubbard operators given that
other terms will not contribute. Eq. \ref{eq:pathintegral5} holds
given that the promotion of operators to the compound space is linear.
Eq. \ref{eq:pathintegral6} introduces the shift operator $\barhat[Q]$
which allows one to recast the sum as a trace, and can be recognized
as the integer time version of $\exp(\int_{0}^{\beta}d\tau\bar{\varphi}(\tau)\partial_{\tau}\varphi(\tau))$
from the usual path integral\cite{Negele19980738200522}. Eq. \ref{eq:pathintegral7}
reorders all of the noninteracting projectors to the left, the interacting
projectors in the middle, and the observables on the right, and it
should be recalled that the operators $\hat{P}_{\tau}$ are Bosonic.
Therefore, we see that an integer time ordered correlation function
evaluated under an SPD is equivalent to a corresponding static expectation
value in the compound space under an effective density matrix
\begin{align}
 & \spdcs=\spdcs_{0}\prod_{\tau=1}^{\discn}\barhat[P]_{\tau}^{(\tau)}, & \spdcs_{0}=\barhat[Q]\exp(\sum_{\tau=1}^{\discn}\sppm_{\tau}\cdot\spomcs^{(\tau)}),\label{eq:SDA_compound_space}
\end{align}
and it should be noted that the density matrix is not Hermitian in
general. The previous derivation corresponded to a quantity which
was initially time ordered, and in general we have 

\begin{equation}
\langle\text{T}\hat{O}_{1}(\tau_{1})\dots\hat{O}_{M}(\tau_{M})\rangle_{\spd}=\langle\barhat[O]_{1}^{(\tau_{1})}\dots\barhat[O]_{M}^{(\tau_{M})}\rangle_{\spdcs}.
\end{equation}

Notice that the representation of $\barhat[Q]$ in the compound space
is completely determined from the convention in which we combine the
$\discn$ copies of the original system into the compound system.
Therefore, we can straightforwardly determine $\barhat[Q]$ by studying
a special SPD $\spd_{Q}$ with $\mathcal{\hat{P}}_{\tau}=\hat{1}$,
where we have 
\begin{align}
\frac{\text{Tr}\left(\barhat[Q]\spomcs\right)}{\text{Tr}\left(\barhat[Q]\right)}=\boldsymbol{g}_{Q},\label{eq:gQ}
\end{align}
where $\boldsymbol{g}_{Q}$ is the single-particle integer time Green's
function for $\spd_{Q}$, which is equivalent to the definition in
the Schrodinger representation given in Eq. \ref{eq:gqsrep}. Given
that any correlator within the noninteracting SPD can be evaluated
using the integer time Wick's theorem (see Appendix \ref{appendix:wicks_theorem}),
this indicates that $\barhat[Q]$ must also be a noninteracting density
matrix in the compound space. From the Lie group properties of the
non-interacting density matrix (see Appendix \ref{subsec:Appendix-Lie-group}),
we have 
\begin{equation}
\hat{\rho}_{G}(\boldsymbol{g})=\exp\left(\ln\left(\frac{\boldsymbol{1}}{\bm{g}^{-1}-\boldsymbol{1}}\right)^{T}\cdot\spomcs\right),
\end{equation}
which implies that the integer time Green's function uniquely determines
a non-interacting density matrix in the compound space. Therefore,
we can determine
\begin{align}
 & \barhat[Q]=\hat{\rho}_{G}\left(\boldsymbol{g}_{Q}\right).
\end{align}
Given that $\barhat[Q]$ is independent of the discrete action, this
specific determination of $\barhat[Q]$ applies in general.

Recall that Eq. \ref{eq:pathintegral7} is specific to the case of
an SPD, and we can generalize to the case of a general discrete action
as 
\begin{equation}
\spdcs=\barhat[Q]\barhat[\mathcal{A}],\label{eq:general_discrete_action_compound_space}
\end{equation}
where $\barhat[\mathcal{A}]$ is the promotion of $\hat{\mathcal{A}}$,
given as 
\begin{align}
 & \barhat[\mathcal{A}]=\exp(\boldsymbol{v}_{Q}(\boldsymbol{g}_{0})\cdot\spomcs)P_{\eta_{1}..\eta_{\discn},\eta'_{1}..\eta'_{\discn}}\barhat[X]_{\eta_{1}\eta'_{1}}^{\left(1\right)}...\barhat[X]_{\eta_{\discn}\eta'_{\discn}}^{\left(\discn\right)},
\end{align}
where the Einstein summation convention has been used. We now proceed
to derive the expression for $\boldsymbol{v}_{Q}(\boldsymbol{g}_{0})$
as 
\begin{align}
 & \spdcs_{0}=\barhat[Q]\exp(\boldsymbol{v}_{Q}(\boldsymbol{g}_{0})\cdot\spomcs)=\hat{\rho}_{G}\left(\boldsymbol{g}_{0}\right)\rightarrow\\
 & \boldsymbol{v}_{Q}(\boldsymbol{g}_{0})=\ln((\boldsymbol{g}_{0}^{-1}-1)^{-1}(\boldsymbol{g}_{Q}^{-1}-1))^{T}.
\end{align}
A general integer time correlation function can then be written as
\begin{alignat}{1}
 & \frac{\langle\textrm{T}\hat{\mathcal{A}}\hat{O}_{1,S}(\tau_{1})...\hat{O}_{M,S}(\tau_{M})\rangle_{\hat{1}}}{\langle\textrm{T}\hat{\mathcal{A}}\rangle_{\hat{1}}}=\langle\barhat[O]_{1}^{(\tau_{1})}...\barhat[O]_{M}^{(\tau_{M})}\rangle_{\spdcs}.
\end{alignat}
We see that $\spdcs$ is a representation of the discrete action in
the compound space, and we also refer to $\spdcs$ as the discrete
action. Finally, the expression for the general discrete action in
the compound space can be written as
\begin{equation}
\spdcs=\hat{\rho}_{G}(\boldsymbol{g}_{0})\barhat[P].\label{eq:general_discrete_action_compound_space-1}
\end{equation}
The CDA can also be written in the compound space, given as 

\begin{align}
\spdcs & =\hat{\rho}_{G}(\boldsymbol{g}_{0})\prod_{\tau=1}^{\discn}\barhat[P]_{\tau}^{(\tau)}.
\end{align}

Given the abstract nature of the compound space, it is useful to consider
some simple examples. Consider a non-interacting SPD with a single
degree of freedom and $\discn=2$
\begin{equation}
\spd_{0}=\exp(\gamma_{1}\hat{a}^{\dagger}\hat{a})\exp(\gamma_{2}\hat{a}^{\dagger}\hat{a}),
\end{equation}
where 

\begin{align}
\hat{a}^{\dagger} & =\left(\begin{array}{cc}
0 & 0\\
1 & 0
\end{array}\right), & \hat{a} & =\left(\begin{array}{cc}
0 & 1\\
0 & 0
\end{array}\right), & \hat{n}=\hat{a}^{\dagger}\hat{a}=\left(\begin{array}{cc}
0 & 0\\
0 & 1
\end{array}\right),
\end{align}
assuming an ordering of the basis as $|0\rangle,|1\rangle$. Consider
the time ordered expectation value which can be directly evaluated
as
\begin{equation}
\langle\textrm{T}\hat{a}_{I}^{\dagger}(1)\hat{a}_{I}(2)\rangle_{\spd_{0}}=\frac{\exp(\gamma_{1})}{1+\exp(\gamma_{1}+\gamma_{2})}.
\end{equation}
Alternatively, we can evaluate the same quantity using the compound
space. We begin by promoting all operators to the compound space as
\begin{align}
\barhat[a]^{\dagger(1)} & =\left(\begin{array}{cccc}
0 & 0 & 0 & 0\\
1 & 0 & 0 & 0\\
0 & 0 & 0 & 0\\
0 & 0 & 1 & 0
\end{array}\right), & \barhat[a]^{\dagger(2)}=\left(\begin{array}{cccc}
0 & 0 & 0 & 0\\
0 & 0 & 0 & 0\\
1 & 0 & 0 & 0\\
0 & -1 & 0 & 0
\end{array}\right),
\end{align}
where we have chosen an ordering of the basis as $|00\rangle$, $|01\rangle$,
$|10\rangle$, and $|11\rangle$, where 
\begin{align}
|n_{2}n_{1}\rangle=(\barhat[a]^{\dagger(1)})^{n_{1}}(\barhat[a]^{\dagger(2)})^{n_{2}}|00\rangle.
\end{align}
The operator $\barhat[Q]$ can then be constructed as
\begin{align}
\barhat[Q]=\left(\begin{array}{cccc}
1 & 0 & 0 & 0\\
0 & 0 & -1 & 0\\
0 & 1 & 0 & 0\\
0 & 0 & 0 & 1
\end{array}\right)
\end{align}
and the effective density matrix is then
\begin{equation}
\spdcs_{0}=\barhat[Q]\exp(\gamma_{1}\barhat[a]^{\dagger(1)}\barhat[a]^{(1)}+\gamma_{2}\barhat[a]^{(2)}\barhat[a]^{(2)}).
\end{equation}
Finally, we can compute the desired expectation value as
\begin{equation}
\langle\barhat[a]^{\dagger(1)}\barhat[a]^{(2)}\rangle_{\spdcs_{0}}=\frac{\exp(\gamma_{1})}{1+\exp(\gamma_{1}+\gamma_{2})},
\end{equation}
recovering the previous result. This simple example illustrates how
an integer time ordered correlation function is equivalent to a corresponding
static expectation value in the compound space. 

It is useful to explore the integer time correlation function from
the viewpoint of coherent states. We begin by writing the integer
time correlation function using the standard path integral as

\global\long\def\no#1#2{\lfloor#1\rfloor_{#2}}%

\begin{align}
 & \langle\hat{O}_{1}(1)...\hat{O}_{\discn}(\discn)\rangle_{\spd}\nonumber \\
 & =\frac{\langle\hat{P}_{1,I}(1)\hat{O}_{1,I}(1)...\hat{P}_{\discn,I}(\discn)\hat{O}_{\discn,I}(\discn)\rangle_{\spd_{0}}}{\langle\hat{P}_{1,I}(1)...\hat{P}_{\discn,I}(\discn)\rangle_{\spd_{0}}}\\
 & =\frac{\int D[\bar{\varphi},\varphi]\exp(S_{c})\no{\hat{P}_{1}\hat{O}_{1}}1\dots\no{\hat{P}_{\discn}\hat{O}_{\discn}}{\discn}}{\int D[\bar{\varphi},\varphi]\exp(S_{c})\no{\hat{P}_{1}}1\dots\no{\hat{P}_{\discn}}{\discn}},\label{eq:pathintegralline2}
\end{align}
where the notation $\no{\hat{O}}{\tau}$ indicates that the operator
$\hat{O}$ is first transformed into a normal ordered form and then
all creation/annihilation operators are converted to the corresponding
Grassmann numbers at the time index $\tau$; and the action $S_{c}$
is the non-interacting action which has been separated into $\discn$
pieces

\begin{align}
S_{c} & =\sum_{kk'}\sum_{\tau=1}^{\discn}\int_{\tau-1}^{\tau}d\tau'\bar{\varphi}_{k\tau'}(\frac{\partial}{\partial\tau'}\delta_{kk'}+\boldsymbol{\gamma}_{\tau kk'})\varphi_{k'\tau'},
\end{align}
where $\varphi_{k,\tau}$ is the Grassmann variable corresponding
to the operator $\hat{a}_{k}$ at time $\tau$ (it should be noted
that our time convention is opposite to the usual definition\cite{Mahan20000306463385}).
Given that Eq. \ref{eq:pathintegralline2} only measures the correlations
at integer times, we can trace out the Grassmann numbers in the intervals
between integer times as
\begin{align}
 & \langle\hat{O}_{1}(1)\dots\hat{O}_{\discn}(\discn)\rangle_{\spd}=\nonumber \\
 & \frac{\int(\prod_{k\tau}d\bar{\varphi}_{k,\tau}d\varphi_{k,\tau})\exp(S_{d})\no{\hat{P}_{1}\hat{O}_{1}}1...\no{\hat{P}_{\discn}\hat{O}_{\discn}}{\discn}}{\int(\prod_{k\tau}d\bar{\varphi}_{k,\tau}d\varphi_{k,\tau})\exp(S_{d})\no{\hat{P}_{1}}1...\no{\hat{P}_{\discn}}{\discn}},\label{eq:discrete_path_integral}
\end{align}
where
\begin{align}
S_{d} & =\sum_{kk'}\sum_{\tau,\tau'=1}^{\discn}\bar{\varphi}_{k\tau}[\bm{g}_{0}^{-1}]_{k'\tau',k\tau}\varphi_{k'\tau'}=\bar{\boldsymbol{\varphi}}^{T}(\bm{g}_{0}^{-1})^{T}\boldsymbol{\varphi}
\end{align}
and $\bm{g}_{0}$ is the non-interacting integer time Green's function
defined in Eq. \ref{eq:g0interactionrep}. Eq. \ref{eq:discrete_path_integral}
is an exact evaluation, and this is an alternate viewpoint to the
discrete action in the compound space. However, due to the requirement
of the normal ordering procedure in Eq. \ref{eq:discrete_path_integral},
one cannot directly apply standard techniques such as the generating
functional. We can now see that a great advantage of the compound
space is that it circumvents the need to transform an operator into
normal ordered form. 

\subsection{Discrete generating function and discrete Dyson equation\label{subsec:Discrete-Generating-Function-and-Dyson}}

Two key concepts in many-body physics are the interacting single-particle
Green's function and the self-energy, which relates the non-interacting
Green's function to the the interacting Green's function via the Dyson
equation. Here we will further generalize these quantities to the
integer time case corresponding to a GDA. The interacting integer
time Green's function under the GDA is given as
\begin{align}
[\boldsymbol{g}]_{k\tau,k'\tau'} & =\langle\barhat[a]_{k}^{(\tau)\dagger}\barhat[a]_{k'}^{(\tau')}\rangle_{\spdcs},
\end{align}
where $\tau=1,\dots,\discn$ and $k=1,\dots,L$ (see Eq. \ref{eq:heisenberg_rep_def}
for the definition of the integer time Heisenberg representation).
Therefore, $\boldsymbol{g}$ is a matrix of dimension $L\discn\times L\discn$,
and plays a similar role to the usual many-particle Green's function.
Furthermore, the two particle integer time Green's function is given
as
\begin{align}
 & \langle\barhat[a]_{k_{1}}^{(\tau_{1})\dagger}\barhat[a]_{k_{2}}^{(\tau_{2})\dagger}\barhat[a]_{k_{3}}^{(\tau_{3})}\barhat[a]_{k_{4}}^{(\tau_{4})}\rangle_{\spdcs}.
\end{align}
More generally, our goal in this subsection is to compute an arbitrary
M-particle integer time Green's function from the generating function,
yielding a generalization of the Dyson equation, Bethe-Salpeter equation,
etc.

To proceed, we introduce the generating function to generate $M$-particle
integer time Green's functions for a given SPD 
\begin{align}
 & \tilde{Z}(\boldsymbol{v}_{1},\dots,\boldsymbol{v}_{M})=\nonumber \\
 & \frac{\langle\barhat[P]\exp([\boldsymbol{v}_{1}]_{ij}\barhat[n]_{ij})\dots\exp([\boldsymbol{v}_{M}]_{lk}\barhat[n]_{lk})\rangle_{\spdcs_{0}}}{\langle\exp([\boldsymbol{v}_{1}]_{ij}\barhat[n]_{ij})\dots\exp([\boldsymbol{v}_{M}]_{lk}\barhat[n]_{lk})\rangle_{\spdcs_{0}}},\label{eq:general_generating_function}
\end{align}
where $\boldsymbol{v}$ is the source, $\spdcs_{0}=\hat{\rho}_{G}(\boldsymbol{g}_{0})$
is the non-interacting discrete action, $\barhat[P]$ is a general
interacting projector, the indices $i,j,l,k$ are all two-tuples containing
both an orbital and a time index such that $\barhat[n]_{(k,\tau)(k',\tau')}=\barhat[a]_{k}^{\dagger(\tau)}\barhat[a]_{k'}^{(\tau')}$,
and Einstein notation for summations has been employed (and will be
used throughout this subsection). It should be noted that the Lie
group properties of the non-interacting density matrix (see Section
\ref{subsec:Appendix-Lie-group}) demand that
\begin{align}
 & \tilde{Z}(\boldsymbol{v}_{1},..,\boldsymbol{v}_{M})=\tilde{Z}\left(\boldsymbol{v}\right),\\
 & \boldsymbol{v}=\ln(\exp(\boldsymbol{v}_{1})\dots\exp(\boldsymbol{v}_{M})).
\end{align}

The first step is to derive the M-particle integer time Green's functions
in terms of derivatives of $\tilde{Z}\left(\boldsymbol{v}_{1},\dots,\boldsymbol{v}_{M}\right)$.
Given that we are only concerned with the single-particle and two-particle
integer time Green's function in this paper, we restrict ourselves
to $M=2$. Substituting into the expression for $\tilde{Z}$, we find
\begin{align}
 & \tilde{Z}(\boldsymbol{v}_{1},\boldsymbol{v}_{2})=\nonumber \\
 & \langle\barhat[P]\rangle_{\spdcs_{0}}+[\boldsymbol{v}_{1}]_{ij}(\langle\barhat[P]\barhat[n]_{ij}\rangle_{\spdcs_{0}}-\langle\barhat[P]\rangle_{\spdcs_{0}}\langle\barhat[n]_{ij}\rangle_{\spdcs_{0}})+\nonumber \\
 & [\boldsymbol{v}_{2}]_{lk}(\langle\barhat[P]\barhat[n]_{lk}\rangle_{\spdcs_{0}}-\langle\barhat[P]\rangle_{\spdcs_{0}}\langle\barhat[n]_{lk}\rangle_{\spdcs_{0}})\nonumber \\
 & [\boldsymbol{v}_{1}]_{ij}[\boldsymbol{v}_{2}]_{lk}\bigg(\langle\barhat[P]\barhat[n]_{ij}\barhat[n]_{lk}\rangle_{\spdcs_{0}}-\langle\barhat[P]\rangle_{\spdcs_{0}}\langle\barhat[n]_{ij}\barhat[n]_{lk}\rangle_{\spdcs_{0}}+\nonumber \\
 & 2\langle\barhat[P]\rangle_{\spdcs_{0}}\langle\barhat[n]_{ij}\rangle_{\spdcs_{0}}\langle\barhat[n]_{lk}\rangle_{\spdcs_{0}}-\nonumber \\
 & \langle\barhat[n]_{ij}\rangle_{\spdcs_{0}}\langle\barhat[P]\barhat[n]_{lk}\rangle_{\spdcs_{0}}-\langle\barhat[P]\barhat[n]_{ij}\rangle_{\spdcs_{0}}\langle\barhat[n]_{lk}\rangle_{\spdcs_{0}}\bigg)+\dots
\end{align}
We can now evaluate the first and second derivatives of $\tilde{Z}\left(\boldsymbol{v}_{1},\boldsymbol{v}_{2}\right)$
divided by $\tilde{Z}(\boldsymbol{0},\boldsymbol{0})$ 
\begin{equation}
\frac{1}{\tilde{Z}(\boldsymbol{0},\boldsymbol{0})}\left.\frac{\partial\tilde{Z}\left(\boldsymbol{v}_{1},\boldsymbol{v}_{2}\right)}{\partial[\boldsymbol{v}_{1}]_{ij}}\right|_{\boldsymbol{v}_{\ell}=\boldsymbol{0}}=\langle\barhat[n]_{ij}\rangle_{\spdcs}-\langle\barhat[n]_{ij}\rangle_{\spdcs_{0}},\label{eq:ztilde1}
\end{equation}

\begin{align}
 & \frac{1}{\tilde{Z}(\boldsymbol{0},\boldsymbol{0})}\left.\frac{\partial^{2}\tilde{Z}\left(\boldsymbol{v}_{1},\boldsymbol{v}_{2}\right)}{\partial[\boldsymbol{v}_{1}]_{ij}\partial[\boldsymbol{v}_{2}]_{lk}}\right|_{\boldsymbol{v}_{\ell}=\boldsymbol{0}}=\nonumber \\
 & \langle\barhat[n]_{ij}\barhat[n]_{lk}\rangle_{\spdcs}-\langle\barhat[n]_{ij}\barhat[n]_{lk}\rangle_{\spdcs_{0}}+2\langle\barhat[n]_{ij}\rangle_{\spdcs_{0}}\langle\barhat[n]_{lk}\rangle_{\spdcs_{0}}-\nonumber \\
 & \langle\barhat[n]_{ij}\rangle_{\spdcs_{0}}\langle\barhat[n]_{lk}\rangle_{\spdcs}-\langle\barhat[n]_{ij}\rangle_{\spdcs}\langle\barhat[n]_{lk}\rangle_{\spdcs_{0}},\label{eq:ztilde2}
\end{align}
where $\boldsymbol{v}_{\ell}=\boldsymbol{0}$ for $\ell=1,2$, and
we used $\langle\barhat[O]\rangle_{\spdcs}=\langle\barhat[P]\barhat[O]\rangle_{\spdcs_{0}}/\langle\barhat[P]\rangle_{\spdcs_{0}}$.

We now have all of the information we need to construct arbitrary
one and two particle integer time Green's functions. However, this
formulation is somewhat inconvenient given that we will use Wick's
theorem to evaluate the generating function which necessitates the
use of $\boldsymbol{g}_{0}$ instead of $\boldsymbol{v}$. Therefore,
it is convenient to perform a change of variables into the non-interacting
single-particle integer time Green's function
\begin{align}
 & Z(\boldsymbol{g}_{0})\equiv\langle\barhat[P]\rangle_{\hat{\rho}_{G}(\boldsymbol{g}_{0})}=\tilde{Z}\left(\boldsymbol{v}(\boldsymbol{g}_{0})\right),\\
 & \boldsymbol{v}(\boldsymbol{g}_{0})\equiv\ln\left(\left(\frac{\boldsymbol{1}}{\bm{g}_{0}^{-1}-\boldsymbol{1}}\right)^{T}\left(\bm{g}_{0}^{\star-1}-\boldsymbol{1}\right)^{T}\right),
\end{align}
where $\boldsymbol{v}(\boldsymbol{g}_{0})$ was obtained by inverting
the following relation 
\begin{equation}
\boldsymbol{g}_{0}(\boldsymbol{v})=\langle\boldsymbol{\barhat[n]}\rangle_{\exp(\boldsymbol{v}\cdot\spomcs)\spdcs_{0}}
\end{equation}
and $\bm{g}_{0}^{\star}$ is the single-particle integer time Green's
function under $\spd_{0}$ (i.e. $\bm{g}_{0}^{\star}=\bm{g}_{0}(\boldsymbol{0})$). 

In order to convert equations \ref{eq:ztilde1} and \ref{eq:ztilde2}
to functions of $\boldsymbol{\boldsymbol{g}}_{0}$, we translate the
derivatives using the chain rule. First, we need the derivatives of
$\boldsymbol{g}_{0}(\boldsymbol{v})$ with respect to $\boldsymbol{v}_{n}$
\begin{align}
\left.\frac{\partial[\boldsymbol{g}_{0}]_{ij}}{\partial[\boldsymbol{v}_{n}]_{lk}}\right|_{\boldsymbol{v}_{\ell}=\boldsymbol{0}} & =-\left.\frac{\partial[\boldsymbol{p}_{0}]_{ij}}{\partial[\boldsymbol{v}_{n}]_{lk}}\right|_{\boldsymbol{v}_{\ell}=\boldsymbol{0}}=[\boldsymbol{g}_{0}]_{ik}[\boldsymbol{p}_{0}]_{lj},
\end{align}
where $\boldsymbol{p}_{0}=\boldsymbol{1}-\boldsymbol{g}_{0}$. We
can now construct the first derivative of $\tilde{Z}$ as

\begin{equation}
\left.\frac{\partial\tilde{Z}}{\partial[\boldsymbol{v}_{1}]_{ij}}\right|_{\boldsymbol{v}_{\ell}=\boldsymbol{0}}=\frac{\partial Z}{\partial[\boldsymbol{g}_{0}]_{mn}}[\boldsymbol{g}_{0}]_{mj}[\boldsymbol{p}_{0}]_{in}.
\end{equation}
Similarly, for the second derivative 
\begin{align}
 & \left.\frac{\partial\tilde{Z}}{\partial[\boldsymbol{v}_{1}]_{ij}[\boldsymbol{v}_{2}]_{lk}}\right|_{\boldsymbol{v}_{\ell}=\boldsymbol{0}}=\nonumber \\
 & \frac{\partial Z}{\partial[\boldsymbol{g}_{0}]_{mn}\partial[\boldsymbol{g}_{0}]_{st}}[\boldsymbol{g}_{0}]_{sk}[\boldsymbol{p}_{0}]_{lt}[\boldsymbol{g}_{0}]_{mj}[\boldsymbol{p}_{0}]_{in}+\nonumber \\
 & \frac{\partial Z}{\partial[\boldsymbol{g}_{0}]_{mn}}\left([\boldsymbol{g}_{0}]_{mk}[\boldsymbol{p}_{0}]_{lj}[\boldsymbol{p}_{0}]_{in}-[\boldsymbol{g}_{0}]_{mj}[\boldsymbol{g}_{0}]_{ik}[\boldsymbol{p}_{0}]_{ln}\right).
\end{align}
We now have all derivatives of $\tilde{Z}$ up to second order in
terms of derivatives of $Z$. 

Given that we will always be evaluating the derivative for $\boldsymbol{g}_{0}=\boldsymbol{g}_{0}^{\star}$,
we will suppress the star superscript hereafter. We can then write
an equation for $\boldsymbol{g}$ as 
\begin{align}
[\boldsymbol{g}]_{ij} & =[\boldsymbol{g}_{0}]_{ij}+\frac{1}{Z}\frac{\partial Z}{\partial[\boldsymbol{g}_{0}]_{mn}}[\boldsymbol{g}_{0}]_{mj}[\boldsymbol{1}-\boldsymbol{g}_{0}]_{in}.\label{eq:dyson}
\end{align}
Similarly, for the two-particle quantities, we can write the interacting
single-particle and two-particle integer time Green's functions in
terms of the non-interacting integer time Green's function and derivatives
of the generating function as

\begin{align}
 & \left\langle \barhat[n]_{ij}\barhat[n]_{lk}\right\rangle _{\spdcs}=[\boldsymbol{g}_{0}]_{ij}[\boldsymbol{g}_{0}]_{lk}+[\boldsymbol{g}_{0}]_{ik}[\boldsymbol{p}_{0}]_{lj}+\nonumber \\
 & \frac{1}{Z}\frac{\partial Z}{\partial[\boldsymbol{g}_{0}]_{mn}}\bigg([\boldsymbol{g}_{0}]_{mk}[\boldsymbol{g}_{0}]_{ij}[\boldsymbol{p}_{0}]_{ln}+[\boldsymbol{g}_{0}]_{mj}[\boldsymbol{g}_{0}]_{lk}[\boldsymbol{p}_{0}]_{in}+\nonumber \\
 & [\boldsymbol{g}_{0}]_{mk}[\boldsymbol{p}_{0}]_{lj}[\boldsymbol{p}_{0}]_{in}-[\boldsymbol{g}_{0}]_{mj}[\boldsymbol{g}_{0}]_{ik}[\boldsymbol{p}_{0}]_{ln}\bigg)+\nonumber \\
 & \frac{1}{Z}\frac{\partial Z}{\partial[\boldsymbol{g}_{0}]_{mn}\partial[\boldsymbol{g}_{0}]_{st}}[\boldsymbol{g}_{0}]_{sk}[\boldsymbol{p}_{0}]_{lt}[\boldsymbol{g}_{0}]_{mj}[\boldsymbol{p}_{0}]_{in}.\label{eq:bethesalpeter}
\end{align}
Equation \ref{eq:dyson} can be rewritten in a more convenient form
motivated from the Lie group structure of the non-interacting density
matrix (see Section \ref{subsec:Appendix-Lie-group}). Introducing
the integer time self-energy $\boldsymbol{\Sigma}$ and its exponential
form $\boldsymbol{S}=\exp\left(-\boldsymbol{\Sigma}^{T}\right)$ as
\begin{equation}
\boldsymbol{S}=\left(Z\boldsymbol{1}+\left(\frac{\partial Z}{\partial\boldsymbol{g}_{0}^{T}}\right)\boldsymbol{p}_{0}\right)^{-1}\left(Z\boldsymbol{1}-\left(\frac{\partial Z}{\partial\boldsymbol{g}_{0}^{T}}\right)\boldsymbol{g}_{0}\right),
\end{equation}
we arrive at 
\begin{align}
\left(\frac{\boldsymbol{1}}{\boldsymbol{g}^{-1}-\boldsymbol{1}}\right)^{T}=\left(\frac{\boldsymbol{1}}{\boldsymbol{g}_{0}^{-1}-\boldsymbol{1}}\right)^{T}\exp\left(\boldsymbol{\Sigma}\right),
\end{align}
which can be further rearranged to our preferred form of the discrete
Dyson equation 
\begin{align}
\left(\bm{g}^{-1}-\boldsymbol{1}\right)=\left(\bm{g}_{0}^{-1}-\boldsymbol{1}\right)\boldsymbol{S}.\label{eq:preferred_dyson}
\end{align}
This discrete Dyson equation plays an important role in the discrete
action theory, analogous to the usual Dyson equation. While we have
only derived the equations for the one and two particle integer time
Green's functions, it should be clear that the above procedure can
be formally executed for arbitrary M-particle integer time Green's
functions. It should be emphasized that $Z$ can be written as a finite
polynomial of $\boldsymbol{g}_{0}$ if $\barhat[P]$ contains a finite
number of terms, which is why it is beneficial to perform the above
change in variables. 

It is useful to illustrate how the discrete Dyson equation connects
with the usual Dyson equation, and we make this connection in two
steps. We begin by rewriting the discrete Dyson equation as
\begin{align}
\left(\frac{\boldsymbol{1}}{\boldsymbol{g}^{-1}-\boldsymbol{1}}\right)^{T}=\left(\frac{\boldsymbol{1}}{\boldsymbol{g}_{Q}^{-1}-\boldsymbol{1}}\right)^{T}\exp(\boldsymbol{v}_{0})\exp\left(\boldsymbol{\Sigma}\right),\label{eq:dysonwithv0}
\end{align}
where $\boldsymbol{v}_{0}$ is defined from
\begin{align}
\left(\frac{\boldsymbol{1}}{\boldsymbol{g}_{0}^{-1}-\boldsymbol{1}}\right)^{T}=\left(\frac{\boldsymbol{1}}{\boldsymbol{g}_{Q}^{-1}-\boldsymbol{1}}\right)^{T}\exp(\boldsymbol{v}_{0}).
\end{align}
We now consider the limit of small $\boldsymbol{\Sigma}$ and $\boldsymbol{v}_{0}$,
where we can Taylor series expand the above two equations and retain
only leading order terms as
\begin{align}
 & \boldsymbol{g}^{-1}-\boldsymbol{1}=(\boldsymbol{g}_{Q}^{-1}-\boldsymbol{1})(\boldsymbol{1}-\boldsymbol{v}_{0}^{T}-\boldsymbol{\Sigma}^{T}),\\
 & \boldsymbol{g}_{0}^{-1}-\boldsymbol{1}=(\boldsymbol{g}_{Q}^{-1}-\boldsymbol{1})(\boldsymbol{1}-\boldsymbol{v}_{0}^{T}).
\end{align}
We now additionally consider the large $\discn$ limit. Given that
\begin{equation}
\left[\boldsymbol{g}_{Q}^{-1}\right]_{ij}=\begin{cases}
-1 & i=j+1\\
1 & i=j\\
0 & \text{otherwise}
\end{cases},
\end{equation}
we have $\boldsymbol{g}_{Q}^{-1}\boldsymbol{v}_{0}^{T}\approx0$ and
$\boldsymbol{g}_{Q}^{-1}\boldsymbol{\Sigma}^{T}\approx0$ for large
$\discn$. Subtracting the above two Taylor series, we then have 
\begin{align}
 & \boldsymbol{g}^{-1}=\boldsymbol{g}_{0}^{-1}+\boldsymbol{\Sigma}^{T}.
\end{align}
If we select an SPD corresponding to the Trotter-Suzuki decomposition
in the large $\discn$ limit, the above equation will recover the
usual Dyson equation. 

\section{The Canonical Discrete Action \label{sec:Categorizing-Discrete-Action}}

The CDA will prove to be relevant in the context of several different
SPD's. In particular, the SPD-l can be evaluated using the CDA. Additionally,
the SPD-d can be evaluated in $d=\infty$ using a CDA with a self-consistently
determined non-interacting integer time Green's function. Therefore,
there is utility in first studying the CDA in its own right, and we
will later illustrate how it can be used to solve the AIM and the
Hubbard model. The key step to evaluating the CDA is to compute the
discrete generating function as
\begin{equation}
Z\left(\boldsymbol{g}_{0}\right)=\langle\prod_{\tau=1}^{\discn}\barhat[P]_{\tau}^{\left(\tau\right)}\rangle_{\hat{\rho}_{G}\left(\boldsymbol{g}_{0}\right)}.
\end{equation}
In the following subsections, we evaluate the generating function
for the CDA in the special case of $\discn=3$ with a single orbital.
Subsequently, we show how the CDA can be used to evaluate the SPD-l.
This particular case can be used to solve the single band AIM\cite{Cheng2020short}.

\subsection{CDA for $\discn=3$ with a single orbital}

\begin{figure}
\includegraphics[width=1\columnwidth]{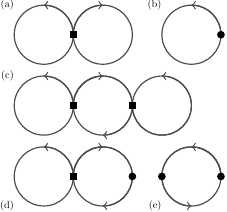}

\caption{\label{fig:schem_feynman_diagram}The connected integer time Feynman
diagrams for the generating function of the CDA with interacting projectors
defined in Eq. \ref{eq:int_proj_cda} at $\discn=3$. Panels $a-e$
illustrate connected diagrams which appear in equations \ref{eq:zc2}-\ref{eq:zc6},
respectively. Lines represent the non-interacting integer time Green's
function $\boldsymbol{g}_{0}$, squares represent the two-particle
vertex associated with the variational parameter $u$, and circles
represent the single-particle vertex associated with the variational
parameter $\mu$.}
\end{figure}

We now consider the CDA for two degenerate spin orbitals with $\discn=3$
for interacting projectors $\hat{P}_{3}=\hat{1}$ and
\begin{align}
 & \hat{P}_{1}=\hat{P}_{2}=(1-\mu-\frac{1}{4}u)\hat{1}+\mu(\hat{n}_{\uparrow}+\hat{n}_{\downarrow})+u\hat{n}_{\uparrow}\hat{n}_{\downarrow},\label{eq:int_proj_cda}
\end{align}
where $\mu$ and $u$ are variational parameters. The spin dependent
$\boldsymbol{g}_{0}$ can be parameterized as
\begin{equation}
\boldsymbol{g}_{0}=\left(\begin{array}{cccccc}
c_{11\uparrow} & 0 & c_{12\uparrow} & 0 & c_{13\uparrow} & 0\\
0 & c_{11\downarrow} & 0 & c_{12\downarrow} & 0 & c_{13\downarrow}\\
c_{21\uparrow} & 0 & c_{22\uparrow} & 0 & c_{23\uparrow} & 0\\
0 & c_{21\downarrow} & 0 & c_{22\downarrow} & 0 & c_{23\downarrow}\\
c_{31\uparrow} & 0 & c_{32\uparrow} & 0 & c_{33\uparrow} & 0\\
0 & c_{31\downarrow} & 0 & c_{32\downarrow} & 0 & c_{33\downarrow}
\end{array}\right),
\end{equation}
where we have used the time major indexing scheme (see Subsection
\ref{subsec:ex-AIM-nle2}), and the parameters $c_{ij\sigma}$ are
arbitrary. The generating function can then be evaluated using the
integer time Wick's theorem, resulting in a polynomial of the form
\begin{align}
Z\left(\boldsymbol{g}_{0}\right) & =\left[Z\left(\boldsymbol{g}_{0}\right)\right]_{1}+\left[Z\left(\boldsymbol{g}_{0}\right)\right]_{2}u+\left[Z\left(\boldsymbol{g}_{0}\right)\right]_{3}u^{2}+\nonumber \\
 & \left[Z\left(\boldsymbol{g}_{0}\right)\right]_{4}\mu+\left[Z\left(\boldsymbol{g}_{0}\right)\right]_{5}\mu u+\left[Z\left(\boldsymbol{g}_{0}\right)\right]_{6}\mu^{2},
\end{align}
where

\begin{alignat}{2}
\left[Z\left(\boldsymbol{g}_{0}\right)\right]_{1}= &  &  & 1,\\
\left[Z\left(\boldsymbol{g}_{0}\right)\right]_{2}= &  &  & c_{11\downarrow}c_{11\uparrow}+c_{22\downarrow}c_{22\uparrow}-\frac{1}{2},\label{eq:zc2}\\
\left[Z\left(\boldsymbol{g}_{0}\right)\right]_{3}= &  &  & \frac{1}{16}\bigg(16c_{12\downarrow}c_{21\downarrow}\left(c_{12\uparrow}c_{21\uparrow}-c_{11\uparrow}c_{22\uparrow}\right)+\nonumber \\
 &  &  & 4c_{11\downarrow}\left(c_{11\uparrow}\left(4c_{22\downarrow}c_{22\uparrow}-1\right)-4c_{22\downarrow}c_{12\uparrow}c_{21\uparrow}\right)+\nonumber \\
 &  &  & -4c_{22\downarrow}c_{22\uparrow}+1\bigg),\label{eq:zc3}\\
\left[Z\left(\boldsymbol{g}_{0}\right)\right]_{4}= &  &  & c_{11\downarrow}+c_{22\downarrow}+c_{11\uparrow}+c_{22\uparrow}-2,\label{eq:zc4}\\
\left[Z\left(\boldsymbol{g}_{0}\right)\right]_{5}= &  &  & \frac{1}{4}\bigg(-c_{22\downarrow}\left(4c_{12\uparrow}c_{21\uparrow}+4c_{22\uparrow}+1\right)+\nonumber \\
 &  &  & c_{11\uparrow}\left(4c_{22\downarrow}c_{22\uparrow}-4c_{12\downarrow}c_{21\downarrow}-1\right)+\nonumber \\
 &  &  & c_{11\downarrow}(4c_{22\downarrow}c_{22\uparrow}-4c_{12\uparrow}c_{21\uparrow}-1)+\nonumber \\
 &  &  & 4c_{11\uparrow}c_{11\downarrow}\left(c_{22\downarrow}+c_{22\uparrow}-1\right)\nonumber \\
 &  &  & 2-\left(4c_{12\downarrow}c_{21\downarrow}+1\right)c_{22\uparrow}\bigg),\label{eq:zc5}\\
\left[Z\left(\boldsymbol{g}_{0}\right)\right]_{6}= &  &  & c_{11\downarrow}\left(c_{22\downarrow}+c_{22\uparrow}-1\right)+c_{11\uparrow}\left(c_{22\downarrow}+c_{22\uparrow}-1\right)\nonumber \\
 &  &  & -c_{12\downarrow}c_{21\downarrow}-c_{22\downarrow}-c_{12\uparrow}c_{21\uparrow}-c_{22\uparrow}+1.\label{eq:zc6}
\end{alignat}
Each connected term in the above polynomial can be identified with
an integer time Feynman diagram (see Figure \ref{fig:schem_feynman_diagram}
for a schematic). We now have the complete solution for this particular
CDA, and any M-particle integer time correlation function can be evaluated
via differentiation. For example, the single particle and two-particle
integer time Green's functions can be obtained by plugging $Z$ into
equation \ref{eq:dyson} and \ref{eq:bethesalpeter}, respectively
(see \cite{supplementary} for explicit results). 

\subsection{Evaluating the SPD-l using the CDA\label{subsec:Evaluating-the-SPD-l-w-CDA}}

We now explore how to use the CDA in the context of the SPD-l. First
recall that in Section \ref{subsec:Evaluating-a-Local-SPD-via-Wick}
we explored how to evaluate the SPD-l using a diagrammatic approach.
Now we approach the same problem using the generating function and
the CDA. Starting from the sequential discrete action of the SPD-l,
we can trace out all of the orbitals that are not in the space of
the interacting projector, which we denote as bath orbitals, and obtain
a local discrete action as
\begin{align}
\spdcs_{loc} & =\textrm{Tr}_{bath}\spdcs=\textrm{Tr}_{bath}(\spdcs_{0})\barhat[P]_{1}^{(1)}\dots\barhat[P]_{\discn}^{(\discn)}\\
 & =\hat{\rho}_{G}(\boldsymbol{g}_{loc;0})\barhat[P]_{1}^{(1)}\dots\barhat[P]_{\discn}^{(\discn)},
\end{align}
and we see that $\spdcs_{loc}$ is indeed a CDA. If we study the generating
function of the SPD-l, we find
\begin{equation}
Z(\boldsymbol{g}_{0})=\langle\barhat[P]\rangle_{\hat{\rho}_{G}(\boldsymbol{g}_{0})}=\langle\barhat[P]\rangle_{\hat{\rho}_{G}(\boldsymbol{g}_{loc;0})}=Z_{c}\left(\boldsymbol{g}_{loc;0}\right),\label{eq:spdltocda}
\end{equation}
where $Z_{c}$ is the generating function of the CDA, the interacting
projector is $\barhat[P]=\barhat[P]_{1}^{(1)}\dots\barhat[P]_{\discn}^{(\discn)}$,
and $\boldsymbol{g}_{loc;0}$ is the local impurity sub-block of the
noninteracting integer time Green's function, given as
\begin{equation}
\boldsymbol{g}_{0}=\left(\begin{array}{cc}
\boldsymbol{g}_{loc;0} & \bm{g}_{lb;0}\\
\bm{g}_{bl;0} & \bm{g}_{bath;0}
\end{array}\right),
\end{equation}
and the latter equality in Eq. \ref{eq:spdltocda} holds given that
$\barhat[P]$ is local. Therefore, we can see that evaluating the
SPD-l amounts to evaluating a CDA with the corresponding interacting
projectors and a non-interacting integer time Green's function $\boldsymbol{g}_{loc;0}$.
It is useful to note that the integer time self-energy of the SPD-l
is local 
\begin{align}
\boldsymbol{\Sigma}\left(\bm{g}_{0}\right)=\bm{\Sigma}_{loc}\left(\bm{g}_{loc;0}\right)\oplus\bm{0},
\end{align}
where $\bm{\Sigma}_{loc}$ is completely determined from $Z_{c}$,
and this implies that 
\begin{align}
\bm{S}=\bm{S}_{loc}\oplus\boldsymbol{1}.
\end{align}
In summary, we have shown that by tracing out the bath orbitals of
the SPD-l, one obtains a CDA which can be used to evaluate the SPD-l.
In analogy to the traditional many-body Green's function approach,
the CDA is analogous to the action obtained by integrating out all
bath states from the AIM. More precisely, for the particular case
of the Trotter SPD-l in the large $\discn$ limit, the Trotter SPD-l
yields the exact density matrix of the corresponding Anderson impurity
model, and the corresponding CDA is equivalent to the effective action
of the impurity obtained by integrating out the bath sites.

\section{Self-consistent Canonical Discrete Action (SCDA)\label{sec:Self-consistent-Canonical-Discre}}

\subsection{Defining the SCDA algorithm}

In the preceding sections, we have built a complete formalism for
evaluating integer time correlation functions under a discrete action,
which can then be used to evaluate an SPD. For a general SPD, one
is still faced with a formidable problem, and therefore we need to
develop approximations and search for relevant scenarios where an
appropriate SPD can be exactly evaluated. Fortunately, all of the
usual approaches from many-body physics can be generalized to our
discrete action formalism. 

A common scenario for models of interacting electrons is where the
interaction is local, but not restricted to a single subspace; prominent
examples include the Hubbard model and the periodic Anderson impurity
model. In such cases, it is natural to study the SPD-d (see Section
\ref{subsec:Variational-Theory-and-SPD}), where the interacting projectors
are composed of disjoint projectors. To specifically address the SPD-d,
we introduce the  self-consistent canonical discrete action approximation
(SCDA), which is the integer time analogue of the dynamical mean-field
theory (DMFT)\cite{Georges199613,Kotliar200453,Vollhardt20121}. The
key idea for the SCDA is that the integer time self-energy is local,
and this can be determined by mapping the SPD-d to a collection of
CDA's determined from a self-consistency condition. Analogous to DMFT,
we will prove that the SCDA is an exact evaluation of the SPD-d in
infinite dimensions (see Section \ref{subsec:Proof-that-LSA}). 

We begin by outlining the SCDA in the most general case, in the absence
of any symmetry, where we consider an SPD-d with $N$ sites and the
interacting projectors are local within each site. The key idea for
the SCDA is the assumption that the integer time self-energy is local
\begin{align}
\bm{\Sigma}\left(\bm{g}\right)=\oplus_{i=1}^{N}\boldsymbol{\Sigma}_{i}(\boldsymbol{g}_{ii}),
\end{align}
where $i$ labels a given site. The self-consistent procedure can
then be defined, beginning with an initial guess for the non-interacting
integer time Green's function of the CDA and the identification of
the interacting projector of the CDA as that from site $i$ of the
SPD-d, and we have
\begin{align}
\bm{\mathcal{G}}_{i} & =\boldsymbol{g}_{0;ii}, & \barhat[P]_{i}=\prod_{\tau=1}^{\discn}\barhat[P]_{i,\tau}^{(\tau)},
\end{align}
which completely defines the effective CDA for site $i$. The effective
CDA can then be solved by computing the discrete generating function
$Z_{i}$, yielding the exponential integer time self-energy $\boldsymbol{S}_{i}$
as 
\begin{equation}
\boldsymbol{S}_{i}=\left(Z_{i}\boldsymbol{1}+\frac{\partial Z_{i}}{\partial\boldsymbol{\mathcal{G}}_{i}^{T}}(\boldsymbol{1}-\bm{\mathcal{G}}_{i})\right)^{-1}\left(Z_{i}\boldsymbol{1}-\frac{\partial Z_{i}}{\partial\boldsymbol{\mathcal{G}}_{i}^{T}}\bm{\mathcal{G}}_{i}\right).
\end{equation}
In the absence of symmetry, one must solve a CDA for each site, yielding
the total exponential self-energy for the system as
\begin{equation}
\boldsymbol{S}=\oplus_{i=1}^{N}\boldsymbol{S}_{i}.
\end{equation}
The interacting integer time Green's function can then be constructed
as
\begin{align}
\bm{g}=\frac{\boldsymbol{1}}{\boldsymbol{g}_{0}+\left(\boldsymbol{1}-\bm{g}_{0}\right)\bm{S}}\bm{g}_{0}.
\end{align}
Finally, we construct a new non-interacting integer time Green's function,
yielding the updated CDA
\begin{align}
\bm{\mathcal{G}}_{i}=\bm{S}_{i}\frac{\boldsymbol{1}}{\left(\boldsymbol{1}+\bm{g}{}_{ii}\left(\bm{S}_{i}-\boldsymbol{1}\right)\right)}\bm{g}{}_{ii}.
\end{align}
This entire procedure is then iterated until self-consistency is achieved.
Upon achieving self-consistency, one has completed a single evaluation
of the SPD-d. In order to obtain the ground state energy, one needs
to minimize over the variational parameters. 

\begin{figure}
\includegraphics[width=1\columnwidth]{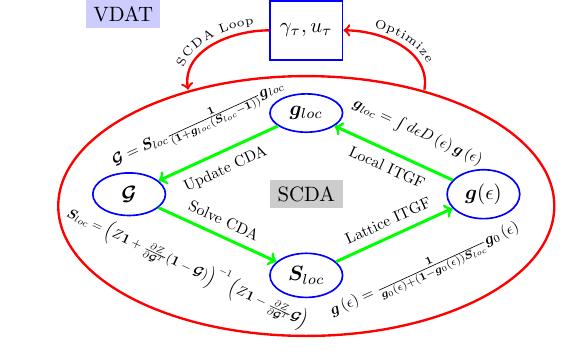}\caption{\label{fig:schem_dmft-1} A schematic of how VDAT is used to solve
for the ground state properties of a Hamiltonian under SPD-d within
the SCDA. ITGF is the acronym for the integer time Green's function.}
\end{figure}

The preceding outline of the SCDA is applied to a generic system without
symmetry, and now we specify the SCDA to the case of translation symmetry
where all types of orbitals have the same density-of-states. We can
begin with a guess for the non-interacting integer time impurity Green's
function $\bm{\mathcal{G}}$ as

\begin{align}
\bm{\mathcal{G}}=\int d\epsilon D\left(\epsilon\right)\bm{g}_{0}\left(\epsilon\right),
\end{align}
where $D(\epsilon)$ is the density-of-states. This defines our effective
CDA for the crystal, which can then be solved by computing the discrete
generating function $Z$, yielding the local exponential integer time
self-energy $\boldsymbol{S}$ as 
\begin{equation}
\boldsymbol{S}_{loc}=\left(Z\boldsymbol{1}+\frac{\partial Z}{\partial\boldsymbol{\mathcal{G}}^{T}}(\boldsymbol{1}-\bm{\mathcal{G}})\right)^{-1}\left(Z\boldsymbol{1}-\frac{\partial Z}{\partial\boldsymbol{\mathcal{G}}^{T}}\bm{\mathcal{G}}\right).
\end{equation}
We then use this integer time self-energy to update the interacting
integer time Green's function for each energy orbital as 
\begin{align}
\bm{g}\left(\epsilon\right)=\frac{\boldsymbol{1}}{\boldsymbol{g}_{0}\left(\epsilon\right)+\left(\boldsymbol{1}-\bm{g}_{0}\left(\epsilon\right)\right)\bm{S}_{loc}}\bm{g}_{0}\left(\epsilon\right).
\end{align}
Then we obtain the new interacting local integer time Green's function
as 
\begin{align}
\bm{g}_{loc}=\int d\epsilon D\left(\epsilon\right)\bm{g}\left(\epsilon\right).
\end{align}
Finally, we construct a new non-interacting integer time Green's function,
yielding the updated CDA as 
\begin{align}
\bm{\mathcal{G}}=\bm{S}_{loc}\frac{\boldsymbol{1}}{\left(\boldsymbol{1}+\bm{g}_{loc}\left(\bm{S}_{loc}-\boldsymbol{1}\right)\right)}\bm{g}{}_{loc}.
\end{align}
This process is then iterated until self-consistency is achieved,
and then the entire procedure is iterated when minimizing over the
variational parameters (see Figure \ref{fig:schem_dmft-1}). 

\subsection{Proof that the SCDA is exact in $d=\infty$\label{subsec:Proof-that-LSA}}

Here we prove that the SCDA exactly evaluates the SPD-d in $d=\infty$
for a general multi-band Hubbard model, where the local interacting
projectors $\hat{P}_{i}$ are confined to site $i$. The main idea
follows the cavity construction method used for proving that DMFT
is exact in infinite dimensions\cite{Georges199613}. We begin by
considering the non-interacting discrete action
\begin{align}
\spdcs_{0}=\barhat[Q]\exp(\sppm_{1}\cdot\spomcs^{(1)})...\exp(\sppm_{\discn}\cdot\spomcs^{(\discn)}),
\end{align}
where
\begin{equation}
\exp(\sppm_{\tau}\cdot\spomcs^{(\tau)})=\exp(\sum_{\vec{k}\sigma}\gamma_{\vec{k}\sigma;\tau}\barhat[n]_{\vec{k}\sigma}^{(\tau)}),
\end{equation}
In the cavity construction, one selects a particular site in the lattice,
denoted site $i$, and traces out all other sites. We can rewrite
the non-interacting discrete action in the following form
\begin{equation}
\spdcs_{0}=\exp\left(\left(\bm{v}_{i}+\bm{v}_{b}+\bm{v}_{ib}\right)\cdot\spomcs\right),
\end{equation}
where $\boldsymbol{v}_{i}$ is a single-particle potential within
site $i$, $\boldsymbol{v}_{b}$ is the single-particle potential
of the remaining sites, and $\boldsymbol{v}_{ib}$ is the off-diagonal
component of the single-particle potential between site $i$ and the
remaining sites. We can then construct the local discrete action for
site $i$ by tracing out all other sites 
\begin{align}
\spdcs_{loc} & =\text{Tr}_{/i}\bigg(\spdcs_{0}\prod_{j}\barhat[P]_{j}\bigg)=\spdcs_{loc;0}\barhat[P]_{i},
\end{align}
where
\begin{align}
\spdcs_{loc;0} & =\text{Tr}_{/i}\bigg(\spdcs_{0}\prod_{j\ne i}\barhat[P]_{j}\bigg), & \barhat[P]_{j}=\prod_{\tau=1}^{\discn}\barhat[P]_{\tau,j}^{(\tau)}.
\end{align}
We now seek to prove that $\spdcs_{loc}$ is a CDA in $d=\infty$.
By expanding $\spdcs_{loc;0}$ in terms of $\boldsymbol{v}_{ib}$,
we prove that the interacting projectors within $\spdcs_{loc;0}$
can be replaced by an effective non-interacting projector as
\begin{align}
\spdcs_{loc;0} & =\text{Tr}_{/i}(\spdcs_{0}\exp(\boldsymbol{\Sigma}_{B;i}\spomcs)),\label{eq:spd0dinf}
\end{align}
where $\boldsymbol{\Sigma}_{B;i}$ is a single-particle potential
for the sites not containing $i$. Recall the general expression for
the expansion of the exponential of a sum of two operators $\barhat[A]$
and $\barhat[B]$ 

\begin{align}
 & \exp(\barhat[A]+\barhat[B])=\exp(\barhat[A])+\nonumber \\
 & \int_{0}^{1}d\lambda\exp(\lambda\barhat[A])\barhat[B]\exp((1-\lambda)\barhat[A])+\nonumber \\
 & \int_{0}^{1}d\lambda_{1}\int_{\lambda_{1}}^{1}d\lambda_{2}\exp(\lambda_{1}\barhat[A])\barhat[B]\exp((\lambda_{2}-\lambda_{1})\barhat[A])\barhat[B]\times\nonumber \\
 & \exp((1-\lambda_{2})\barhat[A])+\dots
\end{align}
and equating 
\begin{align}
\barhat[A] & =(\bm{v}_{i}+\bm{v}_{b})\cdot\spomcs=\barhat[A]_{i}+\barhat[A]_{b},\\
\barhat[A]_{i} & =\bm{v}_{i}\cdot\spomcs=\sum_{\tau\tau'}v^{\tau\tau'}\barhat[a]_{i}^{\dagger\left(\tau\right)}\barhat[a]_{i}^{\left(\tau'\right)}+h.c.,\\
\barhat[A]_{b} & =\bm{v}_{b}\cdot\spomcs=\sum_{j\ne i,j'\ne i,\tau\tau'}v_{jj'}^{\tau\tau'}\barhat[a]_{j}^{\dagger\left(\tau\right)}\barhat[a]_{j'}^{\left(\tau'\right)},\\
\barhat[B] & =\bm{v}_{ib}\cdot\spomcs=\sum_{j\neq i,\tau\tau'}(t_{j}^{\tau\tau'}\barhat[a]_{i}^{\dagger\left(\tau\right)}\barhat[a]_{j}^{\left(\tau'\right)}+h.c.),
\end{align}
we can then consider the expansion in $\boldsymbol{v}_{ib}$ order
by order
\begin{equation}
\hat{\rho}_{loc;0}=\text{Tr}_{/i}\bigg(\exp(\barhat[A])\prod_{j\ne i}\hat{P}_{j}\bigg)+\left[\hat{\rho}_{loc;0}\right]_{1}+\dots
\end{equation}
 where

\begin{widetext}

\begin{align}
\left[\spdcs_{loc;0}\right]_{1} & =\text{Tr}_{/i}\left(\int_{0}^{1}d\lambda_{1}\exp(\lambda_{1}\barhat[A])\sum_{j_{1}\neq i,\tau\tau'}\left(t_{j_{1}}^{\tau\tau'}\barhat[a]_{i}^{\dagger\left(\tau\right)}\barhat[a]_{j_{1}}^{\left(\tau'\right)}+h.c.\right)\exp((1-\lambda_{1})\barhat[A])\prod_{j\ne i}\barhat[P]_{j}\right),\\
\left[\spdcs_{loc;0}\right]_{2} & =\text{Tr}_{/i}\bigg(\int_{0}^{1}d\lambda_{1}\int_{\lambda_{1}}^{1}d\lambda_{2}\exp(\lambda_{1}\barhat[A])\sum_{j_{1}\neq i,\tau_{1}\tau_{1}'}\left(t_{j_{1}}^{\tau_{1}\tau_{1}'}\barhat[a]_{i}^{\dagger\left(\tau_{1}\right)}\barhat[a]_{j_{1}}^{\left(\tau_{1}'\right)}+h.c.\right)\exp((\lambda_{2}-\lambda_{1})\barhat[A])\nonumber \\
 & \times\sum_{j_{2}\neq i,\tau_{2}\tau_{2}'}\left(t_{j_{2}}^{\tau_{2}\tau_{2}'}\barhat[a]_{i}^{\dagger\left(\tau_{2}\right)}\barhat[a]_{j_{2}}^{\left(\tau_{2}'\right)}+h.c.\right)\exp((1-\lambda_{2})\barhat[A])\prod_{j\ne i}\barhat[P]_{j}\bigg).
\end{align}
\end{widetext}We observe that a cavity Green's function emerges as
a key quantity to evaluate
\begin{align}
 & G_{j_{1}j_{2}}^{\tau_{1}\tau_{2}}(\lambda_{1},\lambda_{2})=\text{Tr}_{/i}\bigg(\exp(\lambda_{1}\barhat[A]_{b})\barhat[a]_{j_{1}}^{\left(\tau_{1}\right)}\times\nonumber \\
 & \exp((\lambda_{2}-\lambda_{1})\barhat[A]_{b})\barhat[a]_{j_{2}}^{\dagger\left(\tau_{2}\right)}\exp((1-\lambda_{2})\barhat[A]_{b})\prod_{j\ne i}\barhat[P]_{j}\bigg).
\end{align}
Given that the projectors $\barhat[P]_{j}$ are local, the scaling
of the cavity Green's function is\cite{Georges199613}
\begin{align}
G_{j_{1}j_{2}}^{\tau_{1}\tau_{2}}(\lambda_{1},\lambda_{2}) & \sim\left(\frac{1}{\sqrt{d}}\right)^{\left|j-j'\right|},
\end{align}
where $d$ is the dimension of the lattice. Analogous to the case
of DMFT, the local discrete action $\spdcs_{loc;0}$ only depends
on $G_{j_{1}j_{2}}^{\tau_{1}\tau_{2}}(0,0)$ and $\barhat[A]_{b}$.
To illustrate this, consider the second order contribution 
\begin{align}
\text{Tr}_{/i}\bigg( & \exp(\lambda_{1}\barhat[A])t_{j_{1}}^{\tau_{1}\tau_{1}'}\barhat[a]_{i}^{\dagger\left(\tau_{1}\right)}\barhat[a]_{j_{1}}^{\left(\tau_{1}'\right)}\exp((\lambda_{2}-\lambda_{1})\barhat[A])\times\nonumber \\
 & t_{j_{2}}^{\tau_{2}\tau_{2}'}\barhat[a]_{j_{2}}^{\dagger\left(\tau_{2}\right)}\barhat[a]_{i}^{\left(\tau_{2}'\right)}\exp((1-\lambda_{2})\barhat[A])\prod_{j\ne i}\barhat[P]_{j}\bigg)\nonumber \\
= & t_{j_{1}}^{\tau_{1}\tau_{1}'}t_{j_{2}}^{\tau_{2}\tau_{2}'}G_{j_{1}j_{2}}^{\tau_{1}'\tau_{2}}(\lambda_{1},\lambda_{2})\exp(\lambda_{1}\barhat[A]_{i})\barhat[a]_{i}^{\dagger\left(\tau_{1}\right)}\times\nonumber \\
 & \exp((\lambda_{2}-\lambda_{1})\barhat[A]_{i})\barhat[a]_{i}^{\left(\tau_{2}'\right)}\exp((1-\lambda_{2})\barhat[A]_{i}),
\end{align}
where the Einstein summation convention is assumed for the orbital
and time index. The total scaling of this term will be 
\begin{equation}
d^{|j_{1}-i|}d^{|j_{2}-i|}d^{-\frac{1}{2}|j_{1}-i|}d^{-\frac{1}{2}|j_{2}-i|}d^{-\frac{1}{2}|j_{1}-j_{2}|}\sim1.
\end{equation}
Considering a fourth order contribution 
\begin{align}
 & t_{j_{1}}^{\tau_{1}\tau_{1}'}t_{j_{2}}^{\tau_{2}\tau_{2}'}t_{j_{3}}^{\tau_{3}\tau_{3}'}t_{j_{4}}^{\tau_{4}\tau_{4}'}G_{j_{1}j_{2}j_{3}j_{4}}^{\tau_{1}'\tau_{2}\tau_{3}'\tau_{4}}(\lambda_{1},\lambda_{2},\lambda_{3},\lambda_{4})\exp(\lambda_{1}\barhat[A]_{i})\times\nonumber \\
 & a_{i}^{\dagger\left(\tau_{1}\right)}\exp((\lambda_{2}-\lambda_{1})\barhat[A]_{i})a_{i}^{\left(\tau_{2}'\right)}\exp((\lambda_{3}-\lambda_{2})\barhat[A]_{i})\times\nonumber \\
 & a_{i}^{\dagger\left(\tau_{3}\right)}\exp((\lambda_{4}-\lambda_{3})\barhat[A]_{i})a_{i}^{\left(\tau_{4}'\right)}\exp((1-\lambda_{4})\barhat[A]_{i}),
\end{align}
the scaling for one of the connected portions is 
\begin{alignat}{1}
 & d^{|j_{1}-i|}d^{|j_{2}-i|}d^{|j_{3}-i|}d^{|j_{4}-i|}d^{-\frac{1}{2}|j_{1}-i|}d^{-\frac{1}{2}|j_{2}-i|}\times\nonumber \\
 & d^{-\frac{1}{2}|j_{3}-i|}d^{-\frac{1}{2}|j_{4}-i|}d^{-\frac{1}{2}|j_{1}-j_{2}|}d^{-\frac{1}{2}|j_{2}-j_{3}|}d^{-\frac{1}{2}|j_{3}-j_{4}|}\\
 & \sim d^{-\frac{1}{2}|j_{2}-j_{3}|}\rightarrow0,
\end{alignat}
and all other connected diagrams will scale to zero as well. The same
result holds for higher orders, thus proving Eq. \ref{eq:spd0dinf}. 

We proceed by rewriting $\spdcs_{loc;0}$ as

\begin{align}
 & \spdcs_{loc;0}=\text{Tr}_{/i}\left(\spdcs^{\star}\exp\left(-\boldsymbol{\Sigma}_{loc;i}\cdot\spomcs\right)\right),\\
 & \spdcs^{\star}=\spdcs_{0}\exp\left(\left(\boldsymbol{\Sigma}_{B;i}+\boldsymbol{\Sigma}_{loc;i}\right)\cdot\spomcs\right),
\end{align}
where $\boldsymbol{\Sigma}_{loc;i}$ is the integer time self-energy
for the CDA with a given $\spdcs_{loc;0}$ and thus is a single-particle
potential within the site $i$; and $\spdcs^{\star}$ is a non-interacting
discrete action that has the same integer time Green's function as
$\spdcs$. It should be noted that $\boldsymbol{\Sigma}_{loc;i}$
and $\boldsymbol{\Sigma}_{B;i}$ occupy distinct blocks within the
integer time self-energy matrix and do not mix.

Finally, we prove that $\boldsymbol{\Sigma}$ is the sum of the local
integer time self-energy for all sites. To see this, we notice that
the above construction can be applied to every site $i$, and thus
we have $\boldsymbol{\Sigma}=\boldsymbol{\Sigma}_{B;i}+\boldsymbol{\Sigma}_{loc;i}$
for every site $i$. Recalling the block structure of the self-energy,
we can solve for $\boldsymbol{\Sigma}=\sum_{i}\boldsymbol{\Sigma}_{loc;i}$
, proving the SCDA self-consistency condition, analogous to DMFT. 

For the special case of $\discn=2$ with a G-type SPD-d, the SCDA
recovers the classic observation that the Gutzwiller approximation
exactly evaluates the Gutzwiller wave function in $d=\infty$ \cite{Metzner1987121,Metzner19887382,Metzner1989324,Bunemann19977343}.
For $\discn=2$ with a B-type SPD-d, our proof demonstrates that the
Baeriswyl wave function\cite{Baeriswyl19870} is exactly evaluated
in $d=\infty$ via the SCDA, which was not previously known. For the
case of $\discn=3$, we see that the Gutzwiller-Baeriswyl\cite{Otsuka19921645}
and Baeriswyl-Gutzwiller\cite{Dzierzawa19951993} wave functions can
be exactly evaluated in $d=\infty$, which also was not known. Furthermore,
there are an infinite number of wave functions for $\discn\ge4$ which
have not been considered, but can be exactly evaluated via the SCDA.

\subsection{The SCDA for $\discn=2$ \label{subsec:The-LSA-forn2dinf}}

Here we consider the case of $\discn=2$ for the Hubbard model, which
is of practical importance given that it recovers the Gutzwiller approximation.
We will demonstrate that the SCDA at $\discn=2$ is a very special
case in that the SCDA self-consistency condition can be achieved a
priori by choosing $\boldsymbol{\mathcal{G}}$ as the non-interacting
local integer time Green's function and constraining the interacting
projector such that $\spd_{loc}$ and $\spd_{loc;0}$ have the same
local single-particle density matrix. Given this particular implementation
of the SCDA at $\discn=2$, we can explicitly derive the Gutzwiller
equations. Alternatively, we could choose not follow such constraints,
and then the SCDA self-consistency condition cannot be fulfulled a
priori, though we are gauranteed to reach the same ground state properties.
This latter scenario could be numerically beneficial given that we
will not need to satisfy any constraint when minimizing over the variational
parameters, though we will need to execute the SCDA self-consistency
condition instead.

In the remainder of this subsection, we will explicitly derive the
Gutzwiller equations in a sequence of increasingly complex scenarios:
the single-band Hubbard model at half-filling, the multi-band Hubbard
model at arbitrary filling but with symmtery dictating that the local
single-particle density matrix is diagonal, and finally the most general
possible case for the multi-band Hubbard model. The last case results
in rotationally invariant Gutzwiller equations in an arbitrary basis,
which have not yet been presented in the literature; and they recover
the particular case of a ``mixed original-natural basis''\cite{Lanata2012035133}. 

\subsubsection{Single-band Hubbard model at half-filling}

To understand how the SCDA works, we study the case of $\discn=2$
for a G-type SPD-d, which recovers the Gutzwiller approximation. We
first focus on the case of the one band model at half-filling, and
we later generalize to the multi-orbital case at arbitrary density.
For the former, the SPD-d is 
\begin{align}
\spd=\exp(\sum_{i}u\Delta\hat{d}_{i})\exp(\sum_{\epsilon\sigma}\gamma_{\sigma}(\epsilon)\hat{n}_{\epsilon\sigma})\exp(\sum_{i}u\Delta\hat{d}_{i}),
\end{align}
where $\Delta\hat{d}$ is defined in Eq. \ref{eq:delta_d}, $\hat{n}_{\epsilon\sigma}$
is the number operator for the orbital with energy $\epsilon$ and
spin $\sigma$, and the variational parameters are $u$ and $\gamma_{\sigma}(\epsilon)$.
Thus, the non-interacting integer time Green's function for spin $\sigma$
and energy $\epsilon$ is

\begin{align}
\bm{g}_{0;\sigma}(\epsilon)=\begin{pmatrix}n_{\sigma}(\epsilon) & \left(1-n_{\sigma}(\epsilon)\right)\\
-n_{\sigma}(\epsilon) & n_{\sigma}(\epsilon)
\end{pmatrix},
\end{align}
where
\begin{equation}
n_{\sigma}(\epsilon)=\frac{1}{1+\exp\left(-\gamma_{\sigma}(\epsilon)\right)}\in[0,1],
\end{equation}
is used as a reparameterization of the variational parameter $\gamma_{\sigma}(\epsilon)$.
As an initial guess, we choose the non-interacting integer time Green's
function of the CDA as the local non-interacting integer time Green's
function, given as
\begin{align}
\boldsymbol{\mathcal{G}}_{\sigma}=\int d\epsilon D(\epsilon)\bm{g}_{0;\sigma}(\epsilon)=\begin{pmatrix}\frac{1}{2} & \frac{1}{2}\\
-\frac{1}{2} & \frac{1}{2}
\end{pmatrix},
\end{align}
where $D(\epsilon)$ is the density-of-states per spin. It will also
be necessary to introduce $\boldsymbol{\mathcal{G}}$ for an arbitrary
density and spin as
\begin{equation}
\boldsymbol{\mathcal{G}}=\left(\begin{array}{cccc}
a_{11\uparrow} & 0 & a_{12\uparrow} & 0\\
0 & a_{11\downarrow} & 0 & a_{12\downarrow}\\
a_{21\uparrow} & 0 & a_{22\uparrow} & 0\\
0 & a_{21\downarrow} & 0 & a_{22\downarrow}
\end{array}\right),
\end{equation}
where we have used the time major scheme (see Subsection \ref{subsec:ex-AIM-nle2}
for a definition), and this more general definition is needed given
that we will take the derivative of the discrete generating function.
The discrete generating function is given as

\begin{align}
Z & =1+\frac{1}{16}(uZ_{1}+u^{2}Z_{2}),
\end{align}
where
\begin{align}
 & Z_{1}=8\left(\left(2a_{22\downarrow}-1\right)a_{22\uparrow}-a_{22\downarrow}\right)+\nonumber \\
 & 16a_{11\downarrow}a_{11\uparrow}-8a_{11\downarrow}-8a_{11\uparrow}+8,
\end{align}
and 
\begin{align}
 & Z_{2}=4a_{12\downarrow}a_{21\downarrow}\left(4a_{12\uparrow}a_{21\uparrow}+2a_{22\uparrow}-1\right)+1+\nonumber \\
 & 2\left(\left(2a_{22\downarrow}-1\right)a_{22\uparrow}-a_{22\downarrow}\right)-4a_{22\downarrow}a_{11\uparrow}\left(2a_{22\uparrow}-1\right)+\nonumber \\
 & 4a_{11\downarrow}\left(2\left(1-2a_{22\downarrow}\right)a_{12\uparrow}a_{21\uparrow}+a_{22\downarrow}\left(1-2a_{22\uparrow}\right)\right)+\nonumber \\
 & 4a_{11\downarrow}\left(a_{22\uparrow}-2a_{11\uparrow}a_{22\uparrow}+2a_{22\downarrow}a_{11\uparrow}\left(2a_{22\uparrow}-1\right)\right)+\nonumber \\
 & 4\left(2a_{22\downarrow}-1\right)a_{12\uparrow}a_{21\uparrow}-8a_{12\downarrow}a_{21\downarrow}a_{11\uparrow}\left(2a_{22\uparrow}-1\right)+\nonumber \\
 & a_{11\downarrow}\left(4a_{11\uparrow}-2\right)-2a_{11\uparrow}+4a_{11\uparrow}a_{22\uparrow}.
\end{align}
Evaluating $Z$ and its derivatives for half filling gives 
\begin{align}
Z=\frac{u^{2}}{16}+1, &  & \frac{\partial Z}{\partial\boldsymbol{\mathcal{G}}_{\sigma}^{T}}=\left(\begin{array}{cc}
0 & -\frac{u^{2}}{8}\\
\frac{u^{2}}{8} & 0
\end{array}\right),
\end{align}
and using the discrete Dyson equation, we have
\begin{align}
\boldsymbol{g}_{\sigma} & =\boldsymbol{\mathcal{G}}_{\sigma}+\frac{1}{Z}(\boldsymbol{1}-\boldsymbol{\mathcal{G}}_{\sigma})\frac{\partial Z}{\partial\boldsymbol{\mathcal{G}}_{\sigma}^{T}}\boldsymbol{\mathcal{G}}_{\sigma}\\
 & =\left(\begin{array}{cc}
\frac{1}{2} & \frac{1}{2}z(u)\\
-\frac{1}{2}z(u) & \frac{1}{2}
\end{array}\right),
\end{align}
where 
\begin{equation}
z(u)=\frac{32}{u^{2}+16}-1.
\end{equation}
The exponential integer time self-energy can be constructed as 
\begin{align}
S_{\sigma} & =\left(Z+\frac{\partial Z}{\partial\boldsymbol{\mathcal{G}}_{\sigma}^{T}}(1-\boldsymbol{\mathcal{G}}_{\sigma})\right)^{-1}\left(Z-\frac{\partial Z}{\partial\boldsymbol{\mathcal{G}}_{\sigma}^{T}}\boldsymbol{\mathcal{G}}_{\sigma}\right)\\
 & =\left(\begin{array}{cc}
\frac{512}{u^{4}+256}-1 & \frac{32u^{2}}{u^{4}+256}\\
-\frac{32u^{2}}{u^{4}+256} & \frac{512}{u^{4}+256}-1
\end{array}\right)\\
 & =\left(\begin{array}{cc}
\frac{2z(u)}{z(u)^{2}+1} & \frac{2}{z(u)^{2}+1}-1\\
1-\frac{2}{z(u)^{2}+1} & \frac{2z(u)}{z(u)^{2}+1}
\end{array}\right).
\end{align}
The interacting integer time Green's function for a given $\epsilon\sigma$
is then

\begin{align}
\bm{g}_{\sigma}(\epsilon)=\begin{pmatrix}n_{\sigma}(\epsilon) & \left(1-n_{\sigma}(\epsilon)\right)z(u)\\
-n_{\sigma}(\epsilon)z(u) & \left(n_{\sigma}(\epsilon)-\frac{1}{2}\right)z(u)^{2}+\frac{1}{2}
\end{pmatrix}.
\end{align}
The new interacting local integer time Green's function can then be
constructed as 
\begin{align}
\bm{g}'_{loc;\sigma}=\int d\epsilon D\left(\epsilon\right)\bm{g}_{\sigma}(\epsilon)=\begin{pmatrix}\frac{1}{2} & \frac{1}{2}z(u)\\
-\frac{1}{2}z(u) & \frac{1}{2}
\end{pmatrix},
\end{align}
which is same as the interacting local integer time Green's function
from the initial guess. Therefore, we have already achieved self-consistency.
In order to evaluate the ground state energy, we also need to evaluate
the double occupancy as
\begin{align}
\langle\Delta\hat{d}\rangle_{\spd} & =\frac{2u}{u^{2}+16}\equiv\Delta d,
\end{align}
which is computed by evaluating Eq. \ref{eq:bethesalpeter}. Now we
can proceed to minimize over the variational parameters
\begin{align}
\mathcal{E} & =\min_{u,n_{\epsilon\sigma}\in\left[0,1\right]}\left(\frac{2uU}{u^{2}+16}+\sum_{\sigma}\int d\epsilon D(\epsilon)n_{\sigma}(\epsilon)z(u)^{2}\right)\\
 & =\min_{u}\left(\frac{2uU}{u^{2}+16}+\epsilon_{0}z(u)^{2}\right)\\
 & =\min_{\Delta d\in[-\frac{1}{4},\frac{1}{4}]}(\Delta dU+\epsilon_{0}(1-(4\Delta d)^{2}))\\
 & =\begin{cases}
\epsilon_{0}-\frac{U^{2}}{64\epsilon_{0}} & U\le8\epsilon_{0}\\
0 & U\ge8\epsilon_{0}
\end{cases},
\end{align}
where $\epsilon_{0}=2\int_{-\infty}^{0}d\epsilon D(\epsilon)\epsilon$.
Here we see that we have recovered the Gutzwiller approximation, with
the Brinkman-Rice transition\cite{Brinkman19704302} at $U=8\epsilon_{0}$.

\subsubsection{Multi-band Hubbard model at arbitrary filling with a diagonal local
single-particle density matrix }

Here the preceding analysis is generalized to the multi-band Hubbard
model for the special case where symmetry dictates that the local
single-particle density matrix is diagonal, and we recover the usual
Gutzwilller approximation in this scenario. We begin with the SPD
for this multi-band case 
\begin{align}
\spd=\left(\prod_{i}\hat{P}_{1,i}\right)\exp(\sum_{\epsilon\spinorb}\gamma_{\spinorb}(\epsilon)\hat{n}_{\epsilon\spinorb})\left(\prod_{i}\hat{P}_{2,i}\right),
\end{align}
where $\hat{P}_{1,i}=\sum_{\Gamma\Gamma'}P_{1,\Gamma\Gamma'}\hat{X}_{i;\Gamma\Gamma'}=\hat{P}_{2,i}^{\dagger}$
is the interacting projector of site $i$ in the physical space, and
the variational parameters are $P_{1,\Gamma\Gamma'}$. The non-interacting
integer time Green's function for orbital $\epsilon\spinorb$ is then
\begin{align}
\bm{g}_{0;\spinorb}(\epsilon)=\begin{pmatrix}n_{\spinorb}(\epsilon) & \left(1-n_{\spinorb}(\epsilon)\right)\\
-n_{\spinorb}(\epsilon) & n_{\spinorb}(\epsilon)
\end{pmatrix},
\end{align}
where 
\begin{equation}
n_{\spinorb}(\epsilon)=\frac{1}{1+\exp\left(-\gamma_{\spinorb}(\epsilon)\right)}\in[0,1]
\end{equation}
is used as a reparameterization of the variational parameters $\gamma_{\spinorb}(\epsilon)$.
As an initial guess, we choose the non-interacting integer time Green's
function of the CDA to be the non-interacting local integer time Green's
function
\begin{align}
\boldsymbol{\mathcal{G}}_{\spinorb}=\int d\epsilon D_{\spinorb}\left(\epsilon\right)\bm{g}_{0;\spinorb}(\epsilon)=\begin{pmatrix}\bar{n}_{\spinorb} & \left(1-\bar{n}_{\spinorb}\right)\\
-\bar{n}_{\spinorb} & \bar{n}_{\spinorb}
\end{pmatrix},
\end{align}
where $D_{\spinorb}(\epsilon)$ is the partial density-of-states and
$\bar{n}_{\spinorb}=\int d\epsilon D_{\spinorb}\left(\epsilon\right)n_{\spinorb}(\epsilon)$.
Considering the CDA of site $i$ for this SPD-d, we have
\begin{align}
 & \spdcs_{loc}=\spdcs_{loc;0}\barhat[P]_{1,i}^{(1)}\barhat[P]_{2,i}^{(2)},\\
 & \spdcs_{loc;0}=\hat{\rho}_{G}(\boldsymbol{\mathcal{G}})=\barhat[Q]\exp(\boldsymbol{v}_{1}\cdot\spomcs_{i}^{(1)}+\boldsymbol{v}_{2}\cdot\spomcs_{i}^{(2)}),
\end{align}
where
\begin{align}
\boldsymbol{v}_{1} & =0, & [\boldsymbol{v}_{2}]_{\spinorb,\spinorb'}=-\delta_{\spinorb,\spinorb'}\ln(\bar{n}_{\spinorb}^{-1}-1).
\end{align}
We see that $\spdcs_{loc}$ is an effective discrete action of an
SPD given as
\begin{align}
 & \spd_{loc}=\exp(\boldsymbol{v}_{1}\cdot\spom_{i})\hat{P}_{1,i}\exp(\boldsymbol{v}_{2}\cdot\spom_{i})\hat{P}_{2,i},
\end{align}
and 
\begin{align}
 & \spd_{loc;0}=\exp(\boldsymbol{v}_{1}\cdot\spom_{i})\exp(\boldsymbol{v}_{2}\cdot\spom_{i}).
\end{align}
 We can now directly evaluate the interacting integer time Green's
function as 
\begin{align}
\bm{g}_{loc;\spinorb}=\begin{pmatrix}\bar{n}_{\spinorb} & \left(1-\bar{n}_{\spinorb}\right)z_{\spinorb}\\
-\bar{n}_{\spinorb}z_{\spinorb} & \bar{n}_{\spinorb}
\end{pmatrix},
\end{align}
where
\begin{align}
z_{\spinorb} & =\frac{\text{Tr}\left(\hat{P}_{1,i}\hat{a}_{i\spinorb}^{\dagger}\spd_{loc;0}\hat{P}_{2,i}\hat{a}_{i\spinorb}\right)}{\text{Tr}\left(\hat{a}_{i\spinorb}^{\dagger}\spd_{loc;0}\hat{a}_{i\spinorb}\right)},
\end{align}
and the constraints on the normalization of the SPD require that 
\begin{align}
 & \frac{\text{Tr}\left(\hat{P}_{1,i}\spd_{loc;0}\hat{P}_{2,i}\right)}{\text{Tr}\left(\spd_{loc;0}\right)}=1,\label{eq:mgutzc1}
\end{align}
while the constraint of the local density matrix requires
\begin{equation}
\frac{\text{Tr}\left(\hat{P}_{1,i}\spd_{loc;0}\hat{P}_{2,i}\hat{a}_{i\spinorb}^{\dagger}\hat{a}_{i\spinorb}\right)}{\text{Tr}\left(\spd_{loc;0}\right)}=\bar{n}_{\spinorb}.\label{eq:mgutzc2}
\end{equation}
To connect with the corresponding expression for the multiband Gutzwiller
approximation\cite{Bunemann19986896,Bunemann2007193104,Bunemann20121282,Lanata2012035133}
in this case, we can rewrite $z_{\spinorb}$ as
\begin{align}
z_{\spinorb} & =\frac{\text{Tr}\left(\hat{P}_{1,i}\sqrt{\spd_{loc;0}}\hat{a}_{i\spinorb}^{\dagger}\sqrt{\spd_{loc;0}}\hat{P}_{1,i}^{\dagger}\hat{a}_{i\spinorb}\right)}{\text{Tr}\left(\sqrt{\spd_{loc;0}}\hat{a}_{i\spinorb}^{\dagger}\sqrt{\spd_{loc;0}}\hat{a}_{i\spinorb}\right)},
\end{align}
given that $\boldsymbol{v}_{2}$ is diagonal in the $\alpha,\sigma$
basis. We can also connect with the form presented in the off-shell
effective energy theory\cite{Cheng2020081105} for the $\mathcal{K}$
formulation within the central point expansion as 
\begin{equation}
z_{\spinorb}=\frac{\text{Tr}\left(\sqrt{\spd_{loc}}\hat{a}_{i\spinorb}^{\dagger}\sqrt{\spd_{loc}}\hat{a}_{i\spinorb}\right)}{\text{Tr}\left(\sqrt{\spd_{loc;0}}\hat{a}_{i\spinorb}^{\dagger}\sqrt{\spd_{loc;0}}\hat{a}_{i\spinorb}\right)},
\end{equation}
where we have assumed that $\hat{P}_{1}$ commutes with $\spd_{loc;0}$.
We can now compute the exponential integer time self-energy as 
\begin{align}
\bm{S}_{loc;\spinorb}=\left(\begin{array}{cc}
\frac{z_{\spinorb}}{-\bar{n}_{\spinorb}z_{\spinorb}^{2}+\bar{n}_{\spinorb}+z_{\spinorb}^{2}} & \frac{1}{-\bar{n}_{\spinorb}z_{\spinorb}^{2}+\bar{n}_{\spinorb}+z_{\spinorb}^{2}}-1\\
\frac{\bar{n}_{\spinorb}-\bar{n}_{\spinorb}z_{\spinorb}^{2}}{(\bar{n}_{\spinorb}-1)z_{\spinorb}^{2}-\bar{n}_{\spinorb}} & \frac{z_{\spinorb}}{-\bar{n}_{\spinorb}z_{\spinorb}^{2}+\bar{n}_{\spinorb}+z_{\spinorb}^{2}}
\end{array}\right).
\end{align}
Now, we have 
\begin{align}
\bm{g}_{\spinorb}(\epsilon) & =\left(\begin{array}{cc}
n_{\spinorb}(\epsilon) & z_{\spinorb}(1-n_{\spinorb}(\epsilon))\\
-z_{\spinorb}n_{\spinorb}(\epsilon) & z_{\spinorb}^{2}(n_{\spinorb}(\epsilon)-\bar{n}_{\spinorb})+\bar{n}_{\spinorb}
\end{array}\right),
\end{align}
and the new local interacting integer time Green's function can then
be computed as
\begin{equation}
\boldsymbol{g}_{loc;\spinorb}=\int d\epsilon D_{\spinorb}(\epsilon)\bm{g}_{\spinorb}(\epsilon)=\left(\begin{array}{cc}
\bar{n}_{\spinorb} & z_{\spinorb}(1-\bar{n}_{\spinorb})\\
-z_{\spinorb}\bar{n}_{\spinorb} & \bar{n}_{\spinorb}
\end{array}\right),
\end{equation}
thus proving that self-consistency has been achieved. Finally, the
ground state energy can be constructed as \begin{widetext}
\begin{align}
 & \mathcal{E}=\min_{\begin{array}{c}
P_{1,i,\Gamma\Gamma'}\\
n_{\spinorb}(\epsilon)\in[0,1]
\end{array}}\bigg\{\sum_{\spinorb}\int d\epsilon D_{\spinorb}(\epsilon)\epsilon\left(z_{\spinorb}^{2}(n_{\spinorb}(\epsilon)-\bar{n}_{\spinorb})+\bar{n}_{\spinorb}\right)+\langle\hat{H}_{loc}\rangle_{\spd_{loc}}|P_{1,i,\Gamma\Gamma'}\in\mathcal{C}\bigg\},
\end{align}
\end{widetext} where $\mathcal{C}$ denotes that the constraints
in equations \ref{eq:mgutzc1} and \ref{eq:mgutzc2} must be satisfied.
In summary, the above analysis proves that the G-type $\discn=2$
SPD-d recovers the multi-orbital Gutzwiller approximation in this
case. 

\subsubsection{Multi-band Hubbard model: the general case.}

Here we treat the most general case of the multi-orbital Hubbard model,
where the Hamiltonian is defined as
\[
\hat{H}=\sum_{k\alpha\beta}t_{\spinorb\beta}(k)\hat{a}_{k\spinorb}^{\dagger}\hat{a}_{k\beta}+\sum_{i}\hat{H}_{loc,i}
\]
where $k$ is a reciprocal lattice point in a $\mathscr{D}$ dimensional
crystal; the spin orbital indices are labeled by $\alpha,\beta$ (with
a total number of spin orbitals $M$ at a given $k$-point); and $\hat{H}_{loc,i}$
is a completely general local interaction at site $i$. The corresponding
SPD-d is then 

\begin{align}
\spd=\left(\prod_{i}\hat{P}_{1,i}\right)\exp(\sum_{k}\sppm(k)\cdot\spom_{k})\left(\prod_{i}\hat{P}_{2,i}\right),
\end{align}
where $\hat{P}_{1,i}=\sum_{\Gamma\Gamma'}P_{1,i,\Gamma\Gamma'}\hat{X}_{i;\Gamma\Gamma'}=\hat{P}_{2,i}^{\dagger}$
is the interacting projector of site $i$ in the physical space, and
the variational parameters for a given $k$ are encoded in the $M\times M$
matrix $\sppm(k)$. The non-interacting integer time Green's function
for the $k$ point is a $2M\times2M$ matrix given as 
\begin{align}
\bm{g}_{0}(k)=\begin{pmatrix}\boldsymbol{n}(k) & \left(\boldsymbol{1}-\boldsymbol{n}(k)\right)\\
-\boldsymbol{n}(k) & \boldsymbol{n}(k)
\end{pmatrix},
\end{align}
where 
\begin{equation}
\boldsymbol{n}(k)=\left(\frac{\boldsymbol{1}}{\boldsymbol{1}+\exp\left(-\sppm(k)\right)}\right)^{T}
\end{equation}
is used as a reparameterization of the variational parameters $\sppm(k)$,
and $\boldsymbol{n}(k)$ is constrained to be Hermitian with eigenvalues
between zero and one. As an initial guess, we choose the non-interacting
integer time Green's function of the CDA to be the non-interacting
local integer time Green's function 
\begin{align}
\boldsymbol{\mathcal{G}}=\frac{V}{(2\pi)^{\mathscr{D}}}\int d^{\mathscr{D}}k\bm{g}_{0}(k)=\begin{pmatrix}\bar{\boldsymbol{n}} & \left(\boldsymbol{1}-\bar{\boldsymbol{n}}\right)\\
-\bar{\boldsymbol{n}} & \bar{\boldsymbol{n}}
\end{pmatrix},
\end{align}
where $\bar{\boldsymbol{n}}=\frac{V}{(2\pi)^{\mathscr{D}}}\int d^{\mathscr{D}}k\boldsymbol{n}(k)$.
Considering the CDA at site $i$ for this SPD-d, we have
\begin{align}
 & \spdcs_{loc}=\spdcs_{loc;0}\barhat[P]_{1,i}^{(1)}\barhat[P]_{2,i}^{(2)},\\
 & \spdcs_{loc;0}=\hat{\rho}_{G}(\boldsymbol{\mathcal{G}})=\barhat[Q]\exp(\boldsymbol{v}_{1}\cdot\spomcs_{i}^{(1)}+\boldsymbol{v}_{2}\cdot\spomcs_{i}^{(2)}),
\end{align}
where
\begin{align}
\boldsymbol{v}_{1} & =0, & \boldsymbol{v}_{2}=-\ln(\bar{\boldsymbol{n}}^{-1}-1)^{T}.
\end{align}
We see that $\spdcs_{loc}$ is an effective discrete action of an
SPD given as
\begin{align}
 & \spd_{loc}=\exp(\boldsymbol{v}_{1}\cdot\spom_{i})\hat{P}_{1,i}\exp(\boldsymbol{v}_{2}\cdot\spom_{i})\hat{P}_{2,i},
\end{align}
and the non-interacting SPD is then 
\begin{align}
 & \spd_{loc;0}=\exp(\boldsymbol{v}_{1}\cdot\spom_{i})\exp(\boldsymbol{v}_{2}\cdot\spom_{i}).
\end{align}
As before, we enforce the constraints on the interacting projectors
as 
\begin{align}
 & \frac{\text{Tr}\left(\hat{P}_{1,i}\spd_{loc;0}\hat{P}_{2,i}\right)}{\text{Tr}\left(\spd_{loc;0}\right)}=1, & \frac{\text{Tr}\left(\hat{P}_{1,i}\spd_{loc;0}\hat{P}_{2,i}\spom_{i}\right)}{\text{Tr}\left(\spd_{loc;0}\right)}=\bar{\boldsymbol{n}}.\label{eq:gen-gutz-constr}
\end{align}
We can now directly evaluate the interacting integer time Green's
function as 
\begin{align}
\bm{g}_{loc}=\begin{pmatrix}\bar{\boldsymbol{n}} & \boldsymbol{B}\\
-\boldsymbol{A} & \bar{\boldsymbol{n}}
\end{pmatrix},
\end{align}
where
\begin{align}
A_{\alpha\beta} & =\frac{\text{Tr}\left(\hat{P}_{1,i}\hat{a}_{i\beta}\spd_{loc;0}\hat{P}_{2,i}\hat{a}_{i\alpha}^{\dagger}\right)}{\text{Tr}\left(\spd_{loc;0}\right)},\\
B_{\alpha\beta} & =\frac{\text{Tr}\left(\hat{P}_{1,i}\hat{a}_{i\alpha}^{\dagger}\spd_{loc;0}\hat{P}_{2,i}\hat{a}_{i\beta}\right)}{\text{Tr}\left(\spd_{loc;0}\right)}.
\end{align}

\renewcommand\arraystretch{1.5}Using the discrete Dyson equation,
we can evaluate the local exponential integer time self-energy as 

\begin{align}
\boldsymbol{S}_{loc} & =\left(\boldsymbol{\mathcal{G}}^{-1}-\boldsymbol{1}\right)^{-1}\left(\boldsymbol{g}_{loc}^{-1}-\boldsymbol{1}\right)\\
 & =\begin{pmatrix}\begin{array}{cc}
\left[\boldsymbol{g}_{loc}^{-1}\right]_{21} & \left[\boldsymbol{g}_{loc}^{-1}\right]_{22}-\boldsymbol{1}\\
\frac{\boldsymbol{1}}{\boldsymbol{1}-\bar{\boldsymbol{n}}^{-1}}\left(\left[\boldsymbol{g}_{loc}^{-1}\right]_{11}-\boldsymbol{1}\right) & \frac{\boldsymbol{1}}{\boldsymbol{1}-\bar{\boldsymbol{n}}^{-1}}\left[\boldsymbol{g}_{loc}^{-1}\right]_{12}
\end{array}\end{pmatrix}.
\end{align}
Now, for a given $k$ point we can use the discreet Dyson equation
to obtain the interacting integer time Green's function
\begin{align}
 & \boldsymbol{g}\left(k\right)^{-1}-\boldsymbol{1}=\left(\boldsymbol{g}_{0}\left(k\right)^{-1}-\boldsymbol{1}\right)\boldsymbol{S}_{loc}=\\
 & \begin{pmatrix}\begin{array}{cc}
\boldsymbol{C}(k)\left(\left[\boldsymbol{g}_{loc}^{-1}\right]_{11}-\boldsymbol{1}\right) & \boldsymbol{C}(k)\left[\boldsymbol{g}_{loc}^{-1}\right]_{12}\\
\left[\boldsymbol{g}_{loc}^{-1}\right]_{21} & \left[\boldsymbol{g}_{loc}^{-1}\right]_{22}-\boldsymbol{1}
\end{array}\end{pmatrix},
\end{align}
where
\begin{equation}
\boldsymbol{C}(k)=\left(\boldsymbol{1}-\boldsymbol{n}(k)^{-1}\right)\left(\boldsymbol{1}-\bar{\boldsymbol{n}}^{-1}\right)^{-1},
\end{equation}
and finally we have
\begin{equation}
\boldsymbol{g}\left(k\right)=\begin{pmatrix}\boldsymbol{n}(k) & (\boldsymbol{1}-\boldsymbol{n}(k))\frac{\boldsymbol{1}}{\boldsymbol{1}-\bar{\boldsymbol{n}}}\boldsymbol{B}\\
-\boldsymbol{A}\bar{\boldsymbol{n}}^{-1}\boldsymbol{n}(k) & \bar{\boldsymbol{n}}+\boldsymbol{A}\bar{\boldsymbol{n}}^{-1}(\boldsymbol{n}(k)-\bar{\boldsymbol{n}})\frac{\boldsymbol{1}}{\boldsymbol{1}-\bar{\boldsymbol{n}}}\boldsymbol{B}
\end{pmatrix}.
\end{equation}
\renewcommand\arraystretch{1}We therefore have proven that self-consistency
has been achieved. The total energy can then be written as 
\begin{equation}
\mathcal{E}=\min\left\{ K+\langle\hat{H}_{loc}\rangle_{\spd_{loc}}\right\} ,\label{eq:multiorbgutz}
\end{equation}
where the kinetic energy $K$ is given as 
\begin{align}
K & =\frac{V}{(2\pi)^{\mathscr{D}}}\int d^{\mathscr{D}}k\text{Tr}\left(\boldsymbol{t}\left(k\right)^{T}\boldsymbol{A}\bar{\boldsymbol{n}}^{-1}\boldsymbol{n}(k)(\boldsymbol{1}-\bar{\boldsymbol{n}})^{-1}\boldsymbol{B}\right),
\end{align}
and the minimization over all variational parameters is performed
under the constraints given in Eq. \ref{eq:gen-gutz-constr} and the
restriction that $\boldsymbol{n}(k)$ is Hermitian with eigenvalues
between zero and one. We emphasize that our expression for the total
energy is fully rotationally invariant and holds for an arbitrary
basis. 

Although it is not immediately obvious, Eq. \ref{eq:multiorbgutz}
recovers the particular rotationally invariant multi-orbital Gutzwiller
equation of Ref. \onlinecite{Lanata2012035133} when we specialize
to the case of a ``mixed original-natural basis''. This can be seen
by inserting the identity as

\begin{widetext}
\begin{align}
K & =\frac{V}{(2\pi)^{D}}\int d^{D}k\text{Tr}\left(\boldsymbol{t}\left(k\right)^{T}\boldsymbol{A}\boldsymbol{U}\boldsymbol{U}^{\dagger}\bar{\boldsymbol{n}}^{-1}\boldsymbol{U}\boldsymbol{U}^{\dagger}\boldsymbol{n}(k)\boldsymbol{U}\boldsymbol{U}^{\dagger}(\boldsymbol{1}-\bar{\boldsymbol{n}})^{-1}\boldsymbol{U}\boldsymbol{U}^{\dagger}\boldsymbol{B}\right)\\
 & =\frac{V}{(2\pi)^{D}}\int d^{D}k\sum_{\beta\alpha\gamma\delta}\left([\boldsymbol{t}\left(k\right)^{T}]_{\beta\alpha}[\boldsymbol{A}\boldsymbol{U}]_{\alpha\gamma}\frac{1}{n_{\gamma}^{0}}[\boldsymbol{U}^{\dagger}\boldsymbol{n}(k)\boldsymbol{U}]_{\gamma\delta}\frac{1}{1-n_{\delta}^{0}}[\boldsymbol{U}^{\dagger}\boldsymbol{B}]_{\delta\beta}\right)
\end{align}
\end{widetext}where $\boldsymbol{U}$ is a unitary transformation
that diagonalizes the local single-particle density matrix
\begin{align}
[\boldsymbol{U}^{\dagger}\bar{\boldsymbol{n}}\boldsymbol{U}]_{\alpha\beta} & =\langle\hat{d}_{R\alpha}^{\dagger}\hat{d}_{R\beta}\rangle_{\spd_{loc;0}}=n_{\alpha}^{0}\delta_{\alpha\beta}
\end{align}
and 
\begin{equation}
[\boldsymbol{U}^{\dagger}\boldsymbol{n}(k)\boldsymbol{U}]_{\alpha\beta}=\langle\hat{d}_{k\alpha}^{\dagger}\hat{d}_{k\beta}\rangle_{\spd_{0}}
\end{equation}
where $R$ labels the site of the SCDA and $\hat{d}_{k\delta}=\sum_{l}\hat{a}_{kl}[\boldsymbol{U}]_{l\delta}$
is the annihlation operator of the natural orbital. We can then evaluate
the matrix elements of $\boldsymbol{A}\boldsymbol{U}$ as 

\begin{align}
 & [\boldsymbol{A}\boldsymbol{U}]_{\alpha\gamma}=\sum_{l}\frac{\text{Tr}\left(\hat{P}_{1,R}\hat{a}_{Rl}\spd_{loc;0}\hat{P}_{2,R}\hat{a}_{R\alpha}^{\dagger}\right)}{\text{Tr}\left(\spd_{loc;0}\right)}[\boldsymbol{U}]_{l\gamma}\\
 & =\frac{\text{Tr}\left(\hat{P}_{1,R}\hat{d}_{R\gamma}\spd_{loc;0}\hat{P}_{2,R}\hat{a}_{R\alpha}^{\dagger}\right)}{\text{Tr}\left(\spd_{loc;0}\right)}\\
 & =\frac{\text{Tr}\left(\hat{P}_{1,R}\sqrt{\spd_{loc;0}}\hat{d}_{R\gamma}\sqrt{\spd_{loc;0}}\hat{P}_{2,R}\hat{a}_{R\alpha}^{\dagger}\right)}{\text{Tr}\left(\spd_{loc;0}\right)}\sqrt{\frac{n_{\gamma}^{0}}{1-n_{\gamma}^{0}}}\\
 & =\textrm{Tr}\left(\phi f_{\gamma}\phi^{\dagger}f_{\alpha}^{\dagger}\right)\sqrt{\frac{n_{\gamma}^{0}}{1-n_{\gamma}^{0}}}=\mathcal{R}_{\alpha\gamma}n_{\gamma}^{0},
\end{align}
where 
\begin{align}
[\phi]_{\Gamma n} & =\langle\Gamma;R|\hat{P}_{1,R}\sqrt{\spd_{loc;0}}|n;R\rangle,\\{}
[f_{\gamma}]_{nn'} & =\langle n;R|\hat{d}_{R\gamma}|n';R\rangle,\\{}
[f_{\gamma}]_{\Gamma\Gamma'} & =\langle\Gamma;R|\hat{a}_{R\gamma}|\Gamma';R\rangle,\\
\mathcal{R}_{\alpha\gamma} & =\frac{\textrm{Tr}\left(\phi f_{\gamma}\phi^{\dagger}f_{\alpha}^{\dagger}\right)}{\sqrt{n_{\gamma}^{0}(1-n_{\gamma}^{0})}},
\end{align}
and it is critical to establish a consistent ordering of the original
and natural states for evaluating $f_{\gamma}$\cite{Lanata2012035133}.
Similarly, we can construct the matrix elements of $\boldsymbol{U}^{\dagger}\boldsymbol{B}$
as
\begin{align}
 & [\boldsymbol{U}^{\dagger}\boldsymbol{B}]_{\delta\beta}=\sum_{l}[\boldsymbol{U}^{\dagger}]_{\delta l}\frac{\text{Tr}\left(\hat{P}_{1,R}\hat{a}_{Rl}^{\dagger}\spd_{loc;0}\hat{P}_{2,R}\hat{a}_{R\beta}\right)}{\text{Tr}\left(\spd_{loc;0}\right)}\\
 & =\frac{\text{Tr}\left(\hat{P}_{1,R}\hat{d}_{R\delta}^{\dagger}\spd_{loc;0}\hat{P}_{2,R}\hat{a}_{R\beta}\right)}{\text{Tr}\left(\spd_{loc;0}\right)}\\
 & =\frac{\text{Tr}\left(\hat{P}_{1,R}\sqrt{\spd_{loc;0}}\hat{d}_{R\delta}^{\dagger}\sqrt{\spd_{loc;0}}\hat{P}_{2,R}\hat{a}_{R\beta}\right)}{\text{Tr}\left(\spd_{loc;0}\right)}\sqrt{\frac{1-n_{\delta}^{0}}{n_{\delta}^{0}}}\\
 & =\text{Tr}\left(\phi f_{\delta}^{\dagger}\phi^{\dagger}f_{\beta}\right)\sqrt{\frac{1-n_{\delta}^{0}}{n_{\delta}^{0}}}=\mathcal{R}_{\beta\delta}^{*}(1-n_{\delta}^{0}).
\end{align}
Finally, we can express the kinetic energy as 
\begin{equation}
K=\frac{V}{(2\pi)^{\mathscr{D}}}\int d^{\mathscr{D}}k\sum_{\alpha\beta\gamma\delta}[\boldsymbol{t}\left(k\right)]_{\alpha\beta}\mathcal{R}_{\alpha\gamma}\mathcal{R}_{\beta\delta}^{*}\langle\hat{d}_{k\gamma}^{\dagger}\hat{d}_{k\delta}\rangle_{\spd_{0}},
\end{equation}
which is simply the Fourier transform of the kinetic energy in Eq.
27 of Ref. \onlinecite{Lanata2012035133}, while the potential energy
is straightforwardly equivalent. 

\section{VDAT Workflow\label{sec:Parameterizing-the-SPD}}

\subsection{General considerations}

\begin{figure}
\includegraphics[width=1\columnwidth]{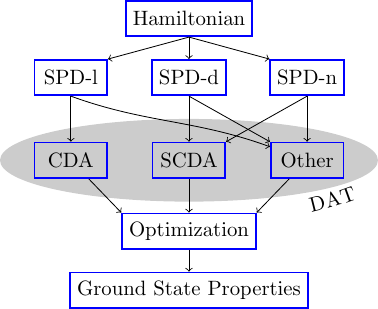}

\caption{\label{fig:schem_vdat_workflow}A schematic of the VDAT workflow.}
\end{figure}

Having presented the entire VDAT formalism, we now discuss the overall
execution of the theory (see Figure \ref{fig:schem_vdat_workflow}
for a schematic). We begin with some Hamiltonian for which we need
to solve the ground state properties. The first step is to choose
an appropriate SPD for the given Hamiltonian, and the best choice
will not be a priori obvious given the competition between the complexity
of the interacting projector versus the number of integer time steps
(see discussion in Subsection \ref{subsec:Classification-of-SPD}).
Broadly speaking, it will be clear that SPD-l would be used for a
model with strictly local interactions, SPD-d would be used for lattice
models with interactions restricted to some range, and SPD-2 would
be natural for a general model with long range Coulomb interactions.
The details of the interacting projectors in each case may be tailored
to the problem at hand. Given that the projective SPD appears to converge
faster with $\discn$ as compared to the unitary SPD, the former is
recommended (see Subsection \ref{subsec:ex-AIM-nle2} for a comparison).

Having selected an SPD, our approach is to use the discrete action
theory to evaluate it. In the case of SPD-l, we can always use the
CDA to evaluate it. In the case of SPD-d, a possible choice would
be to use the SCDA, though for a finite dimensional lattice this would
only be an approximate evaluation of the SPD-d. Other choices would
involve a stochastic evaluation of the integer time Feynman diagrams,
which would provide a numerically exact evaluation, though we have
not explored this in the present manuscript. In the case of SPD-2,
there could be various possibilities. First, a stochastic evaluation
would be possible. Second, a diagrammatic evaluation based on some
class of integer time Feynman diagrams, such as the GW approach\cite{Aryasetiawan1998237},
would be possible, though this would be an approximate evaluation. 

Having evaluated the SPD within the discrete action theory, the final
step is to minimize the ground state energy over the variational parameters,
which involves reevaluating the SPD at different sets of variational
parameters. In general, one can obtain the gradient of the total energy
with respect to the variational parameters in terms of the integer
time correlation functions, which is critical for an efficient minimization
(see Subsection \ref{subsec:Minimization-of-the-SPD}). For $\discn=1$
and $\discn=2$, there are cases where the total energy can be written
in a closed form in terms of the variational parameters, such that
the gradient can be trivially evaluated. In any case, it useful to
contemplate how much variational freedom is actually needed to achieve
precise ground state properties, and we explore the parameterization
of the SPD in the following sections.

\subsection{Parameterization of the SPD}

An SPD will normally contain a large number of variational parameters,
potentially even infinite. Our definition of an SPD dictates that
the non-interacting SPD has full variational freedom, and then in
practice one can decide whether or not to exploit all of it. We emphasize
that this philosophy sometimes departs with common practices in related
variational wave functions, such as in the case of the Baeriswyl wave
function (i.e. $\discn=2$, B-type SPD), which typically only has
a single variational parameter for the kinetic projectors\cite{Baeriswyl19870,Baeriswyl2009075010}.
Alternatively, for Hartree-Fock (i.e. $\discn=1$, G-type SPD), full
variational freedom for the non-interacting projectors is exploited.
In any case, we have proven that for $\discn\ge3$, even very naive
schemes for parameterizing the space of non-interacting variational
parameters in terms of a small number of variables can give highly
precise results\cite{Cheng2020short}. We will present examples in
the context of the Hubbard model and AIM to illustrate these ideas.

In terms of the interacting projector, the number of variational parameters
could be as large as the Fock space that the interacting projectors
span, which can be impractical even in principle; or it could be as
small as one variational parameter, which would still recover the
exact solution in the large $\discn$ limit. In the examples considered
below, we will only be evaluating a single interacting orbital, such
that there is only a single variational parameter at each integer
time. The remainder of this section focuses purely on parameterizing
the non-interacting projector for applications with $\discn\ge3$,
which have been used in our accompanying applications\cite{Cheng2020short}. 

\subsubsection{Parameterization of SPD-l for the AIM on a ring\label{subsec:Parameterization-AIM-ring}}

Here we consider the Anderson Impurity model (AIM) on a ring\cite{Barcza2019165130},
given by
\begin{align}
 & \hat{H}=\hat{H}_{0}+\hat{V},\hspace{1em}\hat{V}=U\left(\hat{f}_{\uparrow}^{\dagger}\hat{f}_{\uparrow}-\frac{1}{2}\right)\left(\hat{f}_{\downarrow}^{\dagger}\hat{f}_{\downarrow}-\frac{1}{2}\right),\\
 & \hat{H}_{0}=v\sum_{\sigma}(\hat{f}_{\sigma}^{\dagger}\hat{c}_{0,\sigma}+h.c.)-\frac{W}{4}\sum_{\sigma,n=0}^{L-1}\left(\hat{c}_{n,\sigma}^{\dagger}\hat{c}_{n+1,\sigma}+h.c\right).
\end{align}
The interacting projector of the SPD-l is given as 
\begin{align}
\hat{P}_{\tau} & =P_{\tau,0}\hat{X}_{0}+P_{\tau,\downarrow}\hat{X}_{\downarrow}+P_{\tau,\uparrow}\hat{X}_{\uparrow}+P_{\tau,\uparrow\downarrow}\hat{X}_{\uparrow\downarrow}\\
 & =(1-\mu_{\tau}-\frac{1}{4}u_{\tau})\hat{1}+\mu_{\tau}\sum_{\sigma}\hat{f}_{\sigma}^{\dagger}\hat{f}_{\sigma}+u_{\tau}\hat{f}_{\uparrow}^{\dagger}\hat{f}_{\uparrow}\hat{f}_{\downarrow}^{\dagger}\hat{f}_{\downarrow},\label{eq:AIMintproj}
\end{align}
where $\hat{X}_{\Gamma}$ are diagonal Hubbard operators and $P_{\tau,\Gamma}$
are variational parameters, which can then be constrained using the
density and normalization. For this case of spin symmetry, we formally
have two variational parameters $u_{\tau}$ and $\mu_{\tau}$ for
each time step, though the latter should be considered as a parameter
to constrain the local density. 

For the non-interacting projector, we allow for three variational
parameters: one for each independent parameter in the non-interacting
Hamiltonian, though one parameter will be fixed by the total density.
In order to implement this, it is most convenient to use a diagonal
form, so we construct an effective basis which is a function of the
three variational parameters, which then allows us to enforce a semi-definite
matrix. The non-interacting projector is given by

\begin{equation}
\exp\left(\sppm_{\tau}\cdot\spom\right)=\prod_{j\sigma}\big((1+h(\epsilon_{j\sigma;\tau})(\hat{b}_{j\sigma}^{\dagger}\hat{b}_{j\sigma}-\frac{1}{2})\big),
\end{equation}
where $\epsilon_{j\sigma;\tau}$ are functions of the variational
parameters given by
\begin{align}
 & \sum_{j\sigma}\epsilon_{j\sigma;\tau}\hat{b}_{j\sigma}^{\dagger}\hat{b}_{j\sigma}=\eta_{1;\tau}\sum_{\sigma}\left(\hat{f}_{\sigma}^{\dagger}\hat{c}_{0,\sigma}+h.c.\right)-\nonumber \\
 & \eta_{2;\tau}\sum_{\sigma,n=0}^{L-1}\left(\hat{c}_{n,\sigma}^{\dagger}\hat{c}_{n+1,\sigma}+h.c\right)+\eta_{3;\tau}\hat{N},
\end{align}
and $\eta_{1;\tau},\eta_{2;\tau},\eta_{3;\tau}$ are the variational
parameters, the indices $j,\sigma$ label the eigenstates of the above
diagonalized form, and the function $h$ is defined as
\begin{align}
 & h(x)=\begin{cases}
2, & x>2\\
x, & -2<x\le2\\
-2, & x\le-2.
\end{cases}\label{eq:hofx}
\end{align}

\subsubsection{Parameterization of the SPD-d for the Hubbard model\label{subsec:Parameterization-Hub}}

Here we consider the single-band Hubbard model in an arbitrary dimension,
defined as 

\begin{align}
 & \hat{H}=\sum_{ij\sigma}t_{ij}\hat{a}_{i\sigma}^{\dagger}\hat{a}_{j\sigma}+U\sum_{i}\hat{d}_{i},
\end{align}
where $\hat{d}_{i}=\hat{n}_{i\uparrow}\hat{n}_{i\downarrow}$, and
we employ the SPD-d with local interacting projectors 
\begin{align}
\hat{P}_{\tau} & =\prod_{i}\big((1-\mu_{\tau}-\frac{1}{4}u_{\tau})\hat{1}+\mu_{\tau}\sum_{\sigma}\hat{n}_{i\sigma}+u_{\tau}\hat{d}_{i}\big),
\end{align}
where $\hat{P}_{\tau}$ is a product of local projectors with the
same form as Eq. \ref{eq:AIMintproj}, and translation symmetry has
been assumed. 

The non-interacting projector for this SPD-d is simpler than the case
of the SPD-l for the AIM, presented in the preceding section, given
that translation symmetry fully diagonalizes the non-interacting Hamiltonian.
In this case, we choose four variational parameters, allowing more
freedom than the non-interacting form of the Hamiltonian, which only
has nearest neighbor hopping. We have the following form for the non-interacting
projector 

\begin{equation}
\exp\left(\sppm_{\tau}\cdot\spom\right)=\prod_{\vec{k}\sigma}\left(1+h(w_{\vec{k}\sigma;\tau})(\hat{n}_{\vec{k}\sigma}-\frac{1}{2})\right),
\end{equation}
where 
\begin{align}
 & w_{\vec{k}\sigma;\tau}=\eta_{1}\theta(\epsilon_{\vec{k}\sigma}-\eta_{2})+\eta_{3}f(\epsilon_{\vec{k}\sigma}-\eta_{2},\eta_{4})+\eta_{2},\\
 & f(x,\alpha)=\textrm{sign}(x)(1-\exp(-\alpha|x|))/(1-\exp(-\alpha)).
\end{align}
where $h(x)$ is defined in Eq. \ref{eq:hofx} and the four variational
parameters are $\eta_{i}$. The form of $w_{\vec{k}\sigma;\tau}$
is chosen such that it recovers the Baeriswyl kinetic projector when
$\eta_{1}=0$ and $\eta_{4}\rightarrow0$, thus allowing extra variational
freedom. This particular form was motivated by examining the density
distribution within perturbation theory. However, even when restricting
to $\eta_{2}$ and $\eta_{3}$, the results for the ground state energy
are nearly unchanged. 

\section{Summary and Conclusions}

In this work, we have introduced the variational discrete action theory
(VDAT), which is a systematic variational approach for solving the
ground state of quantum many-body Hamiltonians. VDAT consists of two
important components: a variational ansatz referred to as the sequential
product density matrix (SPD) and a general formalism to compute observables
under the SPD, where the latter is referred to as the discrete action
theory (DAT). It should be emphasized that when DAT exactly evaluates
the energy of some Hamiltonian under the SPD, the result is a rigorous
upper bound to the exact ground state energy, and minimization over
the variational parameters within the SPD will provide the optimal
solution for the SPD at a given $\discn$. While for typical discretized
Green's function approaches, such as auxiliary field quantum Monte-Carlo,
a large number of time steps are required, VDAT can provide remarkably
accurate ground state results even for very small $\discn$.

The SPD is clearly inspired by the Trotter-Suzuki decomposition, and
the essence of the idea in the context of variational wave functions
can be traced back to the work of Baeriswyl and coworkers decades
ago\cite{Dzierzawa19951993}. However, without a clear prescription
for evaluation, it is a tool of limited applicability; and the discrete
action theory put forward in this paper greatly extends the reach
of the SPD. In this paper, we put forward the most generic notion
of an SPD, which can be partitioned into unitary, projective, and
general cases. There are always two ways to constrain the SPD to be
Hermitian and semi-definite: Gutzwiller-type and Baeriswyl-type, denoted
G-type and B-type. We introduced three important classes of SPD in
this work: the local SPD (SPD-l), the disjoint SPD (SPD-d), and the
n-particle SPD (SPD-n). The key characteristic of an SPD-l is that
it has a finite number of local interacting projectors, and the SPD-l
is naturally applied to models where the interactions are restricted
to some subspace, like the Anderson impurity model. Alternatively,
the SPD-d has multiple sets of local interacting projectors which
do not overlap, and the SPD-d is naturally applied to interacting
lattice model like the Hubbard model or periodic Anderson model. The
SPD-n has a general n-particle interacting projector, and the n=2
case would naturally apply to models with long range Coulomb repulsion.

In order to evaluate observables under the SPD, we introduced the
integer time Green's function formalism, which can be generally characterized
by a discrete action. We demonstrate that for the SPD-l, one can sum
all local diagrams, allowing for the exact evaluation of observables
under the SPD-l, and illustrations for $\discn\le2$ are provided
for the AIM. To evaluate the SPD-d, the more advanced tools of many-body
physics had to be generalized to our integer time formalism. We first
generalized the path integral to integer time, demonstrating that
integer time correlation functions under the SPD are equivalent to
a static correlation function in a compound Fock space under an effective
density matrix. Just as in the case of the usual path integral, the
integer time case unifies the role of spatial and temporal degrees
of freedom, allowing a straightforward generalization of the generating
functional, Dyson equation, and Bethe-Salpeter equation to integer
time.

We introduced a hierarchical scheme to categorize three types of discrete
actions: the sequential discrete action (SDA), the canonical discrete
action (CDA), and the general discrete action (GDA); where we the
former is always a subset of the latter. The SDA is the discrete action
of an SPD, which is a product of three operators in the compound space:
a shift matrix $\barhat[Q]$, a promoted sequential product of non-interacting
projectors, and a promoted sequential product of interacting projectors.
For the SDA, only $\barhat[Q]$ introduces correlations between different
integer times, while the other two components are integer time blocked.
The CDA, is a product of two operators in the compound space: a general
non-interacting projector and a promoted sequential product of interacting
projectors. For the CDA, only the non-interacing projector introduces
correlations between different integer times, while the interacting
projectors are integer time blocked. The GDA is a general operator
in the compound space. 

A key step forward in this paper is the self-consistent canonical
discrete action approximation (SCDA), which is the integer time generalization
of the dynamical mean-field theory. In general, the SPD-d cannot be
exactly evaluated. However, we prove that in $d=\infty$, the SPD-d
can be exactly mapped to a CDA with a self-consistently determined
non-interacting integer time Green's function. Therefore, the SPD-d
can be exactly evaluated for the Hubbard model in $d=\infty$, making
VDAT a potent tool for solving the multi-band Hubbard model. For $\discn=2$,
this recovers the classic result that the Gutzwiller approximation
exactly evaluates the Gutzwiller wave function in $d=\infty$\cite{Metzner1987121,Metzner19887382,Metzner1989324,Bunemann19977343}.
For $\discn=3$, this proves that both the Gutzwiller-Baeriswyl and
the Baeriswyl-Gutzwiller wave functions can be exactly evaluate in
$d=\infty$. For $\discn\ge4$, this yields new variational wave functions
that can be exactly evaluated in $d=\infty$ (see Reference \onlinecite{Cheng2020short}
for applications). The VDAT gives a long awaited variational understanding
of $d=\infty$, which traditionally necessitated the use of Green's
function based approaches. The SPD-d within the SCDA can also be applied
as an approximation in finite dimensions, improving upon well characterized
theories such as the Gutzwiller approximation. 

The VDAT formalism unifies many seemingly disparate variational wave
functions, connecting Hartree-Fock, the Gutzwiller wave function,
etc, all into the same framework, while putting forward an infinite
number of extensions. Equally importantly, the integer time Green's
function and the discrete action theory provides a paradigm for generalizing
concepts from the standard many-body Green's functions methods. For
example, once the general formalism was identified, it was obvious
how to realize the dynamical mean-field theory in the integer time
formalism, allowing for remarkably accurate results at $\discn=3$\cite{Cheng2020short}.

There are many near term and far term directions to consider for VDAT.
In the near term, the multi-band Hubbard model is the most obvious
target. Given that the Gutzwiller approximation is known to produce
reasonable results in the Fermi-liquid regime of the multi-band Hubbard
model\cite{Deng2009075114,Lanata2012035133}, and that we know VDAT($\discn=3$)
precisely captures the metallic and insulating regime of the one band
model in $d=\infty$\cite{Cheng2020short}, VDAT($\discn=3$) within
the SCDA should be sufficiently accurate for the multi-band case.
Further applications could include novel approaches for evaluating
the SPD-d in low dimensions using quantum Monte-Carlo, which would
expand upon the recent work of Baeriswyl in the 2d Hubbard model\cite{Baeriswyl2019235152},
which in our language was a perturbative evaluation of a G-type SPD-d
at $\discn=3$. Other interesting applications include the homogeneous
electron gas, which could require the use of an integer time generalized
auxiliary field QMC to evaluate the SPD-2. In terms of further methodological
developments, it will be useful to consider finite temperatures, where
one must develop approaches for evaluating the entropy. Furthermore,
it will be useful to extend the Landau-Gutzwiller quasiparticle approach\cite{Bunemann2003075103}
in the context of VDAT, such that excited state properties can be
treated.

\section{Acknowledgments}

This work was supported by the grant DE-SC0016507 funded by the U.S.
Department of Energy, Office of Science. This research used resources
of the National Energy Research Scientific Computing Center, a DOE
Office of Science User Facility supported by the Office of Science
of the U.S. Department of Energy under Contract No. DE-AC02-05CH11231. 

\section{Appendix\label{sec:Appendix}}

\subsection{Lie group properties of non-interacting systems\label{subsec:Appendix-Lie-group}}

In this appendix, we study the properties of the density matrices
that are exponentially generated from single particle potentials,
and this applies equally both to the physical Fock space and the compound
space (see Section \ref{subsec:Integer-time-Path-integral}). Understanding
these objects is critical given that they are the noninteracting projectors
of the SPD. We start with a system containing $L$ spin orbitals,
and we define a generalized non-interacting many-body density matrix
as 
\begin{align}
\hat{\rho}\left(\boldsymbol{v}\right)=\exp(\boldsymbol{v}\cdot\spom)=\exp\bigg(\sum_{ij}v_{ij}\hat{a}_{i}^{\dagger}\hat{a}_{j}\bigg),\label{eq:rhononinteracting}
\end{align}
and we refer to this as generalized because $v_{ij}$ forms a complex
matrix that is not necessarily Hermitian. In order to understand that
this operator can be viewed as being generated from a Lie algebra,
we rewrite Eq. \ref{eq:rhononinteracting} in the form
\begin{equation}
\hat{\rho}\left(\boldsymbol{v}\right)=\exp(\sum_{\Gamma}v_{\Gamma}\hat{O}_{\Gamma}),
\end{equation}
where $\Gamma=\left(i,j\right)$ and $\hat{O}_{\left(i,j\right)}=\hat{a}_{i}^{\dagger}\hat{a}_{j}$.
From the anti-commutation relation $\{\hat{a}_{i}^{\dagger},\hat{a}_{j}\}=\delta_{ij}$,
we deduce

\begin{align}
\left[\hat{O}_{\left(i,j\right)},\hat{O}_{\left(k,l\right)}\right] & =\hat{O}_{\left(i,l\right)}\delta_{kj}-\hat{O}_{\left(k,j\right)}\delta_{il}.
\end{align}
Therefore, the operators $\hat{O}_{\Gamma}$ form a Lie algebra and
$\hat{\rho}\left(\boldsymbol{v}\right)$ form a Lie group. This Lie
group structure has long been recognized\cite{Wybourne19731117,Fukutome19771554},
though previous application did not consider non-Hermitian cases.

It is impractical to directly compute the non-interacting density
matrix in the Fock space for a large system, and therefore we need
a more efficient approach, which can be achieved by finding an isomorphism
of the Lie algebra $\hat{O}_{\Gamma}\leftrightarrow A_{\Gamma}$.
The matrices $A_{\Gamma}$ are defined as 
\begin{align}
[A_{(i,j)}]_{ml}=\delta_{im}\delta_{jl},
\end{align}
and have dimension $L\times L$. The commutation relations are given
as
\begin{align}
\left[A_{\left(i,j\right)},A_{\left(k,l\right)}\right] & =A_{\left(i,l\right)}\delta_{k,j}-A_{\left(k,j\right)}\delta_{il},
\end{align}
 which have the same commutation relations as $\hat{O}_{\left(i,j\right)}$,
proving the isomorphism. Therefore, we have the map 
\begin{alignat}{3}
\hat{\rho}\left(\boldsymbol{v}\right) &  & = &  & \exp\left(\sum_{\Gamma}v_{\Gamma}\hat{O}_{\Gamma}\right)\\
 &  & \updownarrow\nonumber \\
\exp\left(\boldsymbol{v}\right) &  & = &  & \exp\left(\sum_{\Gamma}v_{\Gamma}A_{\Gamma}\right).
\end{alignat}
The single-particle density matrix is defined as 
\begin{align}
\boldsymbol{n}\left(\boldsymbol{v}\right) & =\langle\spom\rangle_{\hat{\rho}(\boldsymbol{v})}=\begin{pmatrix}\langle\hat{a}_{1}^{\dagger}\hat{a}_{1}\rangle_{\hat{\rho}(\boldsymbol{v})} & \dots & \langle\hat{a}_{1}^{\dagger}\hat{a}_{L}\rangle_{\hat{\rho}(\boldsymbol{v})}\\
\dots & \dots & \dots\\
\langle\hat{a}_{L}^{\dagger}\hat{a}_{1}\rangle_{\hat{\rho}(\boldsymbol{v})} & \dots & \langle\hat{a}_{L}^{\dagger}\hat{a}_{L}\rangle_{\hat{\rho}(\boldsymbol{v})}
\end{pmatrix}\\
 & =\left(\frac{\boldsymbol{1}}{\boldsymbol{1}+\exp\left(-\boldsymbol{v}\right)}\right)^{T}.
\end{align}
The above equation can be derived as following, starting with 

\begin{align}
\text{Tr}\left(\hat{\rho}\left(\boldsymbol{v}\right)\right)=\text{det}\left(\boldsymbol{1}+\exp\left(\boldsymbol{v}\right)\right),
\end{align}
and writing the single particle density matrix as a derivative, we
have

\begin{align}
\boldsymbol{n}\left(\bm{v}\right) & =\frac{\partial\ln\left(\text{Tr}\left(\hat{\rho}\left(\bm{v}\right)\right)\right)}{\partial\bm{v}}=\frac{\partial\ln\left(\text{det}\left(\boldsymbol{1}+\exp\left(\bm{v}\right)\right)\right)}{\partial\bm{v}}\\
 & =\frac{\partial\text{Tr}\left(\text{\ensuremath{\ln}}\left(\boldsymbol{1}+\exp\left(\bm{v}\right)\right)\right)}{\partial\bm{v}}=\left(\frac{\boldsymbol{1}}{\boldsymbol{1}+\exp\left(-\bm{v}\right)}\right)^{T}.
\end{align}
 Now we consider group multiplication in terms of $\bm{n}$. We begin
by parametrizing the group element in terms of $\bm{n}$ as 
\begin{align}
\exp\left(\bm{v}\right)=\left(\frac{\boldsymbol{1}}{\bm{n}\left(\bm{v}\right)^{-1}-\boldsymbol{1}}\right)^{T},
\end{align}
and recalling the group multiplication
\begin{align}
\exp\left(\bm{v}\right)=\exp\left(\bm{v}_{1}\right)\exp\left(\bm{v}_{2}\right),
\end{align}
we then have 
\begin{align}
\left(\bm{n}^{-1}-\bm{1}\right)=\left(\bm{n}_{1}^{-1}-\bm{1}\right)\left(\bm{n}_{2}^{-1}-\bm{1}\right),
\end{align}
and solving for $\boldsymbol{n}$ we have

\begin{align}
\bm{n}=\bm{n}_{2}\left(\bm{1}-\bm{n}_{1}-\bm{n}_{2}+2\bm{n}_{1}\bm{n}_{2}\right)^{-1}\bm{n}_{1}=\bm{n}_{1}\star\bm{n}_{2}.\label{eq:g_mult_rules}
\end{align}
Finally, we have the group multiplication rules in terms of $\bm{n}$.
This is very useful given that we will normally parameterize the non-interacting
projector in terms of the single particle density matrix instead of
the potential $\boldsymbol{v}$.

Recall that the integer time Green's function is a non-Hermitian single-particle
density matrix in the compound space. Therefore, the preceding derivations
will greatly facilitate the manipulation of the integer time Green's
function. Consider a non-interacting SPD $\spd=\mathcal{\hat{P}}_{1}...\mathcal{\hat{P}}_{\discn}$,
with $\mathcal{\hat{P}}_{\tau}=\exp\left(\sppm_{\tau}\cdot\hat{\bm{n}}\right)$,
which can be reparameterized using 
\begin{align}
\boldsymbol{n}_{\tau}=\left(\frac{\boldsymbol{1}}{\boldsymbol{1}+\exp\left(-\sppm_{\tau}\right)}\right)^{T}.
\end{align}
Using the group notation, we can write the non-interacting integer
time Green's function as
\begin{align}
\bm{g}_{0}=\bm{g}_{Q}\star\textrm{diag(\ensuremath{\bm{n}_{1}},..,\ensuremath{\bm{n}_{\discn}})},
\end{align}
where $\bm{g}_{Q}$ is defined in Eq. \ref{eq:gQ}. For the special
case of $\discn=2$, we have
\begin{align}
\bm{g}_{0}=\begin{pmatrix}\bm{n}_{2}\star\bm{n}_{1} & \left(\bm{n}_{2}^{-1}-\bm{1}\right)\bm{n}_{1}\star\bm{n}_{2}\\
-\left(\bm{n}_{1}^{-1}-\bm{1}\right)\bm{n}_{2}\star\bm{n}_{1} & \bm{n}_{1}\star\bm{n}_{2}
\end{pmatrix}.\label{eq:g02-lie}
\end{align}
The preceding case is very important for a non-interacting SPD at
a general $\discn$ given that one can always split the sequential
integer time blocks into two parts, and iteratively apply Eq. \ref{eq:g02-lie}
using
\begin{align}
\spd'=\left(\mathcal{\hat{P}}_{j+1}...\mathcal{\hat{P}}_{i-1}\mathcal{\hat{P}}_{i}\right)\left(\mathcal{\hat{P}}_{i+1}..\mathcal{\hat{P}}_{j}\right)=\mathcal{\hat{P}}'_{1}\mathcal{\hat{P}}'_{2}.
\end{align}

\subsection{Proof of the integer time Wick's theorem\label{appendix:wicks_theorem}}

Wick's theorem is an important tool for evaluating expectation values
in a non-interacting system, and it is imperative to generalize this
to the integer time formalism. We begin by introducing a unified notation
for creation and annihilation operators as

\begin{equation}
\hat{A}_{i\theta}=\begin{cases}
\hat{a}_{i}^{\dagger}, & \theta=0\\
\hat{a}_{i}, & \theta=1
\end{cases}.
\end{equation}
Next, we consider how to commute $\hat{A}_{i\theta}$ with the non-interacting
density matrix (see Subsection \ref{subsec:Defining-the-SPD}) 
\begin{equation}
\hat{A}_{i\theta}\exp\left(\bm{v}\cdot\spom\right)=M_{i\theta,i'\theta'}\left(\bm{v}\right)\exp\left(\bm{v}\cdot\spom\right)\hat{A}_{i'\theta'},\label{eq:A_commutation}
\end{equation}
where 
\begin{equation}
M_{i\theta,i'\theta'}\left(\bm{v}\right)=\begin{cases}
\left[\exp\left(-\bm{v}^{T}\right)\right]_{ii'}\delta_{\theta,\theta'}, & \theta=0\\
\left[\exp\left(\bm{v}\right)\right]_{ii'}\delta_{\theta,\theta'}, & \theta=1
\end{cases}
\end{equation}
and Einstein notation is employed throughout this section. Note that
the above can be further generalized by considering pairing terms,
which have not presently been included in $\spom$. Recall the integer
time Heisenberg representation for the non-interacting SPD
\begin{equation}
\hat{A}_{i\theta}\left(\tau\right)=U_{\tau}\hat{A}_{i\theta}U_{\tau}^{-1},\label{eq:A_heisenberg}
\end{equation}
where 
\begin{equation}
U_{\tau}=\exp\left(\bm{v}_{1}\cdot\spom\right)...\exp\left(\bm{v}_{\tau}\cdot\spom\right)=\exp\left(V_{1,\tau}\cdot\spom\right).
\end{equation}
We can then use Eq. \ref{eq:A_commutation} and Eq. \ref{eq:A_heisenberg}
to find 

\begin{align}
 & \hat{A}_{i\theta}=M_{i\theta,i'\theta'}(V_{1,\tau})\hat{A}_{i'\theta'}(\tau),\\
 & \hat{A}_{i\theta}(\tau)=M_{i\theta,i'\theta'}(-V_{1,\tau})\hat{A}_{i'\theta'}.
\end{align}
For $\tau_{1}\leq\tau_{2}$, we then have 
\begin{align}
\hat{A}_{i\theta}(\tau_{2}) & =M_{i\theta,i'\theta'}(-V_{1,\tau_{2}})\hat{A}_{i'\theta'}\\
 & =M_{i\theta,i'\theta'}(-V_{1,\tau_{2}})M_{i'\theta',i''\theta''}(V_{1,\tau_{1}})\hat{A}_{i''\theta''}(\tau_{1}).
\end{align}
Recalling the Lie group isomorphism 
\begin{equation}
M\left(V\right)\leftrightarrow\exp\left(V\right)
\end{equation}
and the identity
\begin{equation}
\exp\left(-V_{1,\tau_{2}}\right)\exp\left(V_{1,\tau_{1}}\right)=\exp\left(-V_{\tau_{1}+1,\tau_{2}}\right),
\end{equation}
where 
\begin{equation}
\exp\left(\bm{v}_{\tau_{1}}\right)...\exp\left(\bm{v}_{\tau_{2}}\right)=\exp\left(V_{\tau_{1},\tau_{2}}\right),
\end{equation}
we then have 

\begin{align}
 & M_{i\theta,i'\theta'}(-V_{1,\tau_{2}})M_{i'\theta',i''\theta''}\left(V_{1,\tau_{1}}\right)\nonumber \\
 & =M_{i\theta,i''\theta'',}(-V_{\tau_{1}+1,\tau_{2}}),
\end{align}
which yields

\begin{align}
 & \hat{A}_{i\theta}(\tau_{2})=M_{i\theta,i'\theta'}(-V_{\tau_{1}+1,\tau_{2}})\hat{A}_{i'\theta'}(\tau_{1}),\\
 & \hat{A}_{i\theta}\left(\tau_{1}\right)=M_{i\theta,i'\theta'}(V_{\tau_{1}+1,\tau_{2}})\hat{A}_{i'\theta'}(\tau_{2}).
\end{align}
We also have

\begin{equation}
\left\{ \hat{A}_{i\theta}\left(\tau\right),\hat{A}_{i'\theta'}\left(\tau\right)\right\} =\delta_{ii'}\delta_{\theta+\theta',1}\equiv\Delta_{i\theta,i'\theta'}.
\end{equation}
Considering $\tau_{1}\leq\tau_{2}$ and $\bm{v}_{i+\discn}=\bm{v}_{i}$,
we have

\begin{align}
 & \langle\hat{A}_{i_{1}\theta_{1}}(\tau_{1})\hat{A}_{i_{2}\theta_{2}}(\tau_{2})\rangle\nonumber \\
 & =M_{i_{1}\theta_{1},i'_{1}\theta'_{1}}(V_{\tau_{1}+1,\tau_{2}})\langle\hat{A}_{i'_{1}\theta'_{1}}(\tau_{2})\hat{A}_{i_{2}\theta_{2}}(\tau_{2})\rangle\\
 & =M_{i_{1}\theta_{1},i'_{1}\theta'_{1}}(V_{\tau_{1}+1,\tau_{2}})\Delta_{i'_{1}\theta'_{1},i'_{2}\theta'_{2}}-\nonumber \\
 & M_{i_{1}\theta_{1},i'_{1}\theta'_{1}}(V_{\tau_{1}+1,\tau_{2}})\langle\hat{A}_{i_{2}\theta_{2}}(\tau_{2})\hat{A}_{i'_{1}\theta'_{1}}(\tau_{2})\rangle\\
 & =M_{i_{1}\theta_{1},i'_{1}\theta'_{1}}(V_{\tau_{1}+1,\tau_{2}})\Delta_{i'_{1}\theta'_{1},i_{2}\theta_{2}}-\nonumber \\
 & M_{i_{1}\theta_{1},i'_{1}\theta'_{1}}(V_{\tau_{1}+1,\tau_{1}+\discn})\langle\hat{A}_{i_{2}\theta_{2}}(\tau_{2})\hat{A}_{i'_{1}\theta'_{1}}(\tau_{1}+\discn)\rangle\\
 & =M_{i_{1}\theta_{1},i'_{1}\theta'_{1}}(V_{\tau_{1}+1,\tau_{2}})\Delta_{i'_{1}\theta'_{1},i_{2}\theta_{2}}-\nonumber \\
 & M_{i_{1}\theta_{1},i'_{1}\theta'_{1}}(V_{\tau_{1}+1,\tau_{1}+\discn})\langle\hat{A}_{i'_{1}\theta'_{1}}(\tau_{1})\hat{A}_{i_{2}\theta_{2}}(\tau_{2})\rangle.
\end{align}
Therefore, we have 
\begin{align}
 & (\delta_{i_{1}\theta_{1},i'_{1}\theta'_{1}}+M_{i_{1}\theta_{1},i'_{1}\theta'_{1}}(V_{\tau_{1}+1,\tau_{1}+\discn}))\langle\hat{A}_{i'_{1}\theta'_{1}}(\tau_{1})\hat{A}_{i_{2}\theta_{2}}(\tau_{2})\rangle\nonumber \\
 & =M_{i_{1}\theta_{1},i'_{1}\theta'_{1}}(V_{\tau_{1}+1,\tau_{2}})\Delta_{i'_{1}\theta'_{1},i_{2}\theta_{2}},
\end{align}
and we can solve for the integer time Green's function as

\begin{align}
 & \langle\hat{A}_{i_{1}\theta{}_{1}}(\tau_{1})\hat{A}_{i_{2}\theta_{2}}(\tau_{2})\rangle=\nonumber \\
 & \left[\frac{1}{1+M(V_{\tau_{1}+1,\tau_{1}+N})}M(V_{\tau_{1}+1,\tau_{2}})\Delta\right]_{i_{1}\theta_{1},i_{2}\theta_{2}}.
\end{align}

Now we consider how to evaluate the general case of an M-particle
integer time Green's function. 

\global\long\def\A#1{\hat{A}_{i_{#1}\theta_{#1}}(\tau_{#1})}%

\global\long\def\Ap#1{\hat{A}_{i'_{1}\theta'_{1}}(\tau_{#1})}%

\global\long\def\M#1#2{M_{i_{1}\theta_{1},i'_{1}\theta'_{1}}(V_{#1+1,#2})}%

\global\long\def\D#1{\Delta_{i'_{1}\theta'_{1},i_{#1}\theta_{#1}}}%
Considering the case where the $\tau_{i}$ are ordered and $N=2M$,
we have

\begin{align}
 & \langle\A 1\A 2...\A N\rangle\nonumber \\
 & =\M{\tau_{1}}{\tau_{2}}\langle\Ap 2\A 2...\A N\rangle\nonumber \\
 & =\M{\tau_{1}}{\tau_{2}}\times\nonumber \\
 & \langle\big(\D 2-\A 2\Ap 2\big)...\A N\rangle\nonumber \\
 & =\M{\tau_{1}}{\tau_{2}}\D 2\langle\A 3...\A N\rangle\nonumber \\
 & -\M{\tau_{1}}{\tau_{2}}\times\nonumber \\
 & \langle\A 2\Ap 2\A 3...\A N\rangle\nonumber \\
 & =\M{\tau_{1}}{\tau_{2}}\D 2\langle\A 3...\A N\rangle\nonumber \\
 & -\M{\tau_{1}}{\tau_{3}}\times\nonumber \\
 & \langle\A 2(\D 3-\A 3\Ap 3)...\A N\rangle\nonumber \\
 & =\sum_{\ell=2}^{N}\left(-1\right)^{\ell}\M{\tau_{1}}{\tau_{\ell}}\D{\ell}\times\nonumber \\
 & \langle\A 2...\A{\ell-1}\A{\ell+1}...\A N\rangle\nonumber \\
 & -\M{\tau_{1}}{\tau_{1}+N}\times\nonumber \\
 & \langle\Ap 1\A 2...\A N\rangle.
\end{align}
Therefore, the M-particle Green's function can be written as a summation
of the product of a single-particle and $(M-1)$-particle integer
time Green's functions. 
\begin{align}
 & \langle\A 1\A 2...\A N\rangle\nonumber \\
 & =\sum_{\ell=2}^{N}\left(-1\right)^{\ell}\langle\A 1\A{\ell}\rangle\times\nonumber \\
 & \langle\A 2..\A{\ell-1}\A{\ell+1}...\A N\rangle.
\end{align}
Iteratively applying the preceding equation proves Wick's theorem.


\begin{thebibliography}{58}%
\makeatletter
\providecommand \@ifxundefined [1]{%
 \@ifx{#1\undefined}
}%
\providecommand \@ifnum [1]{%
 \ifnum #1\expandafter \@firstoftwo
 \else \expandafter \@secondoftwo
 \fi
}%
\providecommand \@ifx [1]{%
 \ifx #1\expandafter \@firstoftwo
 \else \expandafter \@secondoftwo
 \fi
}%
\providecommand \natexlab [1]{#1}%
\providecommand \enquote  [1]{``#1''}%
\providecommand \bibnamefont  [1]{#1}%
\providecommand \bibfnamefont [1]{#1}%
\providecommand \citenamefont [1]{#1}%
\providecommand \href@noop [0]{\@secondoftwo}%
\providecommand \href [0]{\begingroup \@sanitize@url \@href}%
\providecommand \@href[1]{\@@startlink{#1}\@@href}%
\providecommand \@@href[1]{\endgroup#1\@@endlink}%
\providecommand \@sanitize@url [0]{\catcode `\\12\catcode `\$12\catcode
  `\&12\catcode `\#12\catcode `\^12\catcode `\_12\catcode `\%12\relax}%
\providecommand \@@startlink[1]{}%
\providecommand \@@endlink[0]{}%
\providecommand \url  [0]{\begingroup\@sanitize@url \@url }%
\providecommand \@url [1]{\endgroup\@href {#1}{\urlprefix }}%
\providecommand \urlprefix  [0]{URL }%
\providecommand \Eprint [0]{\href }%
\providecommand \doibase [0]{http://dx.doi.org/}%
\providecommand \selectlanguage [0]{\@gobble}%
\providecommand \bibinfo  [0]{\@secondoftwo}%
\providecommand \bibfield  [0]{\@secondoftwo}%
\providecommand \translation [1]{[#1]}%
\providecommand \BibitemOpen [0]{}%
\providecommand \bibitemStop [0]{}%
\providecommand \bibitemNoStop [0]{.\EOS\space}%
\providecommand \EOS [0]{\spacefactor3000\relax}%
\providecommand \BibitemShut  [1]{\csname bibitem#1\endcsname}%
\let\auto@bib@innerbib\@empty
\bibitem [{\citenamefont {Jastrow}(1955)}]{Jastrow19551479}%
  \BibitemOpen
  \bibfield  {author} {\bibinfo {author} {\bibfnamefont {R.}~\bibnamefont
  {Jastrow}},\ }\href {\doibase 10.1103/PhysRev.98.1479} {\bibfield  {journal}
  {\bibinfo  {journal} {Phys. Rev.}\ }\textbf {\bibinfo {volume} {98}},\
  \bibinfo {pages} {1479} (\bibinfo {year} {1955})}\BibitemShut {NoStop}%
\bibitem [{\citenamefont {Capello}\ \emph {et~al.}(2005)\citenamefont
  {Capello}, \citenamefont {Becca}, \citenamefont {Fabrizio}, \citenamefont
  {Sorella},\ and\ \citenamefont {Tosatti}}]{Capello2005026406}%
  \BibitemOpen
  \bibfield  {author} {\bibinfo {author} {\bibfnamefont {M.}~\bibnamefont
  {Capello}}, \bibinfo {author} {\bibfnamefont {F.}~\bibnamefont {Becca}},
  \bibinfo {author} {\bibfnamefont {M.}~\bibnamefont {Fabrizio}}, \bibinfo
  {author} {\bibfnamefont {S.}~\bibnamefont {Sorella}}, \ and\ \bibinfo
  {author} {\bibfnamefont {E.}~\bibnamefont {Tosatti}},\ }\href@noop {}
  {\bibfield  {journal} {\bibinfo  {journal} {Phys. Rev. Lett.}\ }\textbf
  {\bibinfo {volume} {94}},\ \bibinfo {pages} {026406} (\bibinfo {year}
  {2005})}\BibitemShut {NoStop}%
\bibitem [{\citenamefont {Bartlett}\ \emph {et~al.}(1989)\citenamefont
  {Bartlett}, \citenamefont {Kucharski},\ and\ \citenamefont
  {Noga}}]{Bartlett1989133}%
  \BibitemOpen
  \bibfield  {author} {\bibinfo {author} {\bibfnamefont {R.~J.}\ \bibnamefont
  {Bartlett}}, \bibinfo {author} {\bibfnamefont {S.~A.}\ \bibnamefont
  {Kucharski}}, \ and\ \bibinfo {author} {\bibfnamefont {J.}~\bibnamefont
  {Noga}},\ }\href@noop {} {\bibfield  {journal} {\bibinfo  {journal} {Chemical
  Physics Letters}\ }\textbf {\bibinfo {volume} {155}},\ \bibinfo {pages} {133}
  (\bibinfo {year} {1989})}\BibitemShut {NoStop}%
\bibitem [{\citenamefont {Kutzelnigg}(1991)}]{Kutzelnigg1991349}%
  \BibitemOpen
  \bibfield  {author} {\bibinfo {author} {\bibfnamefont {W.}~\bibnamefont
  {Kutzelnigg}},\ }\href@noop {} {\bibfield  {journal} {\bibinfo  {journal}
  {Theoretica Chimica Acta}\ }\textbf {\bibinfo {volume} {80}},\ \bibinfo
  {pages} {349} (\bibinfo {year} {1991})}\BibitemShut {NoStop}%
\bibitem [{\citenamefont {Taube}\ and\ \citenamefont
  {Bartlett}(2006)}]{Taube20063393}%
  \BibitemOpen
  \bibfield  {author} {\bibinfo {author} {\bibfnamefont {A.~G.}\ \bibnamefont
  {Taube}}\ and\ \bibinfo {author} {\bibfnamefont {R.~J.}\ \bibnamefont
  {Bartlett}},\ }\href@noop {} {\bibfield  {journal} {\bibinfo  {journal}
  {International Journal Of Quantum Chemistry}\ }\textbf {\bibinfo {volume}
  {106}},\ \bibinfo {pages} {3393} (\bibinfo {year} {2006})}\BibitemShut
  {NoStop}%
\bibitem [{\citenamefont {Trotter}(1959)}]{Trotter1959545}%
  \BibitemOpen
  \bibfield  {author} {\bibinfo {author} {\bibfnamefont {H.~F.}\ \bibnamefont
  {Trotter}},\ }\href@noop {} {\bibfield  {journal} {\bibinfo  {journal}
  {Proceedings of the American Mathematical Society}\ }\textbf {\bibinfo
  {volume} {10}},\ \bibinfo {pages} {545} (\bibinfo {year} {1959})}\BibitemShut
  {NoStop}%
\bibitem [{\citenamefont {Suzuki}(1976)}]{Suzuki1976183}%
  \BibitemOpen
  \bibfield  {author} {\bibinfo {author} {\bibfnamefont {M.}~\bibnamefont
  {Suzuki}},\ }\href@noop {} {\bibfield  {journal} {\bibinfo  {journal}
  {Communications In Mathematical Physics}\ }\textbf {\bibinfo {volume} {51}},\
  \bibinfo {pages} {183} (\bibinfo {year} {1976})}\BibitemShut {NoStop}%
\bibitem [{\citenamefont {Suzuki}(1993)}]{Suzuki1993432}%
  \BibitemOpen
  \bibfield  {author} {\bibinfo {author} {\bibfnamefont {M.}~\bibnamefont
  {Suzuki}},\ }\href@noop {} {\bibfield  {journal} {\bibinfo  {journal}
  {Physica A-statistical Mechanics And Its Applications}\ }\textbf {\bibinfo
  {volume} {194}},\ \bibinfo {pages} {432} (\bibinfo {year}
  {1993})}\BibitemShut {NoStop}%
\bibitem [{\citenamefont {Hubbard}(1959)}]{Hubbard195977}%
  \BibitemOpen
  \bibfield  {author} {\bibinfo {author} {\bibfnamefont {J.}~\bibnamefont
  {Hubbard}},\ }\href@noop {} {\bibfield  {journal} {\bibinfo  {journal} {Phys.
  Rev. Lett.}\ }\textbf {\bibinfo {volume} {3}},\ \bibinfo {pages} {77}
  (\bibinfo {year} {1959})}\BibitemShut {NoStop}%
\bibitem [{\citenamefont {Blankenbecler}\ \emph {et~al.}(1981)\citenamefont
  {Blankenbecler}, \citenamefont {Scalapino},\ and\ \citenamefont
  {Sugar}}]{Blankenbecler19812278}%
  \BibitemOpen
  \bibfield  {author} {\bibinfo {author} {\bibfnamefont {R.}~\bibnamefont
  {Blankenbecler}}, \bibinfo {author} {\bibfnamefont {D.~J.}\ \bibnamefont
  {Scalapino}}, \ and\ \bibinfo {author} {\bibfnamefont {R.~L.}\ \bibnamefont
  {Sugar}},\ }\href@noop {} {\bibfield  {journal} {\bibinfo  {journal}
  {Physical Review D}\ }\textbf {\bibinfo {volume} {24}},\ \bibinfo {pages}
  {2278} (\bibinfo {year} {1981})}\BibitemShut {NoStop}%
\bibitem [{\citenamefont {Gutzwiller}(1963)}]{Gutzwiller1963159}%
  \BibitemOpen
  \bibfield  {author} {\bibinfo {author} {\bibfnamefont {M.~C.}\ \bibnamefont
  {Gutzwiller}},\ }\href@noop {} {\bibfield  {journal} {\bibinfo  {journal}
  {Phys. Rev. Lett.}\ }\textbf {\bibinfo {volume} {10}},\ \bibinfo {pages}
  {159} (\bibinfo {year} {1963})}\BibitemShut {NoStop}%
\bibitem [{\citenamefont {Gutzwiller}(1964)}]{Gutzwiller1964923}%
  \BibitemOpen
  \bibfield  {author} {\bibinfo {author} {\bibfnamefont {M.~C.}\ \bibnamefont
  {Gutzwiller}},\ }\href@noop {} {\bibfield  {journal} {\bibinfo  {journal}
  {Physical Review}\ }\textbf {\bibinfo {volume} {134}},\ \bibinfo {pages}
  {923} (\bibinfo {year} {1964})}\BibitemShut {NoStop}%
\bibitem [{\citenamefont {Gutzwiller}(1965)}]{Gutzwiller19651726}%
  \BibitemOpen
  \bibfield  {author} {\bibinfo {author} {\bibfnamefont {M.~C.}\ \bibnamefont
  {Gutzwiller}},\ }\href@noop {} {\bibfield  {journal} {\bibinfo  {journal}
  {Physical Review}\ }\textbf {\bibinfo {volume} {137}},\ \bibinfo {pages}
  {1726} (\bibinfo {year} {1965})}\BibitemShut {NoStop}%
\bibitem [{\citenamefont {Metzner}\ and\ \citenamefont
  {Vollhardt}(1987)}]{Metzner1987121}%
  \BibitemOpen
  \bibfield  {author} {\bibinfo {author} {\bibfnamefont {W.}~\bibnamefont
  {Metzner}}\ and\ \bibinfo {author} {\bibfnamefont {D.}~\bibnamefont
  {Vollhardt}},\ }\href {\doibase 10.1103/PhysRevLett.59.121} {\bibfield
  {journal} {\bibinfo  {journal} {Phys. Rev. Lett.}\ }\textbf {\bibinfo
  {volume} {59}},\ \bibinfo {pages} {121} (\bibinfo {year} {1987})}\BibitemShut
  {NoStop}%
\bibitem [{\citenamefont {Metzner}\ and\ \citenamefont
  {Vollhardt}(1988)}]{Metzner19887382}%
  \BibitemOpen
  \bibfield  {author} {\bibinfo {author} {\bibfnamefont {W.}~\bibnamefont
  {Metzner}}\ and\ \bibinfo {author} {\bibfnamefont {D.}~\bibnamefont
  {Vollhardt}},\ }\href@noop {} {\bibfield  {journal} {\bibinfo  {journal}
  {Phys. Rev. B}\ }\textbf {\bibinfo {volume} {37}},\ \bibinfo {pages} {7382}
  (\bibinfo {year} {1988})}\BibitemShut {NoStop}%
\bibitem [{\citenamefont {Metzner}\ and\ \citenamefont
  {Vollhardt}(1989)}]{Metzner1989324}%
  \BibitemOpen
  \bibfield  {author} {\bibinfo {author} {\bibfnamefont {W.}~\bibnamefont
  {Metzner}}\ and\ \bibinfo {author} {\bibfnamefont {D.}~\bibnamefont
  {Vollhardt}},\ }\href {\doibase 10.1103/PhysRevLett.62.324} {\bibfield
  {journal} {\bibinfo  {journal} {Phys. Rev. Lett.}\ }\textbf {\bibinfo
  {volume} {62}},\ \bibinfo {pages} {324} (\bibinfo {year} {1989})}\BibitemShut
  {NoStop}%
\bibitem [{\citenamefont {Bunemann}\ \emph {et~al.}(1997)\citenamefont
  {Bunemann}, \citenamefont {Gebhard},\ and\ \citenamefont
  {Weber}}]{Bunemann19977343}%
  \BibitemOpen
  \bibfield  {author} {\bibinfo {author} {\bibfnamefont {J.}~\bibnamefont
  {Bunemann}}, \bibinfo {author} {\bibfnamefont {F.}~\bibnamefont {Gebhard}}, \
  and\ \bibinfo {author} {\bibfnamefont {W.}~\bibnamefont {Weber}},\
  }\href@noop {} {\bibfield  {journal} {\bibinfo  {journal} {Journal Of
  Physics-condensed Matter}\ }\textbf {\bibinfo {volume} {9}},\ \bibinfo
  {pages} {7343} (\bibinfo {year} {1997})}\BibitemShut {NoStop}%
\bibitem [{\citenamefont {Otsuka}(1992)}]{Otsuka19921645}%
  \BibitemOpen
  \bibfield  {author} {\bibinfo {author} {\bibfnamefont {H.}~\bibnamefont
  {Otsuka}},\ }\href@noop {} {\bibfield  {journal} {\bibinfo  {journal} {J.
  Phys. Soc. Jpn.}\ }\textbf {\bibinfo {volume} {61}},\ \bibinfo {pages} {1645}
  (\bibinfo {year} {1992})}\BibitemShut {NoStop}%
\bibitem [{\citenamefont {Dzierzawa}\ \emph {et~al.}(1995)\citenamefont
  {Dzierzawa}, \citenamefont {Baeriswyl},\ and\ \citenamefont
  {Distasio}}]{Dzierzawa19951993}%
  \BibitemOpen
  \bibfield  {author} {\bibinfo {author} {\bibfnamefont {M.}~\bibnamefont
  {Dzierzawa}}, \bibinfo {author} {\bibfnamefont {D.}~\bibnamefont
  {Baeriswyl}}, \ and\ \bibinfo {author} {\bibfnamefont {M.}~\bibnamefont
  {Distasio}},\ }\href@noop {} {\bibfield  {journal} {\bibinfo  {journal}
  {Phys. Rev. B}\ }\textbf {\bibinfo {volume} {51}},\ \bibinfo {pages} {1993}
  (\bibinfo {year} {1995})}\BibitemShut {NoStop}%
\bibitem [{\citenamefont {Cheng}\ and\ \citenamefont
  {Marianetti}(2020{\natexlab{a}})}]{Cheng2020short}%
  \BibitemOpen
  \bibfield  {author} {\bibinfo {author} {\bibfnamefont {Z.}~\bibnamefont
  {Cheng}}\ and\ \bibinfo {author} {\bibfnamefont {C.~A.}\ \bibnamefont
  {Marianetti}},\ }\href@noop {} {\bibfield  {journal} {\bibinfo  {journal}
  {Phys. Rev. Lett. [jointly submitted with current manuscript]}\ }\textbf
  {\bibinfo {volume} {0}},\ \bibinfo {pages} {0} (\bibinfo {year}
  {2020}{\natexlab{a}})}\BibitemShut {NoStop}%
\bibitem [{\citenamefont {Georges}\ \emph {et~al.}(1996)\citenamefont
  {Georges}, \citenamefont {Kotliar}, \citenamefont {Krauth},\ and\
  \citenamefont {Rozenberg}}]{Georges199613}%
  \BibitemOpen
  \bibfield  {author} {\bibinfo {author} {\bibfnamefont {A.}~\bibnamefont
  {Georges}}, \bibinfo {author} {\bibfnamefont {G.}~\bibnamefont {Kotliar}},
  \bibinfo {author} {\bibfnamefont {W.}~\bibnamefont {Krauth}}, \ and\ \bibinfo
  {author} {\bibfnamefont {M.~J.}\ \bibnamefont {Rozenberg}},\ }\href@noop {}
  {\bibfield  {journal} {\bibinfo  {journal} {Rev. Mod. Phys.}\ }\textbf
  {\bibinfo {volume} {68}},\ \bibinfo {pages} {13} (\bibinfo {year}
  {1996})}\BibitemShut {NoStop}%
\bibitem [{\citenamefont {Kotliar}\ and\ \citenamefont
  {Vollhardt}(2004)}]{Kotliar200453}%
  \BibitemOpen
  \bibfield  {author} {\bibinfo {author} {\bibfnamefont {G.}~\bibnamefont
  {Kotliar}}\ and\ \bibinfo {author} {\bibfnamefont {D.}~\bibnamefont
  {Vollhardt}},\ }\href@noop {} {\bibfield  {journal} {\bibinfo  {journal}
  {Physics Today}\ }\textbf {\bibinfo {volume} {57}},\ \bibinfo {pages} {53}
  (\bibinfo {year} {2004})}\BibitemShut {NoStop}%
\bibitem [{\citenamefont {Vollhardt}(2012)}]{Vollhardt20121}%
  \BibitemOpen
  \bibfield  {author} {\bibinfo {author} {\bibfnamefont {D.}~\bibnamefont
  {Vollhardt}},\ }\href@noop {} {\bibfield  {journal} {\bibinfo  {journal}
  {Annalen Der Physik}\ }\textbf {\bibinfo {volume} {524}},\ \bibinfo {pages}
  {1} (\bibinfo {year} {2012})}\BibitemShut {NoStop}%
\bibitem [{\citenamefont {Kotliar}\ \emph {et~al.}(2006)\citenamefont
  {Kotliar}, \citenamefont {Savrasov}, \citenamefont {Haule}, \citenamefont
  {Oudovenko}, \citenamefont {Parcollet},\ and\ \citenamefont
  {Marianetti}}]{Kotliar2006865}%
  \BibitemOpen
  \bibfield  {author} {\bibinfo {author} {\bibfnamefont {G.}~\bibnamefont
  {Kotliar}}, \bibinfo {author} {\bibfnamefont {S.~Y.}\ \bibnamefont
  {Savrasov}}, \bibinfo {author} {\bibfnamefont {K.}~\bibnamefont {Haule}},
  \bibinfo {author} {\bibfnamefont {V.~S.}\ \bibnamefont {Oudovenko}}, \bibinfo
  {author} {\bibfnamefont {O.}~\bibnamefont {Parcollet}}, \ and\ \bibinfo
  {author} {\bibfnamefont {C.~A.}\ \bibnamefont {Marianetti}},\ }\href@noop {}
  {\bibfield  {journal} {\bibinfo  {journal} {Rev. Mod. Phys.}\ }\textbf
  {\bibinfo {volume} {78}},\ \bibinfo {pages} {865} (\bibinfo {year}
  {2006})}\BibitemShut {NoStop}%
\bibitem [{\citenamefont {Wang}\ \emph {et~al.}(2010)\citenamefont {Wang},
  \citenamefont {He}, \citenamefont {Wang}, \citenamefont {Wang}, \citenamefont
  {Wang},\ and\ \citenamefont {Zhang}}]{Wang2010125105}%
  \BibitemOpen
  \bibfield  {author} {\bibinfo {author} {\bibfnamefont {W.~S.}\ \bibnamefont
  {Wang}}, \bibinfo {author} {\bibfnamefont {X.~M.}\ \bibnamefont {He}},
  \bibinfo {author} {\bibfnamefont {D.}~\bibnamefont {Wang}}, \bibinfo {author}
  {\bibfnamefont {Q.~H.}\ \bibnamefont {Wang}}, \bibinfo {author}
  {\bibfnamefont {Z.~D.}\ \bibnamefont {Wang}}, \ and\ \bibinfo {author}
  {\bibfnamefont {F.~C.}\ \bibnamefont {Zhang}},\ }\href@noop {} {\bibfield
  {journal} {\bibinfo  {journal} {Phys. Rev. B}\ }\textbf {\bibinfo {volume}
  {82}},\ \bibinfo {pages} {125105} (\bibinfo {year} {2010})}\BibitemShut
  {NoStop}%
\bibitem [{\citenamefont {Sandri}\ \emph {et~al.}(2013)\citenamefont {Sandri},
  \citenamefont {Capone},\ and\ \citenamefont {Fabrizio}}]{Sandri2013205108}%
  \BibitemOpen
  \bibfield  {author} {\bibinfo {author} {\bibfnamefont {M.}~\bibnamefont
  {Sandri}}, \bibinfo {author} {\bibfnamefont {M.}~\bibnamefont {Capone}}, \
  and\ \bibinfo {author} {\bibfnamefont {M.}~\bibnamefont {Fabrizio}},\
  }\href@noop {} {\bibfield  {journal} {\bibinfo  {journal} {Phys. Rev. B}\
  }\textbf {\bibinfo {volume} {87}},\ \bibinfo {pages} {205108} (\bibinfo
  {year} {2013})}\BibitemShut {NoStop}%
\bibitem [{\citenamefont {Lanata}\ \emph {et~al.}(2015)\citenamefont {Lanata},
  \citenamefont {Deng},\ and\ \citenamefont {Kotliar}}]{Lanata2015081108}%
  \BibitemOpen
  \bibfield  {author} {\bibinfo {author} {\bibfnamefont {N.}~\bibnamefont
  {Lanata}}, \bibinfo {author} {\bibfnamefont {X.~Y.}\ \bibnamefont {Deng}}, \
  and\ \bibinfo {author} {\bibfnamefont {G.}~\bibnamefont {Kotliar}},\
  }\href@noop {} {\bibfield  {journal} {\bibinfo  {journal} {Phys. Rev. B}\
  }\textbf {\bibinfo {volume} {92}},\ \bibinfo {pages} {081108} (\bibinfo
  {year} {2015})}\BibitemShut {NoStop}%
\bibitem [{\citenamefont {Bunemann}\ \emph {et~al.}(2003)\citenamefont
  {Bunemann}, \citenamefont {Gebhard},\ and\ \citenamefont
  {Thul}}]{Bunemann2003075103}%
  \BibitemOpen
  \bibfield  {author} {\bibinfo {author} {\bibfnamefont {J.}~\bibnamefont
  {Bunemann}}, \bibinfo {author} {\bibfnamefont {F.}~\bibnamefont {Gebhard}}, \
  and\ \bibinfo {author} {\bibfnamefont {R.}~\bibnamefont {Thul}},\ }\href@noop
  {} {\bibfield  {journal} {\bibinfo  {journal} {Physical Review B}\ }\textbf
  {\bibinfo {volume} {67}},\ \bibinfo {pages} {075103} (\bibinfo {year}
  {2003})}\BibitemShut {NoStop}%
\bibitem [{sup()}]{supplementary}%
  \BibitemOpen
  \href@noop {} {}\bibinfo {note} {See Supplemental Material at [URL will be
  inserted by publisher] for analytic results of the CDA for one orbital at
  $\discn=3$.}\BibitemShut {Stop}%
\bibitem [{\citenamefont {Hirsch}(1983)}]{Hirsch19834059}%
  \BibitemOpen
  \bibfield  {author} {\bibinfo {author} {\bibfnamefont {J.~E.}\ \bibnamefont
  {Hirsch}},\ }\href@noop {} {\bibfield  {journal} {\bibinfo  {journal} {Phys.
  Rev. B}\ }\textbf {\bibinfo {volume} {28}},\ \bibinfo {pages} {4059}
  (\bibinfo {year} {1983})}\BibitemShut {NoStop}%
\bibitem [{\citenamefont {Hirsch}\ and\ \citenamefont
  {Fye}(1986)}]{Hirsch19862521}%
  \BibitemOpen
  \bibfield  {author} {\bibinfo {author} {\bibfnamefont {J.~E.}\ \bibnamefont
  {Hirsch}}\ and\ \bibinfo {author} {\bibfnamefont {R.~M.}\ \bibnamefont
  {Fye}},\ }\href@noop {} {\bibfield  {journal} {\bibinfo  {journal} {Phys.
  Rev. Lett.}\ }\textbf {\bibinfo {volume} {56}},\ \bibinfo {pages} {2521}
  (\bibinfo {year} {1986})}\BibitemShut {NoStop}%
\bibitem [{bae()}]{baeriswylisdif}%
  \BibitemOpen
  \href@noop {} {}\bibinfo {note} {Baeriswyl \emph{et al.} have also presented
  pedagogical results for the Hubbard plaquette using what is labeled as the
  Gutwziller wavefunction\cite{Baeriswyl2019235152}. In this specific case,
  they choose an unprojected wavefunction which is a mixture of several Slater
  determinants; in contrast to usual applications.}\BibitemShut {Stop}%
\bibitem [{\citenamefont {Cheng}\ and\ \citenamefont
  {Marianetti}(2020{\natexlab{b}})}]{Cheng2020081105}%
  \BibitemOpen
  \bibfield  {author} {\bibinfo {author} {\bibfnamefont {Z.}~\bibnamefont
  {Cheng}}\ and\ \bibinfo {author} {\bibfnamefont {C.~A.}\ \bibnamefont
  {Marianetti}},\ }\href {\doibase 10.1103/PhysRevB.101.081105} {\bibfield
  {journal} {\bibinfo  {journal} {Phys. Rev. B}\ }\textbf {\bibinfo {volume}
  {101}},\ \bibinfo {pages} {081105} (\bibinfo {year}
  {2020}{\natexlab{b}})}\BibitemShut {NoStop}%
\bibitem [{\citenamefont {Fazekas}\ and\ \citenamefont
  {Penc}(1988)}]{fazekas19881021}%
  \BibitemOpen
  \bibfield  {author} {\bibinfo {author} {\bibfnamefont {P.}~\bibnamefont
  {Fazekas}}\ and\ \bibinfo {author} {\bibfnamefont {K.}~\bibnamefont {Penc}},\
  }\href@noop {} {\bibfield  {journal} {\bibinfo  {journal} {International
  Journal of Modern Physics B}\ }\textbf {\bibinfo {volume} {2}},\ \bibinfo
  {pages} {1021} (\bibinfo {year} {1988})}\BibitemShut {NoStop}%
\bibitem [{\citenamefont {Yokoyama}\ and\ \citenamefont
  {Shiba}(1990)}]{Yokoyama19903669}%
  \BibitemOpen
  \bibfield  {author} {\bibinfo {author} {\bibfnamefont {H.}~\bibnamefont
  {Yokoyama}}\ and\ \bibinfo {author} {\bibfnamefont {H.}~\bibnamefont
  {Shiba}},\ }\href@noop {} {\bibfield  {journal} {\bibinfo  {journal} {J.
  Phys. Soc. Jpn.}\ }\textbf {\bibinfo {volume} {59}},\ \bibinfo {pages} {3669}
  (\bibinfo {year} {1990})}\BibitemShut {NoStop}%
\bibitem [{\citenamefont {Grimsley}\ \emph {et~al.}(2019)\citenamefont
  {Grimsley}, \citenamefont {Economou}, \citenamefont {Barnes},\ and\
  \citenamefont {Mayhall}}]{Grimsley20193007}%
  \BibitemOpen
  \bibfield  {author} {\bibinfo {author} {\bibfnamefont {H.~R.}\ \bibnamefont
  {Grimsley}}, \bibinfo {author} {\bibfnamefont {S.~E.}\ \bibnamefont
  {Economou}}, \bibinfo {author} {\bibfnamefont {E.}~\bibnamefont {Barnes}}, \
  and\ \bibinfo {author} {\bibfnamefont {N.~J.}\ \bibnamefont {Mayhall}},\
  }\href@noop {} {\bibfield  {journal} {\bibinfo  {journal} {Nature
  Communications}\ }\textbf {\bibinfo {volume} {10}},\ \bibinfo {pages} {3007}
  (\bibinfo {year} {2019})}\BibitemShut {NoStop}%
\bibitem [{\citenamefont {Bartlett}\ and\ \citenamefont
  {Noga}(1988)}]{Bartlett198829}%
  \BibitemOpen
  \bibfield  {author} {\bibinfo {author} {\bibfnamefont {R.~J.}\ \bibnamefont
  {Bartlett}}\ and\ \bibinfo {author} {\bibfnamefont {J.}~\bibnamefont
  {Noga}},\ }\href@noop {} {\bibfield  {journal} {\bibinfo  {journal} {Chemical
  Physics Letters}\ }\textbf {\bibinfo {volume} {150}},\ \bibinfo {pages} {29}
  (\bibinfo {year} {1988})}\BibitemShut {NoStop}%
\bibitem [{\citenamefont {Ceperley}\ \emph {et~al.}(1977)\citenamefont
  {Ceperley}, \citenamefont {Chester},\ and\ \citenamefont
  {Kalos}}]{Ceperley19773081}%
  \BibitemOpen
  \bibfield  {author} {\bibinfo {author} {\bibfnamefont {D.}~\bibnamefont
  {Ceperley}}, \bibinfo {author} {\bibfnamefont {G.~V.}\ \bibnamefont
  {Chester}}, \ and\ \bibinfo {author} {\bibfnamefont {M.~H.}\ \bibnamefont
  {Kalos}},\ }\href@noop {} {\bibfield  {journal} {\bibinfo  {journal}
  {Physical Review B}\ }\textbf {\bibinfo {volume} {16}},\ \bibinfo {pages}
  {3081} (\bibinfo {year} {1977})}\BibitemShut {NoStop}%
\bibitem [{\citenamefont {Sorella}(2001)}]{Sorella2001024512}%
  \BibitemOpen
  \bibfield  {author} {\bibinfo {author} {\bibfnamefont {S.}~\bibnamefont
  {Sorella}},\ }\href@noop {} {\bibfield  {journal} {\bibinfo  {journal} {Phys.
  Rev. B}\ }\textbf {\bibinfo {volume} {64}},\ \bibinfo {pages} {024512}
  (\bibinfo {year} {2001})}\BibitemShut {NoStop}%
\bibitem [{\citenamefont {Sorella}(2005)}]{Sorella2005241103}%
  \BibitemOpen
  \bibfield  {author} {\bibinfo {author} {\bibfnamefont {S.}~\bibnamefont
  {Sorella}},\ }\href@noop {} {\bibfield  {journal} {\bibinfo  {journal} {Phys.
  Rev. B}\ }\textbf {\bibinfo {volume} {71}},\ \bibinfo {pages} {241103}
  (\bibinfo {year} {2005})}\BibitemShut {NoStop}%
\bibitem [{\citenamefont {Baeriswyl}(1987)}]{Baeriswyl19870}%
  \BibitemOpen
  \bibfield  {author} {\bibinfo {author} {\bibfnamefont {D.}~\bibnamefont
  {Baeriswyl}},\ }in\ \href@noop {} {\emph {\bibinfo {booktitle} {Nonlinearity
  in Condensed Matter}}},\ \bibinfo {editor} {edited by\ \bibinfo {editor}
  {\bibfnamefont {A.~R.}\ \bibnamefont {Bishop}}, \bibinfo {editor}
  {\bibfnamefont {D.~K.}\ \bibnamefont {Campbell}}, \ and\ \bibinfo {editor}
  {\bibfnamefont {S.~E.}\ \bibnamefont {Trullinger}}}\ (\bibinfo  {publisher}
  {Springer-Verlag, Berlin},\ \bibinfo {year} {1987})\ \bibinfo {edition}
  {1st}\ ed.,\ pp.\ \bibinfo {pages} {183--193}\BibitemShut {NoStop}%
\bibitem [{\citenamefont {Farhi}\ \emph {et~al.}(2014)\citenamefont {Farhi},
  \citenamefont {Goldstone},\ and\ \citenamefont {Gutmann}}]{Farhi1411.4028}%
  \BibitemOpen
  \bibfield  {author} {\bibinfo {author} {\bibfnamefont {E.}~\bibnamefont
  {Farhi}}, \bibinfo {author} {\bibfnamefont {J.}~\bibnamefont {Goldstone}}, \
  and\ \bibinfo {author} {\bibfnamefont {S.}~\bibnamefont {Gutmann}},\
  }\href@noop {} {\bibfield  {journal} {\bibinfo  {journal} {arXiv:1411.4028}\
  } (\bibinfo {year} {2014})}\BibitemShut {NoStop}%
\bibitem [{\citenamefont {Wecker}\ \emph {et~al.}(2015)\citenamefont {Wecker},
  \citenamefont {Hastings},\ and\ \citenamefont {Troyer}}]{Wecker2015042303}%
  \BibitemOpen
  \bibfield  {author} {\bibinfo {author} {\bibfnamefont {D.}~\bibnamefont
  {Wecker}}, \bibinfo {author} {\bibfnamefont {M.~B.}\ \bibnamefont
  {Hastings}}, \ and\ \bibinfo {author} {\bibfnamefont {M.}~\bibnamefont
  {Troyer}},\ }\href@noop {} {\bibfield  {journal} {\bibinfo  {journal}
  {Physical Review A}\ }\textbf {\bibinfo {volume} {92}},\ \bibinfo {pages}
  {042303} (\bibinfo {year} {2015})}\BibitemShut {NoStop}%
\bibitem [{\citenamefont {Mahan}(2000)}]{Mahan20000306463385}%
  \BibitemOpen
  \bibfield  {author} {\bibinfo {author} {\bibfnamefont {G.~D.}\ \bibnamefont
  {Mahan}},\ }\href {https://www.xarg.org/ref/a/0306463385/} {\emph {\bibinfo
  {title} {Many-Particle Physics (Physics of Solids and Liquids)}}}\ (\bibinfo
  {publisher} {Springer},\ \bibinfo {year} {2000})\BibitemShut {NoStop}%
\bibitem [{\citenamefont {Baeriswyl}(2019)}]{Baeriswyl2019235152}%
  \BibitemOpen
  \bibfield  {author} {\bibinfo {author} {\bibfnamefont {D.}~\bibnamefont
  {Baeriswyl}},\ }\href@noop {} {\bibfield  {journal} {\bibinfo  {journal}
  {Phys. Rev. B}\ }\textbf {\bibinfo {volume} {99}},\ \bibinfo {pages} {235152}
  (\bibinfo {year} {2019})}\BibitemShut {NoStop}%
\bibitem [{\citenamefont {Barcza}\ \emph {et~al.}(2019)\citenamefont {Barcza},
  \citenamefont {Gebhard}, \citenamefont {Linneweber},\ and\ \citenamefont
  {Legeza}}]{Barcza2019165130}%
  \BibitemOpen
  \bibfield  {author} {\bibinfo {author} {\bibfnamefont {G.}~\bibnamefont
  {Barcza}}, \bibinfo {author} {\bibfnamefont {F.}~\bibnamefont {Gebhard}},
  \bibinfo {author} {\bibfnamefont {T.}~\bibnamefont {Linneweber}}, \ and\
  \bibinfo {author} {\bibfnamefont {O.}~\bibnamefont {Legeza}},\ }\href@noop {}
  {\bibfield  {journal} {\bibinfo  {journal} {Phys. Rev. B}\ }\textbf {\bibinfo
  {volume} {99}},\ \bibinfo {pages} {165130} (\bibinfo {year}
  {2019})}\BibitemShut {NoStop}%
\bibitem [{\citenamefont {Harsha}\ \emph {et~al.}(2018)\citenamefont {Harsha},
  \citenamefont {Shiozaki},\ and\ \citenamefont {Scuseria}}]{Harsha2018044107}%
  \BibitemOpen
  \bibfield  {author} {\bibinfo {author} {\bibfnamefont {G.}~\bibnamefont
  {Harsha}}, \bibinfo {author} {\bibfnamefont {T.}~\bibnamefont {Shiozaki}}, \
  and\ \bibinfo {author} {\bibfnamefont {G.~E.}\ \bibnamefont {Scuseria}},\
  }\href@noop {} {\bibfield  {journal} {\bibinfo  {journal} {Journal Of
  Chemical Physics}\ }\textbf {\bibinfo {volume} {148}},\ \bibinfo {pages}
  {044107} (\bibinfo {year} {2018})}\BibitemShut {NoStop}%
\bibitem [{\citenamefont {Negele}\ and\ \citenamefont
  {Orland}(1998)}]{Negele19980738200522}%
  \BibitemOpen
  \bibfield  {author} {\bibinfo {author} {\bibfnamefont {J.~W.}\ \bibnamefont
  {Negele}}\ and\ \bibinfo {author} {\bibfnamefont {H.}~\bibnamefont
  {Orland}},\ }\href@noop {} {\emph {\bibinfo {title} {Quantum Many-particle
  Systems}}}\ (\bibinfo  {publisher} {Perseus Books},\ \bibinfo {year}
  {1998})\BibitemShut {NoStop}%
\bibitem [{\citenamefont {Lanata}\ \emph {et~al.}(2012)\citenamefont {Lanata},
  \citenamefont {Strand}, \citenamefont {Dai},\ and\ \citenamefont
  {Hellsing}}]{Lanata2012035133}%
  \BibitemOpen
  \bibfield  {author} {\bibinfo {author} {\bibfnamefont {N.}~\bibnamefont
  {Lanata}}, \bibinfo {author} {\bibfnamefont {H.}~\bibnamefont {Strand}},
  \bibinfo {author} {\bibfnamefont {X.}~\bibnamefont {Dai}}, \ and\ \bibinfo
  {author} {\bibfnamefont {B.}~\bibnamefont {Hellsing}},\ }\href@noop {}
  {\bibfield  {journal} {\bibinfo  {journal} {Phys. Rev. B}\ }\textbf {\bibinfo
  {volume} {85}},\ \bibinfo {pages} {035133} (\bibinfo {year}
  {2012})}\BibitemShut {NoStop}%
\bibitem [{\citenamefont {Brinkman}\ and\ \citenamefont
  {Rice}(1970)}]{Brinkman19704302}%
  \BibitemOpen
  \bibfield  {author} {\bibinfo {author} {\bibfnamefont {W.~F.}\ \bibnamefont
  {Brinkman}}\ and\ \bibinfo {author} {\bibfnamefont {T.~M.}\ \bibnamefont
  {Rice}},\ }\href@noop {} {\bibfield  {journal} {\bibinfo  {journal} {Physical
  Review B-solid State}\ }\textbf {\bibinfo {volume} {2}},\ \bibinfo {pages}
  {4302} (\bibinfo {year} {1970})}\BibitemShut {NoStop}%
\bibitem [{\citenamefont {Bunemann}\ \emph {et~al.}(1998)\citenamefont
  {Bunemann}, \citenamefont {Weber},\ and\ \citenamefont
  {Gebhard}}]{Bunemann19986896}%
  \BibitemOpen
  \bibfield  {author} {\bibinfo {author} {\bibfnamefont {J.}~\bibnamefont
  {Bunemann}}, \bibinfo {author} {\bibfnamefont {W.}~\bibnamefont {Weber}}, \
  and\ \bibinfo {author} {\bibfnamefont {F.}~\bibnamefont {Gebhard}},\
  }\href@noop {} {\bibfield  {journal} {\bibinfo  {journal} {Phys. Rev. B}\
  }\textbf {\bibinfo {volume} {57}},\ \bibinfo {pages} {6896} (\bibinfo {year}
  {1998})}\BibitemShut {NoStop}%
\bibitem [{\citenamefont {Bunemann}\ and\ \citenamefont
  {Gebhard}(2007)}]{Bunemann2007193104}%
  \BibitemOpen
  \bibfield  {author} {\bibinfo {author} {\bibfnamefont {J.}~\bibnamefont
  {Bunemann}}\ and\ \bibinfo {author} {\bibfnamefont {F.}~\bibnamefont
  {Gebhard}},\ }\href@noop {} {\bibfield  {journal} {\bibinfo  {journal} {Phys.
  Rev. B}\ }\textbf {\bibinfo {volume} {76}},\ \bibinfo {pages} {193104}
  (\bibinfo {year} {2007})}\BibitemShut {NoStop}%
\bibitem [{\citenamefont {Bunemann}\ \emph {et~al.}(2012)\citenamefont
  {Bunemann}, \citenamefont {Gebhard}, \citenamefont {Schickling},\ and\
  \citenamefont {Weber}}]{Bunemann20121282}%
  \BibitemOpen
  \bibfield  {author} {\bibinfo {author} {\bibfnamefont {J.}~\bibnamefont
  {Bunemann}}, \bibinfo {author} {\bibfnamefont {F.}~\bibnamefont {Gebhard}},
  \bibinfo {author} {\bibfnamefont {T.}~\bibnamefont {Schickling}}, \ and\
  \bibinfo {author} {\bibfnamefont {W.}~\bibnamefont {Weber}},\ }\href@noop {}
  {\bibfield  {journal} {\bibinfo  {journal} {Physica Status Solidi B-basic
  Solid State Physics}\ }\textbf {\bibinfo {volume} {249}},\ \bibinfo {pages}
  {1282} (\bibinfo {year} {2012})}\BibitemShut {NoStop}%
\bibitem [{\citenamefont {Aryasetiawan}\ and\ \citenamefont
  {Gunnarsson}(1998)}]{Aryasetiawan1998237}%
  \BibitemOpen
  \bibfield  {author} {\bibinfo {author} {\bibfnamefont {F.}~\bibnamefont
  {Aryasetiawan}}\ and\ \bibinfo {author} {\bibfnamefont {O.}~\bibnamefont
  {Gunnarsson}},\ }\href@noop {} {\bibfield  {journal} {\bibinfo  {journal}
  {Reports On Progress In Physics}\ }\textbf {\bibinfo {volume} {61}},\
  \bibinfo {pages} {237} (\bibinfo {year} {1998})}\BibitemShut {NoStop}%
\bibitem [{\citenamefont {Baeriswyl}\ \emph {et~al.}(2009)\citenamefont
  {Baeriswyl}, \citenamefont {Eichenberger},\ and\ \citenamefont
  {Menteshashvili}}]{Baeriswyl2009075010}%
  \BibitemOpen
  \bibfield  {author} {\bibinfo {author} {\bibfnamefont {D.}~\bibnamefont
  {Baeriswyl}}, \bibinfo {author} {\bibfnamefont {D.}~\bibnamefont
  {Eichenberger}}, \ and\ \bibinfo {author} {\bibfnamefont {M.}~\bibnamefont
  {Menteshashvili}},\ }\href@noop {} {\bibfield  {journal} {\bibinfo  {journal}
  {New Journal Of Physics}\ }\textbf {\bibinfo {volume} {11}},\ \bibinfo
  {pages} {075010} (\bibinfo {year} {2009})}\BibitemShut {NoStop}%
\bibitem [{\citenamefont {Deng}\ \emph {et~al.}(2009)\citenamefont {Deng},
  \citenamefont {Wang}, \citenamefont {Dai},\ and\ \citenamefont
  {Fang}}]{Deng2009075114}%
  \BibitemOpen
  \bibfield  {author} {\bibinfo {author} {\bibfnamefont {X.~Y.}\ \bibnamefont
  {Deng}}, \bibinfo {author} {\bibfnamefont {L.}~\bibnamefont {Wang}}, \bibinfo
  {author} {\bibfnamefont {X.}~\bibnamefont {Dai}}, \ and\ \bibinfo {author}
  {\bibfnamefont {Z.}~\bibnamefont {Fang}},\ }\href@noop {} {\bibfield
  {journal} {\bibinfo  {journal} {Phys. Rev. B}\ }\textbf {\bibinfo {volume}
  {79}},\ \bibinfo {pages} {075114} (\bibinfo {year} {2009})}\BibitemShut
  {NoStop}%
\bibitem [{\citenamefont {Wybourne}(1973)}]{Wybourne19731117}%
  \BibitemOpen
  \bibfield  {author} {\bibinfo {author} {\bibfnamefont {B.~G.}\ \bibnamefont
  {Wybourne}},\ }\href@noop {} {\bibfield  {journal} {\bibinfo  {journal}
  {International Journal Of Quantum Chemistry}\ }\textbf {\bibinfo {volume}
  {7}},\ \bibinfo {pages} {1117} (\bibinfo {year} {1973})}\BibitemShut
  {NoStop}%
\bibitem [{\citenamefont {Fukutome}\ \emph {et~al.}(1977)\citenamefont
  {Fukutome}, \citenamefont {Yamamura},\ and\ \citenamefont
  {Nishiyama}}]{Fukutome19771554}%
  \BibitemOpen
  \bibfield  {author} {\bibinfo {author} {\bibfnamefont {H.}~\bibnamefont
  {Fukutome}}, \bibinfo {author} {\bibfnamefont {M.}~\bibnamefont {Yamamura}},
  \ and\ \bibinfo {author} {\bibfnamefont {S.}~\bibnamefont {Nishiyama}},\
  }\href@noop {} {\bibfield  {journal} {\bibinfo  {journal} {Progress Of
  Theoretical Physics}\ }\textbf {\bibinfo {volume} {57}},\ \bibinfo {pages}
  {1554} (\bibinfo {year} {1977})}\BibitemShut {NoStop}%
\end{thebibliography}
%

\end{document}